\documentclass[useAMS,usenatbib,usegraphicx]{mn2e}

\newcommand{\shortitle}{Spatially resolved IFS abundance analysis of NGC\,628}

\usepackage{amssymb}
\usepackage[ps2pdf,bookmarksopen,breaklinks=true]{hyperref}

\hypersetup{citecolor={black},linkcolor={black},urlcolor={black}}  

\hypersetup{
  colorlinks=true,             
  bookmarksnumbered=true,      
  bookmarksopen=false,         
  pdfpagemode=UseOutlines,     
  pdfdisplaydoctitle,          
  pdftitle={\shortitle}
}

\newcommand{\hh}{H\,{\footnotesize II}~}
\newcommand{\hei}{He\,{\footnotesize I}}
\newcommand{\nii}{[N\,{\footnotesize II}]}
\newcommand{\sii}{[S\,{\footnotesize II}]}

\newcommand{\oii}{[O\,{\footnotesize II}]}
\newcommand{\oiii}{[O\,{\footnotesize III}]}
\newcommand{\lam}{$\lambda$}

\newcommand{\ha}{H$\alpha$} 
\newcommand{\hb}{H$\beta$} 
\newcommand{\hg}{H$\gamma$} 
 
\newcommand{\flux}{erg\,s$^{-1}$\,cm$^{-2}$}
\newcommand{\fluxA}{erg\,s$^{-1}$\,cm$^{-2}$\,\AA$^{-1}$}

\newcommand{\degree}{\ensuremath{^\circ}}

\newcommand{\ngc}{NGC\,628}
\newcommand{\fxf}{{\em fibre-by-fibre}}

\defcitealias{RosalesOrtega:2010p3836}{Ros10}
\defcitealias{Sanchez:2011p3844}{Paper~I}
\defcitealias{Fathi:2007p2409}{Fathi07}
\defcitealias{Kobulnicky:2004p1700}{KK04}
\defcitealias{Pettini:2004p315}{PP04}

\defcitealias{Talent:1983p3838}{Tal83}
\defcitealias{McCall:1985p1243}{MRS85}
\defcitealias{Ferguson:1998p224}{FGW98}
\defcitealias{vanZee:1998p3468}{vZ98}
\defcitealias{Bresolin:1999p3837}{BKG99}
\defcitealias{Castellanos:2002p3372}{CDT02}
\defcitealias{McGaugh:1991p314}{M91}

\title[\shortitle]
{
PPAK Wide-field Integral Field Spectroscopy of NGC\,628: 
II. Emission line abundance analysis\thanks{Based on observations collected 
  at the Centro Astron\'omico Hispano Alem\'an (CAHA) at Calar Alto, operated
  jointly by the Max-Planck Institut f\"ur Astronomie and the Instituto de
  Astrof\'isica de Andaluc\'ia (CSIC).}
}

\author[Rosales-Ortega, D\'iaz, Kennicutt \& S\'anchez]
{F.~F.~Rosales-Ortega,$^{1,2}$\thanks{E-mail: frosales@cantab.net}
A.~I.~D\'iaz,$^{2}$ 
R.~C.~Kennicutt$^{1}$
and S.~F.~S\'anchez$^{3}$\\ 
$^{1}$Institute of Astronomy, University of Cambridge, Madingley Road, Cambridge
CB3 0HA, UK.\\
$^{2}$Departamento de F\'isica Te\'orica, Universidad Aut\'onoma de Madrid,
28049 Madrid, Spain.\\
$^{3}$Centro Astron\'omico Hispano Alem\'an, Calar Alto, CSIC-MPG,
  C/Jes\'us Durb\'an Rem\'on 2-2, E-04004 Almeria, Spain.
}

\begin{document}


\date{}

\pagerange{\pageref{firstpage}--\pageref{lastpage}} \pubyear{2010}

\maketitle

\label{firstpage}

\begin{abstract}


In this second paper of the series, we present the 2-dimensional (2D) emission
line abundance analysis of \ngc, the largest object within the PPAK Integral
Field Spectroscopy (IFS) Nearby Galaxies Survey: PINGS.
We introduce the methodology applied to the 2D IFS data in order to extract and
deal with large spectral samples, from which a 2D abundance analysis can be later
performed.
We obtain the most complete and reliable abundance gradient of the galaxy
up-to-date, by using the largest number of spectroscopic points sampled in the
galaxy, and by comparing the statistical significance of different strong-line
metallicity indicators.
We find features not previously reported for this galaxy that imply a
multi-modality of the abundance gradient consistent with a nearly
flat-distribution in the innermost regions of the
galaxy, a steep negative gradient along the disc and a shallow gradient or
nearly-constant metallicity beyond the optical edge of the galaxy. The N/O
ratio seems to follow the same radial behaviour.
We demonstrate that the observed dispersion in metallicity shows no systematic
dependence with the spatial position, signal-to-noise or ionization
conditions, implying that the scatter in abundance for a given radius is
reflecting a true spatial physical variation of the oxygen content.
Furthermore, by exploiting the 2D IFS data, we were able to construct the 2D
metallicity structure of the galaxy, detecting regions of metal enhancement, and
showing that they vary depending on the choice of the metallicity
estimator. The analysis of axisymmetric variations in the disc of \ngc\ suggest
that the physical conditions and the star formation history of
different-symmetric regions of the galaxy have evolved in a different manner.

\end{abstract}


\section{Introduction}
\label{sec:intro}

Nebular emission lines from bright-individual \hh regions have been,
historically, the main tool at our disposal for the direct measurement of the
gas-phase abundance at discrete spatial positions in low redshift galaxies.
A good understanding on the distribution of the chemical abundance across the
surface of nearby galaxies (or through the comparison between them) it is
necessary in order to place observational-based constrains on theories of galactic
chemical evolution, while at the same time, derive accurate star formation
histories and obtain information on the stellar nucleosynthesis of normal spiral
galaxies.

Several factors dictate the chemical evolution in a galaxy, including the
primordial composition, the content and distribution of molecular and neutral
gas, the star formation history (SFH), feedback, the transport and mixing of
gas, the initial mass function (IMF), etc. All these ingredients contribute
through a complex process to the evolutionary histories of the stars and the
galaxies in general. Accurate measurements of the present chemical abundance
constrain the different possible evolutionary scenarios, and thus the importance
of determining the elemental composition in a global approach, among different
galaxy types.

Previous spectroscopic studies have unveil some aspects of the complex processes
at play between the chemical abundances of galaxies and their physical
properties. Although these studies have been successful by determining important
relationships, scaling laws and systematic patterns
(e.g. luminosity-metallicity, mass-metallicity, and surface brightness
vs. metallicity relations
\citealt{Skillman:1989p1592,VilaCostas:1992p322,Zaritsky:1994p333,Tremonti:2004p1138}; 
effective yield vs. luminosity and circular velocity relations
\citealt{Garnett:2002p339}; abundance gradients and the effective radius
of disks \citealt{Diaz:1989p3307}; systematic differences in the gas-phase
abundance gradients between normal and barred spirals
\citealt{Zaritsky:1994p333,Martin:1994p1602}; characteristic vs. integrated
abundances \citealt{Moustakas:2006p313}; etc.), they have been limited by the
number of objects sampled, the number of \hh regions observed and the coverage
of these regions within the galaxy surface.

In order to tackle the problem of the 2D spectroscopic coverage of the whole
galaxy surface, we devised the PPAK Integral-field-spectroscopy Nearby Galaxies
Survey: PINGS \citep[][hereafter Ros10]{RosalesOrtega:2010p3836}, an IFS
survey of nearby ($<$\,100 Mpc) well-resolved spiral galaxies. PINGS was
specially designed to obtain complete maps of the emission-line abundances,
stellar populations, and reddening using an IFS mosaicking {\em imaging},
which takes advantage of what is currently one of the world's widest
field-of-view (FOV) integral field unit (IFU).

\ngc\ (Messier\,74) is a close, bright, grand-design spiral galaxy which
has been extensively studied. With a projected optical size of
10.5\,$\times$\,9.5 arcmin, it is the most extended object of the PINGS
sample (see \autoref{tab:properties}). With such dimensions, this galaxy allow us to study the 2D metallicity
structure of the disk and the second order properties of its abundance
distribution.
In the first paper of this series \citep[][hereafter Paper~I]{Sanchez:2011p3844},
we present a study of the line emission and stellar continuum of \ngc\ by
means of pixel-resolved maps across the disk of the galaxy. This study
included a qualitative description of the 2D distribution of the physical
properties inferred from the line intensity maps, and a comparison of these
properties with both the integrated spectrum of the galaxy and the spatially
resolved spectra.
In this second article, we present a detailed, spatially-resolved
spectroscopic abundance analysis, based on different spectral samples extracted
from the area covered by the IFS observations of \ngc. Particular attention
is paid to the spectra selection technique, considering several quality factors
and performing sanity check tests to the different methods employed, after which
we define a spectra selection methodology specially conceived for the study of
the nebular emission in this galaxy, with a potential application to any
IFU-based spectroscopic observation. This allow us to investigate
the small and intermediate scale-size variation in line emission, and to
derive the gas chemistry distribution across the surface of the galaxy with
unprecedented detail.

The paper is organised as follows: In \autoref{sec:obs} we give an overview of
the observations and data reduction, including a description of the technique
implemented in order to decouple the nebular emission from the observed
spectra. In \autoref{sec:2d} we introduce the methodology applied to the IFS
data in order to extract distinct spectral samples from which a 2D abundance
analysis can be performed. Diagnostic diagrams and radial trends of different
physical parameters are derived in these
section, comparing the results between the different sets of IFS data, and with
previously published results, when appropriate. 
In \autoref{sec:grad} we perform an abundance analysis based on the spectra
samples extracted in the previous section, obtaining the abundance gradient of
the galaxy, comparing different strong-line metallicity indicators, and
discussing their statistical significance. 
In \autoref{sec:angle} we explore variations on the nebular properties across
the surface of the galaxy, related to geometrical and morphological features.
Finally, in \autoref{sec:fin} we present a summary of the main results of the
article.

\section{Observations and data reduction}
\label{sec:obs}

The PINGS observations for \ngc\ were carried out at the 3.5m telescope of
the Calar Alto observatory with the Potsdam Multi Aperture Spectrograph, PMAS,
\citep{Roth:2005p2463} in the PPAK mode
\citep{Verheijen:2004p2481,Kelz:2006p3341}. 
With a field-of-view of 74''$\times$65'', and a filling factor of
$\sim$\,65\%, PPAK is currently one of the widest IFU available worldwide. 
The PPAK fibre bundle consists of 331 science fibres of 2.7 arcsec diameter
each, within a single hexagonal pattern. The sky background is sampled by 36
additional fibres, distributed in 6 bundles of 6 fibres each, positioned
along a circle $\sim$\,90 arcsec from the center of the instrument
FOV. Additionally, 15 fibres are used for internal calibration purposes.

Due to the large size of \ngc\ compared to the FOV of the instrument, a
mosaicking scheme was adopted. The observations of the galaxy spanned a period
of three-years (from 2006 to 2009), with a total of six observing nights and
34 different pointings. The central position was observed in dithering mode to
gain spatial resolution, while the remaining 33 positions were observed
without dithering due to the large size of the mosaic. The V300 grating was
used for all the observations, covering the wavelength range $\sim$\,3700-7100
\AA, with a spectral resolution of $FWHM\sim$\,8 \AA. For the central position
we obtained 2 exposures of 600\,s for each dithering pointing, while for the
rest of the tiles we collected 3 exposures of 600s per pointing.
At least two different spectrophotometric stars per observing night were
observed during the runs in order to perform flux calibration. In addition to
the science pointings, sky exposures of 300s were taken each night in order to
perform a proper subtraction of the sky contribution. Additional details on
the observing strategy can be found in \citetalias{RosalesOrtega:2010p3836}
and \citetalias{Sanchez:2011p3844}.

\begin{figure}
  \includegraphics[width=\hsize]{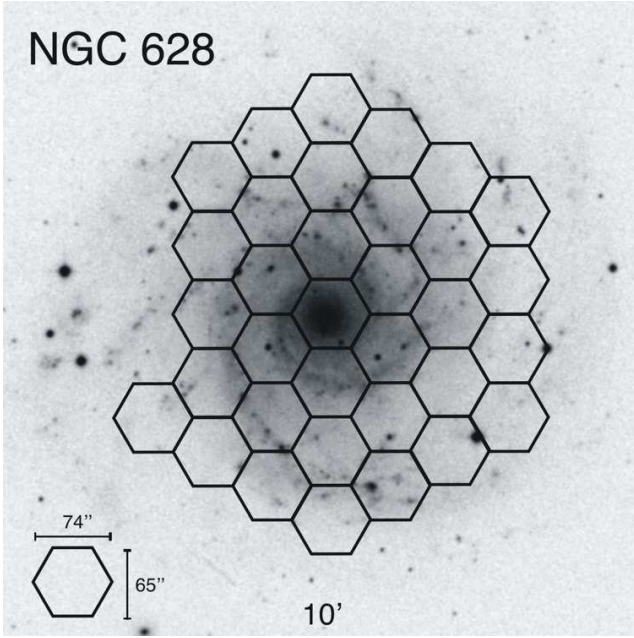}
  \caption[PPAK mosaicking of NGC\,628]
  {
    Area covered by the IFS observations of \ngc. The figure shows a
    $B$-band Digital Sky Survey image of the galaxy with the PPAK mosaic pointings 
    as overlaid hexagons indicating the field-of-view of the
    central fibre-bundle. The image is displayed in top-north, left-east
    standard configuration.
    \label{fig:mosaic}
  }
\end{figure}

\autoref{fig:mosaic} displays a Digital Sky Survey\footnote{The Digitized Sky
  Survey was produced at the Space Telescope Science
  Institute under U.S. Government grant NAG W-2166. The images of these
  surveys are based on photographic data obtained using the Oschin Schmidt
  Telescope on Palomar Mountain and the UK Schmidt Telescope. The plates
  were processed into the present compressed digital form with the
  permission of these institutions.}
image of \ngc, showing the mosaic pattern covered by the IFS observations,
consisting in a central position and consecutive hexagonal concentric
rings. The spectroscopic mosaic contains 11094 individual spectra (considering
overlapping and repeated exposures), the area covered by all the
observed positions accounts approximately for 34 arcmin$^2$, making NGC\,628
the largest area ever covered by a IFU mosaicking so far.

The reduction of the PINGS observations for \ngc\ followed, in
general, the standard steps for fibre-based integral field
spectroscopy. However, given that the observations of individual pointings were
performed on different nights, during different years, with dissimilar
atmospheric conditions, and slightly differing instrument configurations, the
construction of the \ngc\ IFS mosaic required further reduction steps than
for a single, standard IFU observation.
The basic data reduction steps applied to the IFS mosaic of \ngc\ can be
summarised as follows: a) Pre-reduction. b) Identification of the location of
the spectra on the detector. c) Extraction of each individual spectrum. d)
Distortion correction of the extracted spectra. e) Application of wavelength
solution. f) Fibre-to-fibre transmission correction. g) Flux calibration. h)
Allocation of the spectra to the sky position.

Data reduction was performed using {\sc r3d} \citep{Sanchez:2006p331}, in
combination with {\sc iraf}\footnote{IRAF is distributed by the National
  Optical Astronomy Observatories, which are operated by the Association of
  Universities for Research in Astronomy, Inc., under cooperative agreement
  with the National Science Foundation.} packages and {\sc e3d} 
\citep{Sanchez:2004p2632}.
A master bias frame was created by averaging all the bias frames observed
during an observing night and subtracted from the science frames.
The science exposures acquired at the same position on the sky were combined
and the cosmic rays were removed during this process. The location of the
spectra on the CCD were determined using a continuum illuminated exposure
taken before the science exposures, each spectrum was then extracted from the
science frames using an iterative Gaussian-suppression technique which reduces
the effects of the cross-talk to negligible levels
\citep{Sanchez:2006p331,Sanchez:2011p3844}. The extracted flux for each spectrum was
stored in a row-stacked-spectra (RSS) file 
\citep{Sanchez:2004p2632}. Wavelength calibration was performed using HeHgCd
lamp exposures obtained before and after each pointing. Differences in the
fibre-to-fibre transmission throughput were corrected by creating a master
fibre-flat from twilight skyflat exposures taken in every run.

Sky-subtraction was then performed to the extracted, distortion/transmission
corrected and wavelength calibrated spectra. Different sky subtraction schemes
were applied depending on the location of the pointing within the IFS mosaic. 
By construction, in most of the \ngc\ pointings, the sky-fibres of the PPAK
instrument are located within an area containing significant signal from the
galaxy. In those cases, supplementary sky exposures were obtained applying
large offsets from the observed positions. In some other cases, a sufficient
number of sky-fibres were located in regions free from galaxy emission, where 
an accurate sky subtraction of the individual pointings using these sky
spectra was possible.
Relative flux calibration was attained by applying to the science
frames a series of night sensitivity curves, obtained by comparing the
observed flux with calibrated spectrophotometric standard spectra
(reduced following the basic procedure described above), considering
the filling factor of the fibre-bundle, corrections for the airmass and the
optical extinction due to the atmosphere as a function of wavelength.
In order to obtain the most accurate absolute spectrophotometric calibration,
an additional correction was performed by comparing the IFS data with imaging
photometry of the $B$, $V$, $R$ and \ha\ images from the SINGS legacy
survey \citep{Kennicutt:2003p1560}. The estimated spectrophotometric accuracy
of the IFS mosaic is of the order of 0.2 mag. Furthermore, during this
re-normalization process, the astrometry accuracy of the IFS mosaic was
corrected to a $\sim$\,0.3 arcsec level. A complete explanation of the data
processing and IFS mosaic creation is beyond the scope of this paper, but the
reader will find a detailed description of the PINGS reduction process in
\citetalias{RosalesOrtega:2010p3836}, and in particular for the PPAK-IFS
survey of \ngc\ in \citetalias{Sanchez:2011p3844}.

\begin{table}
  \begin{center}
    \caption[Galaxy properties]
    {
      Galaxy properties. Notes:
      (1) Morphological type from the R3C
      catalog \citep{deVaucouleurs:1991p2393}.
      (2) Adopted distance in Mpc; Reference: \citet{Hendry:2005p2408}.
      (3) Projected size, major and minor axes at the $B_{25}$ mag
      arcsec$^{-2}$ from R3C.
      (4) Absolute $B$-band magnitude calculated from the apparent magnitude
      listed in the R3C catalog and the adopted distance to the system. 
      (5) Reference: \citet{Lu:1993p2436}.
      (6) Heliocentric velocities calculated from $v=zc$, with no further
      correction applied.
      (7) Galaxy inclination angle based on the $B_{25}$ mag arcsec$^{-2}$
      from R3C.
      (8) Galaxy position angle, measured positive NE, in the $B_{25}$ mag
      arcsec$^{-2}$.
      (9) Optical radius at the $B_{25}$ mag arcsec$^{-2}$ from R3C.
      (10) Optical radius in kpc assuming the adopted distance to the galaxy.
      \label{tab:properties}
    }
    \begin{tabular}{@{\extracolsep{\fill}} lrc }\hline
      \\[-5pt]
      Property & \ngc\ & Note \\[3pt]\hline
      \\[-5pt]
      Type\,\dotfill                     &  SA(s)c & 1  \\[2pt]
      Adopted D (Mpc)\,\dotfill          &     9.3 & 2   \\[2pt]
      Projected size (arcmin)\,\dotfill  &  10.5\,$\times$\,9.5 & 3 \\[2pt]
      $M_B$\,\dotfill                   &   -19.9 & 4 \\[2pt]
      Redshift\,\dotfill                 & 0.00219 & 5 \\[2pt]
      $V_{\odot}$ (km~s$^{-1}$)\,\dotfill  &     657  & 6 \\[2pt]
      Inclination $i$ (degrees)\,\dotfill&     24  & 7 \\[2pt]
      P.A. (degrees)\,\dotfill           &     25  & 8  \\[2pt]
      $\rho_{25}$ (arcmin)\,\dotfill     &   5.23   & 9  \\[2pt]
      $\rho_{25}$ (kpc)\,\dotfill        &   14.1   & 10  \\[2pt]
      \hline
    \end{tabular}
  \end{center}
\end{table}

The final product of the reduction process for the IFS mosaic of NGC\,628
consist in a RSS file containing 11094 wavelength and flux calibrated spectra,
together with an ASCII position table file, which allocates each individual
spectra to a sky position within the mosaic. However, many regions of the
mosaic present a very low level of signal or do not contain signal at all
(i.e. spectra with a flat continuum consistent with zero-flux intensity). The
reason being that in those regions, the fibres were sampling areas where the
intrinsic flux of the galaxy is low or null (e.g. borders of the mosaic,
intra-arms regions, etc). In order to get rid of spectra where no information
can be derived, we obtained a {\em clean} version of the IFS mosaic of
\ngc\ by applying a flux threshold cut choosing only those fibres with an
average flux along the whole spectral range greater than 10$^{-16}$
\fluxA. Furthermore, bad fibres (due to cosmic rays and CCD cosmetic defects)
and foreground stars (10 within the observed field-of-view of NGC\,628) where
removed from the mosaic. This procedure resulted in a refined mosaic version
which includes only those regions with high-quality spectrophotometric
calibration. The total number of spectra in the {\em clean} IFS mosaic version
of \ngc\ is 6949.

The spectra of the {\em clean} mosaic of \ngc\ consist in a combination of 
continua arising from the different stellar populations, and emission of the
ionized gas present in the interstellar medium of the galaxy.
As the present study is focused on the spectroscopic properties of the
gas-phase of \ngc, the emission lines of the ionized gas were decouple from
the stellar population in each individual spectrum of the IFS mosaic by using
population synthesis to model and subtract the stellar continuum underlying
the nebular emission lines. By applying this technique, the emission-line
measurements are corrected (to a first-order) for stellar absorption.
A thorough explanation of the stellar continuum fitting process (including a
detailed description of the algorithms adopted, estimates of the accuracy of
the SSP-based modeling and the derived parameters based on simulations) can be
found in \citetalias{Sanchez:2011p3844}, here we only present a simplified scheme
describing how the stellar populations and the emission lines in the IFS
mosaic of \ngc\ were decoupled. First, a set of emission lines is identified
from a strong \hh region of the galaxy. For each spectrum in the data
set, the underlying stellar population is fitted by a linear combination of a
grid of Single Stellar Populations (SSP), after correcting for the appropriate
systemic velocity and velocity dispersion, masking all the nebular and sky
emission lines. The template models are selected in order to cover the widest
possible range of ages and metallicities. We consider the effects of dust
extinction by varying A$_V$ from 0 to 1 mag at 0.2 mag steps. We subtract
the fit stellar population from the original spectrum to get a residual pure
emission-line spectrum.

A great deal of debate is found in the literature regarding the drawbacks of the
SSP fitting technique discussed above, especially considering the well-known
degeneracies found in the combination of SSPs \citepalias[see Appendix of][and
references therein]{Sanchez:2011p3844}. However, for this particular
analysis, the only requirement is that that fit model can follow accurately
the underlying continuum in order to decouple it from the emission lines
produced by the ionized gas. Therefore, even in the case that the combination
of SSPs is strongly degenerate, and the created model has no physical meaning,
the latter can still be useful for this specific purpose.
As a result of the SSP fitting procedure described before, we obtain two
additional RSS files: one containing the continuum model fit to each
individual spectra, and other containing the {\em residual} dataset of
gas-free spectra.

Individual emission-line fluxes were measured in each spectrum of the 
{\em residual} RSS mosaic by considering spectral window regions of $\sim$ 200\,\AA. 
We performed a simultaneous multi-component fitting using a single Gaussian
function (for each emission line contained within each window) plus a low
order polynomial (to describe the local continuum and to simplify the fitting
procedure) using {\sc fit3d} \citep{Sanchez:2006p3300}.
The central redshifted wavelengths of the emission lines were fixed and since
the FWHM is dominated by the spectral resolution, the widths of all the lines
were set equal to the width of the brightest line in this spectral region.
Line intensity fluxes were then measured by integrating the observed intensity of
each line. The statistical uncertainty in the measurement of the line flux was
calculated by propagating the error associated to the multi-component fitting
and considering the signal-to-noise of the spectral region.

The final result of all the procedures described above consist of a
set of emission line intensities (and associated errors), per each of the 6949
individual spectra of the {\em residual} mosaic of \ngc. The volume of this
spectroscopic information is quite large, considering the classic view in
which emission-line abundance studies were performed based on a few points
across the surface of the galaxy. The following section describes the
methodology implemented in order to extract meaningful information from this IFS
database, which is then used to perform the 2D abundance analysis of \ngc.

\section{2D spectroscopic analysis of NGC\,628}
\label{sec:2d}

The nearly full IFS coverage of \ngc\ offers the possibility to undertake a
detailed, spatially resolved, spectroscopic analysis of this galaxy based on
thousands of individual spectra (within the limitation of the spatial resolution
of the instrument).
A spectroscopic analysis based on the emission line maps of \ngc\ is presented
in \citetalias{Sanchez:2011p3844}, where the 2D distribution of the physical
properties of the galaxy were studied. However, the conclusions raised from
these maps are based on general trends and depend, to a certain level, on the
interpolation scheme applied in order to derived the pixel-resolved maps.

Classical spectroscopy on this object has typically targeted a handful of
bright individual \hh regions in the galaxy
\citep[e.g.][]{McCall:1985p1243,vanZee:1998p3468,Ferguson:1998p224,Castellanos:2002p3372},
especially at the outer regions and along the spiral arms, where (in general)
the contribution of the stellar population to the observed spectrum is not
significant. However, the spectroscopic dataset presented in this series poses a
challenge with respect to classical spectroscopy, as a right methodology has to be
found in order to handle and analyse --in a homogeneous and meaningful way--
this large spectroscopic database.
In this section we consider two different spectra extraction techniques for
the IFS analysis of \ngc, they take into account the signal-to-noise of the
data, the 2D spatial coverage, the physical sense of the derived results, and
the final number of analysed spectra. In the first one, we consider all the
fibres within the {\em residual} mosaic of the galaxy (i.e. using the maximum
spatial resolution available), while in the second one we define ``classical''
\hh regions by applying an aperture extraction on morphologically-linked
emitting regions. A similar spectroscopic analysis is performed to both
samples, a comparison of the results from the different analyses is also
presented, together with an adopted final methodology.

\subsection{Method I: {\sc fibre-by-fibre analysis}}
\label{sec:fxf}

\begin{figure*}
  \centering
    \includegraphics[width=0.95\textwidth]{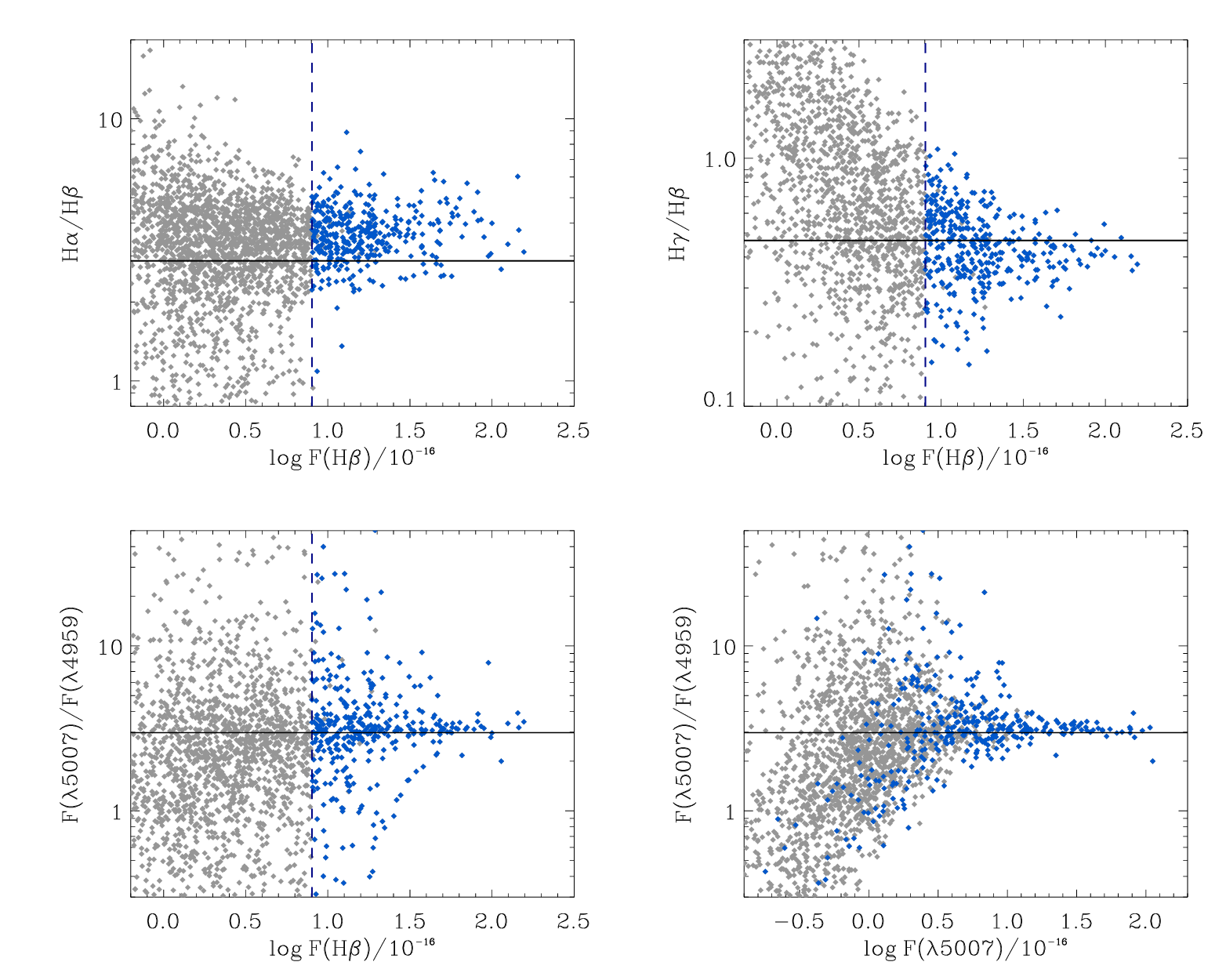}
    \caption[Method I: {\em signal-to-noise} diagrams of NGC\,628]
    {
      \ha/\hb, H$\gamma$/\hb\ and \oiii\ \lam5007/\lam4959 ratios as a
      function of observed flux in \hb\ and F(\lam5007) for the
      \fxf\ selection of \ngc. 
      Grey and blue symbols (in the online version) correspond to the 
      {\em diffuse} and \fxf\ samples, after the 2$^{\rm nd}$ (b) and 
      3$^{\rm rd}$ (c) selection criteria respectively, as described in the
      text. 
      The horizontal lines correspond to the theoretical values for each
      ratio. The vertical dashed line stands for the flux threshold in \hb.
      \label{fig:n628_ratios}
    }
\end{figure*}

The first of the explored analysis methods considers that the 2''.7 aperture of
a single PPAK fibre would sample --in principle-- a large-enough region in
physical scale to subtend a small \hh region and/or a fraction of a larger one
at the adopted distance to the galaxy.
With this assumption as a premise, the method considers every single
fibre of the IFS mosaic as the source of an individual {\em analysable}
spectrum. This method will be referred as the \fxf\ analysis. In the case of
\ngc, at the assumed luminosity distance ($D_{\rm L}$ = 9.3 Mpc), one arcsec
would correspond to a linear scale of $\sim$ 45\,pc, assuming a standard
$\Lambda$CDM cosmology (WMAP 5-years results: $H_0=70.5$, $\Omega=0.27$,
$\Lambda=0.73$, \citealt{Hinshaw:2009p3469}). This linear physical scale implies
that the fibre diameter of PPAK samples $\sim$120\,pc on \ngc, i.e. a region
from which, in principle, one would expect enough signal-to-noise in the
observed spectrum. This scale can be compared to the physical diameter of
a well-known \hh region in our Galaxy, i.e. the Orion nebula ($D \sim$ 8\,pc),
or to the extend of what are considered prototypes of extragalactic giant \hh
regions, such as 30 Doradus ($D \sim$ 200\,pc) or NGC\,604 ($D \sim$ 460\,pc).
Therefore, in the case of the \fxf\ method, the area sampled by an individual
fibre would subtend a fraction of a typical giant \hh region in \ngc, but the
same area would fully encompass a number of small and medium size \hh regions
of the galaxy.

The spectra extraction process for the \fxf\ method was based on the 
{\em residual} RSS file obtained after the {\em clean} mosaic version of the
galaxy, as explained in \autoref{sec:obs}. In the case of \ngc, this mosaic
corresponds to 6949 fibres, i.e. 51\% of the total number of originally
observed fibres. As it can be expected, not all the fibres in this mosaic have
spectra with enough signal-to-noise and/or the right number of detectable
emission lines in order to derive meaningful physical parameters. Following
the experience of \citetalias{Sanchez:2011p3844}, we applied a flux threshold
cut based on the \hb\ line intensity, together with some additional
conditions. As the main focus is to characterise the chemical abundance of
the galaxy, we required the presence in the spectrum of the typical emission
lines from which we could carry out an abundance analysis in the individual
fibres.

The sample selection was split in three different steps, the details of the
criteria conditions in each step and their actual implementation in the IFS
data can be found in Appendix~\ref{app:methods}, they can be summarised as
follows: a) We selected those fibres where the \hb\ and \oiii\
\lam\lam4959,\,5007 emission lines were detected (i.e. line intensities
greater than zero); 
b) We obtain a subsample based on the previous
selection in which the logarithmic extinction coefficient c(\hb) value
(calculated from the \ha/\hb\ ratio, accordingly to the
prescriptions describe in \citetalias{Sanchez:2011p3844})
was a finite-floating number, the total number of fibres
fulfilling the two previous criteria was 2562, i.e. 37\% of the fibres
contained in the {\em clean} mosaic and 23\% of the original number of fibres
in the observed, unprocessed mosaic; 
c) Finally, we selected those fibres where the line
intensity of the \hb\ line was greater than or equal to a given flux limit
threshold, {\em and} the line intensity of \oii\ \lam3727 was greater than zero,
i.e. the emission line was detected. The reasons for dividing the extraction
procedure in these steps are explained in Appendix~\ref{app:methods}, but the
main intention was to extract data sets that could be analysed independently, as
they could potentially trace regions of line emission with different physical
properties, as explained thereinafter.

The value of the \hb\ flux threshold was chosen considering a trade-off between
several factors: 1) The final number of spectra after the flux cut was
applied; 2) The ``quality'' of the spectra as shown by certain line flux
ratios and test diagrams (see below); and 3) The position of the derived spectra
in some of the most common emission-line diagnostic diagrams. 
In the case of \ngc, the flux limit applied in \hb\ was equal
to 8 $\times$ 10$^{-16}$ \flux. The final number of spectra in the \fxf\
sample was 376 fibres, i.e. $\sim$ 6\% of the number of fibres in the 
{\em clean} mosaic and $\sim$ 3\% of the original fibres in the full \ngc\
mosaic.

\begin{figure}
  \centering
    \includegraphics[width=\hsize]{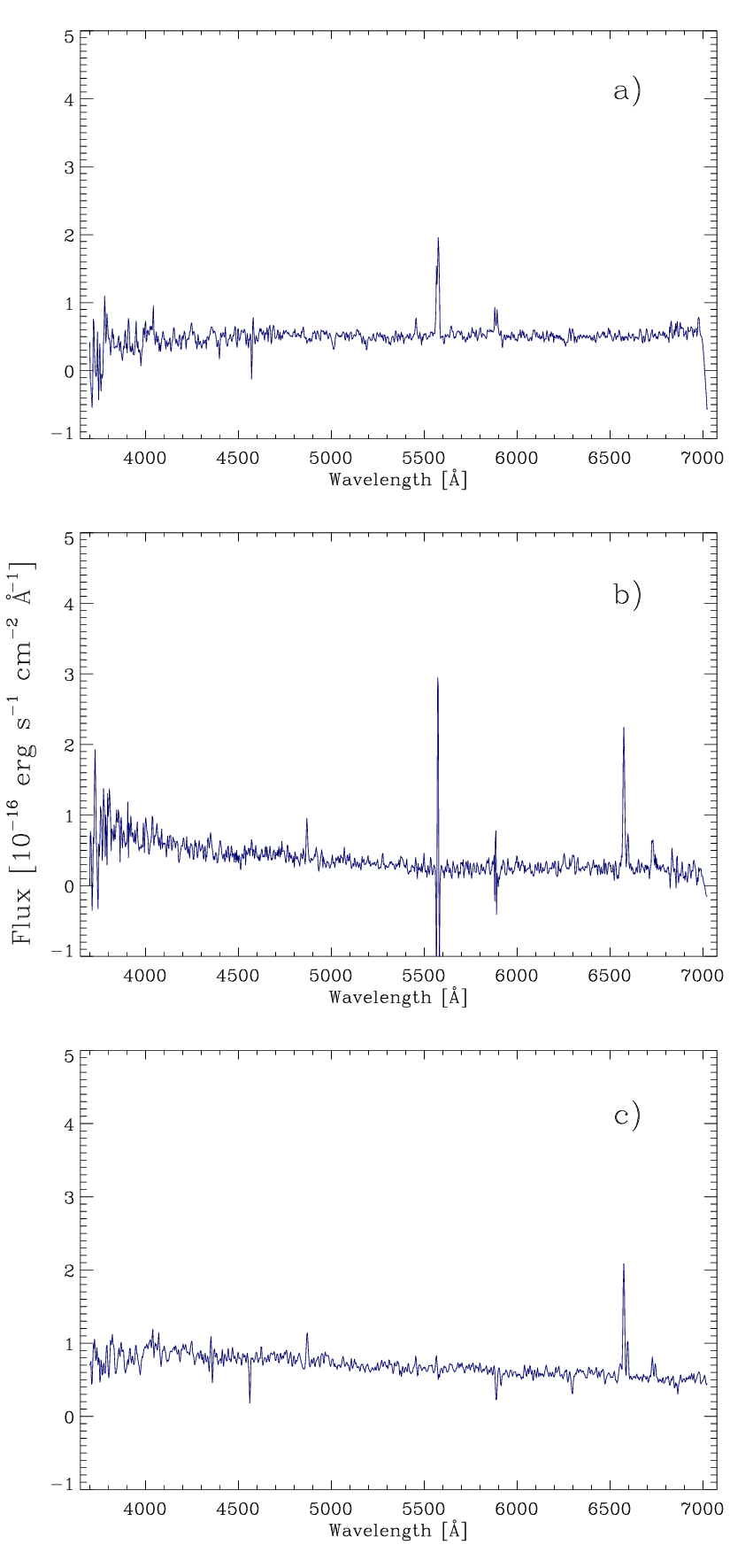}
    \caption[Method I: examples of discarded spectra]
    {
      Examples of discarded spectra after the \fxf\ selection criteria: a)
      spectrum without the presence of \hb; b) spectrum with \ha, \hb, but
      with low S/N and strong sky residuals; c) spectrum without \oii\ \lam3727.
      \label{fig:extra1}
    }
\end{figure}

A better understanding of the quality of the selected spectra can be inferred
from \autoref{fig:n628_ratios}. The top panels show the \ha/\hb\ and
H$\gamma$/\hb\ ratios as a function of observed flux in \hb, i.e. the
variation of these ratios due to the intrinsic signal-to-noise of the data.
The grey symbols correspond to the 2562 spectra selected after applying
the selection criteria b) explained above, the blue symbols are overlaid
on the previous data, showing the position of the selected spectra after the
third selection criteria c) was applied, i.e. the final \fxf\ sample.
These two ratios correspond to the most important
Balmer recombination ratios used to derive the reddening extinction in
spectroscopic studies. Their values should be close to the theoretical ones
(which depend mainly on the characteristic $T_e$) and, in high signal-to-noise
spectra, the deviations from these values correspond to the effect of
interstellar reddening, which tends to increase (\ha/\hb) or
decrease (H$\gamma$/\hb) these ratios depending on the amount of
extinction. For case-B recombination, and assuming a $T_e = 10^4$ K, the
theoretical values for the \ha/\hb\ and H$\gamma$/\hb\ ratios
are 2.87 and 0.466 respectively \citep[Table 4.2,][]{Osterbrock:2006p2331}.
The horizontal lines in each panel correspond to these theoretical
values, while the vertical dashed lines correspond to the \hb\ flux
threshold value.

For larger (observed) fluxes in \hb, the scatter of the \ha/\hb\
and H$\gamma$/\hb\ ratios is smaller. As the signal-to-noise diminishes
(exemplified here by the flux in \hb), the scatter of the ratios
increases to a considerable level. Ideally, in the case of the
\ha/\hb\ ratio, a good signal-to-noise sample would be located above
the theoretical line (consistent with physical reddening) and to the right of
a certain flux limit. Conversely, in the case of the H$\gamma$/\hb\ ratio,
an optimal sample would be located below the theoretical line and to the right
of the flux ratio threshold. The value of the latter has to be chosen in order
to find a good trade-off between the number of physically meaningful selected
spectra, and the point at which the noise starts to dominate the measured
ratios. As the \ha/\hb\ vs. log F(\hb) diagram shows, the flux
threshold is located exactly at the value when the scatter in the
\ha/\hb\ ratio increases significantly for lower values of the
\hb\ flux. A similar behaviour is found in the H$\gamma$/\hb\ diagram,
although the scatter is in general higher, this would be expected as the
H$\gamma$ line is more prone to measurement errors, due to its relatively
low strength and because it is more affected by the correction for underlying
absorption.

The bottom panels of \autoref{fig:n628_ratios} show the distribution of the
\oiii\ \lam5007/\lam4959 ratio as a function of the observed flux in \hb\
(left) and of the observed flux in \oiii\ \lam5007 (right). 
The ratio of the \oiii\ \lam5007/\lam4959 line strengths corresponds to the
magnetic-dipole $^1D_2-^3P_2$ and $^1D_2-^3P_1$ transitions,
which accordingly to theoretical work has a transition probability of
3.01, implying a fixed intensity ratio of 2.98
\citep{Storey:2000p3365}. Therefore, the line ratio of this \oiii\ doublet is
an excellent indicator of the quality of the spectra of an ionized nebular
region. The colour-coding (in the online version) is similar to the upper panels, the horizontal lines
show the theoretical F(\lam5007)/F(\lam4959) ratio value. In both cases, an
ideal spectroscopic sample would lie horizontally along the theoretical value
over most of the intensity range, with a very small scatter around this value. The
F(\lam5007)/F(\lam4959) vs. log\,F(\lam5007) plot shows this behaviour for a range
of observed values log\,F(\lam5007) $\sim$ 1.0--2.0, for lower
log\,F(\lam5007) values (i.e. lower signal-to-noise) the scatter increases
considerably, even for the blue symbols corresponding to the final selected
sample. In the case of the F(\lam5007)/F(\lam4959) vs. F(H$\beta$) diagram,
most of the final selected sample lie near the theoretical value, however,
a significant scatter in the F(\lam5007)/F(\lam4959) ratio is found even for
relatively large \hb\ fluxes (i.e. higher signal-to-noise).
The large dispersion found in the \oiii\ \lam5007/\lam4959 ratio might suggest
that, despite the quality selection criteria and the low final number
of extracted spectra, many of the selected fibres do not correspond to
spectra of a ``physical'' emitting region. However, as it will be discussed
below, the deviations of the \oiii\ ratio from the theoretical value are due
to the effects introduced by subtraction of the stellar continuum in regions
of weak oxygen emission.

\begin{figure*}
  \centering
    \includegraphics[width=0.98\textwidth]{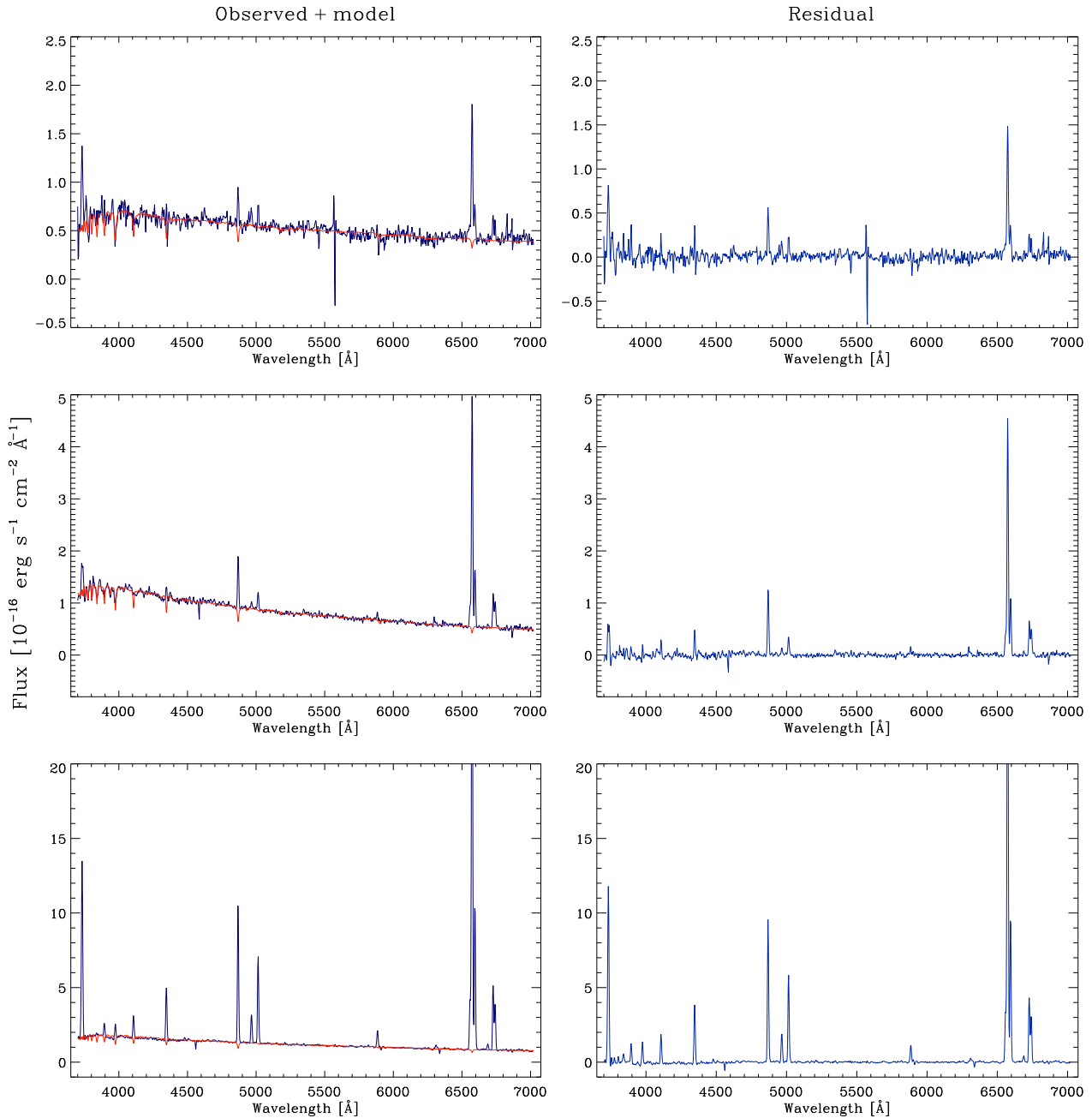}
    \caption[Method I: examples of spectra with different S/N]
    {
      Examples of spectra with different S/N extracted from the final \fxf\ sample. 
      For each fibre, the left column corresponds to
      the observed spectrum, plus the SSP fit model overlaid as a red line,
      the right column shows the residual spectrum to which the selection
      criteria was later applied.
      \label{fig:extra2}
    }
\end{figure*}

\autoref{fig:extra1} shows examples of spectra discarded at different
stages of the selection criteria, the top panel correspond to a spectrum of
nearly null continuum, strong sky residuals and without signatures of \hb\ or
any other important emission line, these sort of spectra were discarded after
the first selection criterion. The rest of the panels show spectra with the
signature of \ha\ and \hb\ but strong sky residuals and low signal-to-noise
(middle panel) or without the presence of the \oii\ \lam3727 line (bottom
panel). These spectra were also discarded after the second and third
selection criteria. On the other hand, \autoref{fig:extra2} shows a series
of three different individual fibres showing different ranges of
signal-to-noise for the final selected \fxf\ sample. For each
fibre, the left column corresponds to the observed spectrum, plus the SSP fit
model overlaid as a red line, the right column shows the residual spectrum
to which the selection criteria was later applied. 
Note that all three spectra show the most important emission lines employed in
the determination of metallicity using strong-line methods, i.e \ha, \hb, 
\oii\ \lam3727,  \oiii\ \lam4959, \oiii\ \lam5007, \nii\ \lam6548, \nii\
\lam6584, \sii\ \lam6717 and \sii\ \lam6731.
Additionally, for those \hh regions with high signal-to-noise we were able to
detect and measure intrinsically fainter lines such as [Ne\,{\footnotesize
  III}] \lam3869, H$\epsilon$ \lam3970, H$\delta$ \lam4101, \hg\ \lam4340,
\hei\ \lam5876, [O\,{\footnotesize I}] \lam6300, and \hei\ \lam6678 (e.g. see
bottom panels of \autoref{fig:extra2}), although they have not been considered
for the present study.

The line intensities of the final \fxf\ sample were corrected by
interstellar reddening using the c(\hb) value together with the
extinction law of \citet{Cardelli:1989p136}, assuming a total to selective
extinction ratio $R=3.1$, following the same procedures as in
\citetalias{Sanchez:2011p3844}. Formal errors were derived by propagating in quadrature
the uncertainty in the flux calibration, the statistical error in the
measurement of the line intensities and the error in the c(H$\beta$) term.

\begin{figure*}
  \centering
    \includegraphics[width=0.95\textwidth]{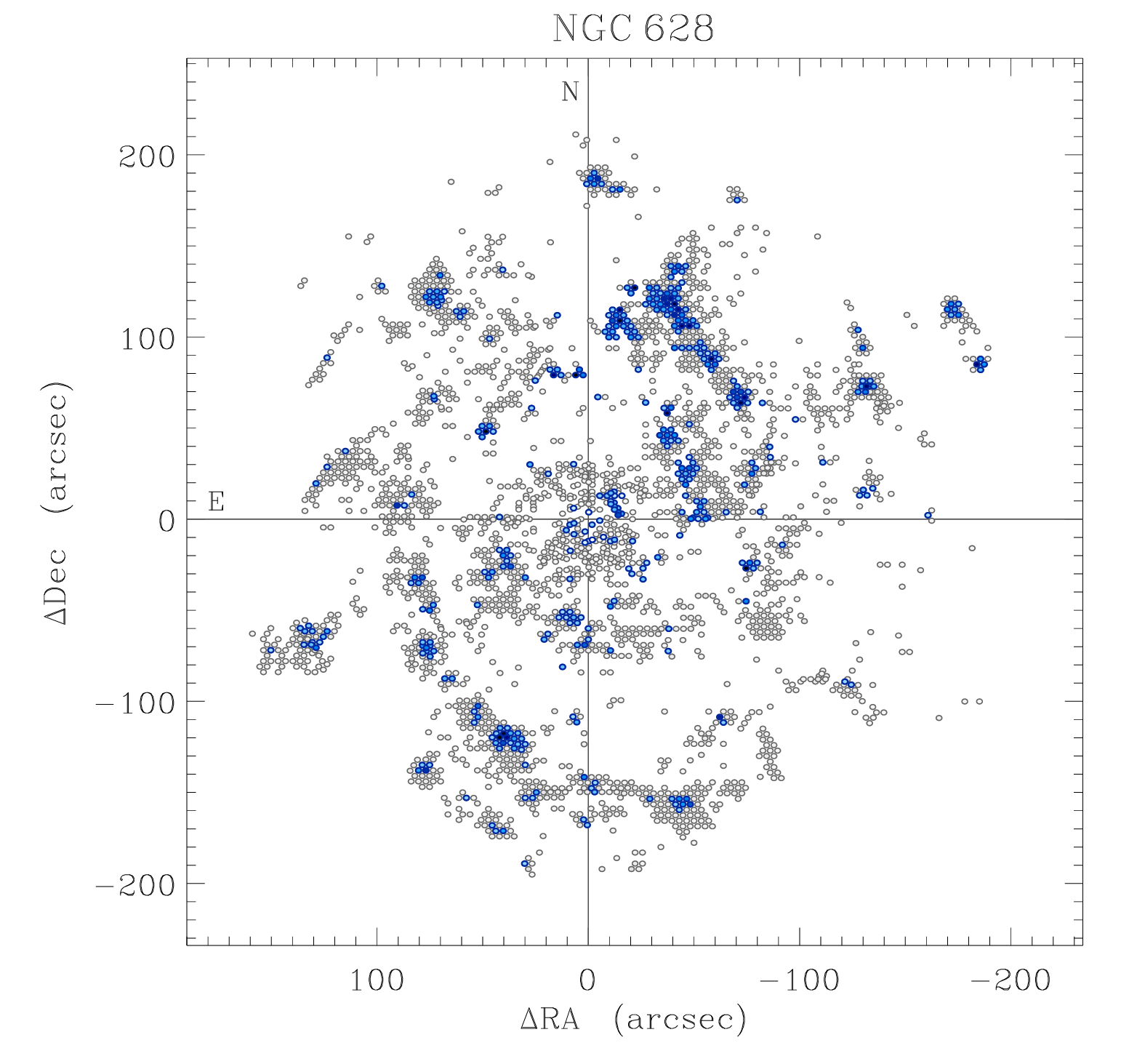}
    \caption[Method I: map of the \fxf\ sample of NGC\,628]
    {
      Spatial map of the \fxf\ (blue) sample and the {\em diffuse} (grey)
      sample for \ngc. The size and position of the fibres (at real scale) are
      displayed in the standard NE-positive orientation. The crosshairs mark
      the central reference point of the IFS mosaic.
      For the \fxf\ sample, the colour intensity of each fibre has been scaled
      to the flux intensity of \ha\ for that particular spectrum.
      \label{fig:n628_offset}
    }
\end{figure*}

\autoref{fig:n628_offset} shows the spatial positions of the different
selected spectra, in a $\Delta$RA--$\Delta$Dec plane in the standard
orientation (north-east positive), for the \fxf\ analysis. The colour-coding is
identical to \autoref{fig:n628_ratios}, i.e. the grey fibres correspond to the
subsample selected after the first and second selection criteria, the blue fibres
are overlaid on the diagram, showing the position of the final selected sample
after applying the third selection criteria. For the \fxf\ sample (blue), the
colour intensity of each fibre has been scaled to the flux intensity of \ha\
for that particular spectrum. In that way, \autoref{fig:n628_offset} would
correspond to a \ha\ {\em emission line map} obtained from the \fxf\ data
sample. The perpendicular lines intersect at the reference point of the IFS
mosaic's position table. A visual comparison with the interpolated H$\alpha$
emission line map of \citetalias{Sanchez:2011p3844} shows that the grey fibres
correspond mainly to the edges of the \hh regions and to regions of diffuse
emission along the spiral arms and in the intra-arms regions.
Some grey fibres are also found as isolated regions all over the surface of
the mosaic. On the other hand, the blue fibres correspond mainly to the central
areas of \hh regions along the spiral arms, as well as some other bright
sources over the surface of the galaxy. This behaviour would be expected,
since the third selection criteria was based mainly on a \hb\ flux
threshold, which separates the brightest fibres (i.e. the centre of \hh
regions) from the weaker ones, which correspond mainly to regions of diffuse
emission. The fibres selected after the first and second selection criteria,
corresponding to the grey fibres in \autoref{fig:n628_offset}, would be
referred as the {\em diffuse} sample.

An additional way of studying the 2D distribution of the galaxy properties
consists in obtaining azimuthally-averaged radial spectra, from which radial
average properties can be derived. Taking the (blue) \fxf
sample shown in \autoref{fig:n628_offset} as a base, radial average spectra
were obtained by co-adding all the spectra of this sample within successive
rings of 10 arcsec, starting from the central reference point.
An average spectrum was obtained for each single annulus at a given
radius. Annulus with less than 5 fibres were excluded and skipped in the
process. The radial average spectra were then analysed using the same fitting
procedures described before. Although the derived spectra present more
signal-to-noise than the single-fibre case, the measured emission lines were
corrected by extinction using only the \ha/\hb\ ratio, for
consistency with the \fxf\ analysis.

\begin{figure*}
  \centering
    \includegraphics[width=\textwidth]{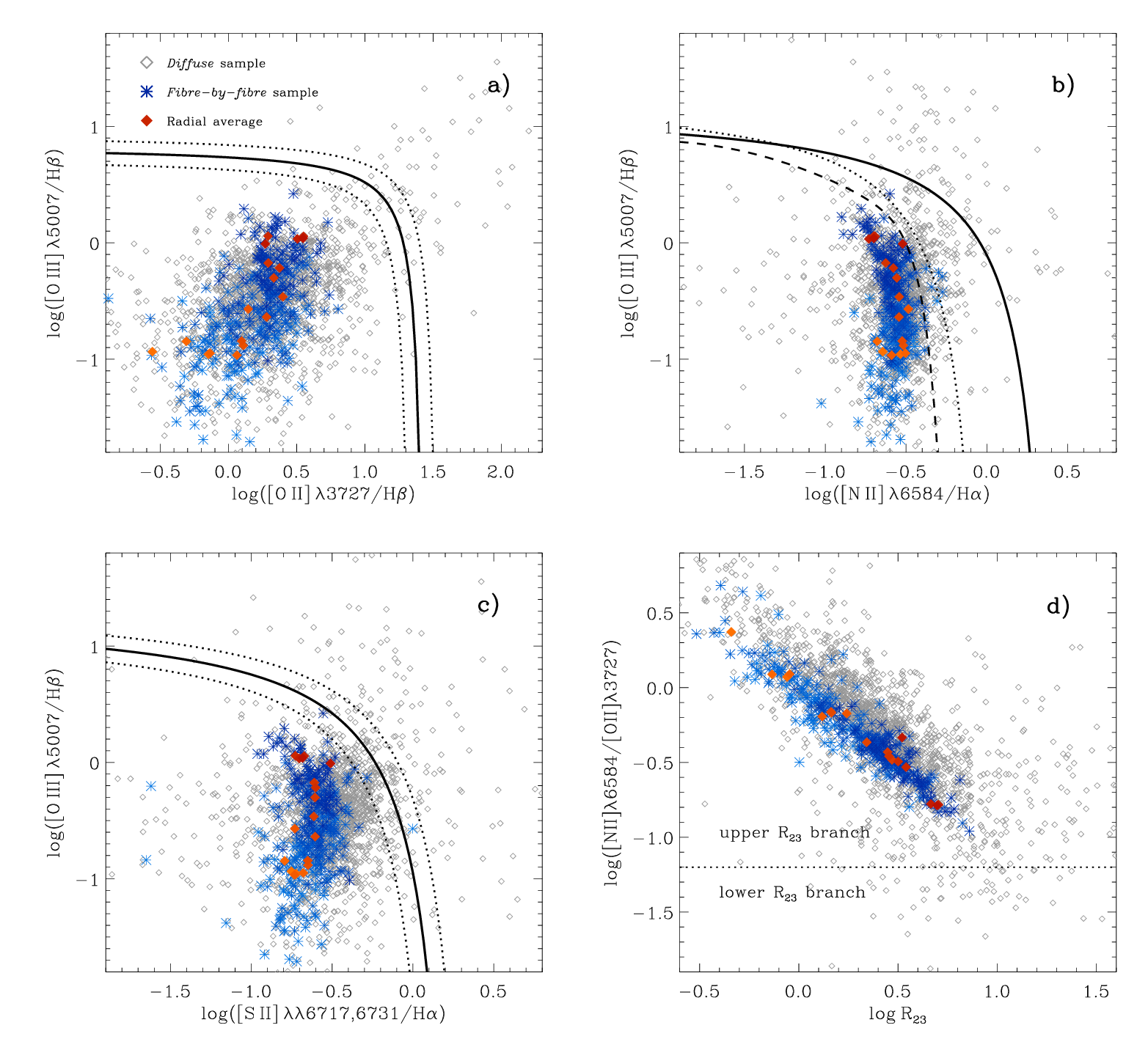}
    \caption[Method I: BPT diagrams for NGC\,628]
    {
      Diagnostic diagrams for \ngc. The grey open diamonds correspond to the
      {\em diffuse} sample, the blue symbols to the final \fxf\ sample and the
      filled diamonds to the azimuthally-averaged radial values. Lighter tones
      correspond to inner regions of the galaxy, darker colours to increasing
      galactocentric distance.
      The different lines correspond to the theoretical boundaries dividing
      star-forming regions from other types of ionization: Panel a)
      \citet{Lamareille:2004p3322};  Panel b) dark-thick line:
      \citet{Kewley:2001p3313}, dotted-line: \citet{Kauffmann:2003p3918},
      dashed-line: \citet{Stasinska:2006p3319}; Panel c) and d)
      \citet{Kewley:2001p3313}. 
      In Panels a) and c) the dashed-lines represent the $\pm$0.1\,dex variation.
      \label{fig:bpt_fxf}
  }
\end{figure*}

Different possible mechanisms can be responsible for the ionization giving rise
to emission line spectra in galaxies. The source driving the ionization in a
given galaxy can be identified by exploring the location of certain line ratios
in the so-called diagnostic diagrams \citep[e.g. BPT,][]{Baldwin:1981p3310,Veilleux:1987p3423}.
These diagrams can be used as a tool to differentiate objects in which the
photoionization is due by hot OB stars (\hh regions), 
from objects in which is due to a non-thermal continuum (e.g. LINERs, AGNs) or
to alternative mechanisms, such as the proposed {\em retired}
galaxies, in which the ionization would be produced by hot post-asymptotic
giant branch stars and white dwarfs \citep{Stasinska:2008p3917}.
\autoref{fig:bpt_fxf} shows a collection of different diagnostic diagrams
for the spectral samples considered above, including the {\em diffuse}
sample (grey open diamonds), the final \fxf\ sample (bluish symbols), and the
radial average sample (reddish diamonds).
Panel a) corresponds to the classic BPT diagram 
\oiii\ \lam5007/\hb\ vs. \oii\ \lam3727/H$\beta$, while panels b) and c)
correspond to \oiii\ \lam5007/\hb\ vs. \nii\ \lam6584/\ha\ and vs. \sii\
\lam\lam 6717,31/H$\alpha$, respectively. 
Only those regions from the {\em diffuse} sample with detected \oii\ \lam3727
are drawn in Panels a) and d).
The diagrams show different demarcation lines corresponding to the theoretical
boundaries dividing the starburst region from other types of ionization. In
Panel a) the delimitation is after \citet{Lamareille:2004p3322}; in Panel
b) the dark-thick line corresponds to the parametrization provided by
\citet{Kewley:2001p3313}, the dotted-line to \citet{Kauffmann:2003p3918},
and the dashed-line to \citet{Stasinska:2006p3319}; in Panel c) the
demarcation is after \citet{Kewley:2001p3313}. In Panels a) and c) the
dashed-lines represent the $\pm$0.1\,dex variation.

From these panels we see that the {\em diffuse}, low signal-to-noise
spectra is scattered all over the diagrams, including those regions outside the
boundaries which correspond to ionization sources different than OB stars. On
the other hand, the line ratios corresponding to the \fxf\ sample are
encompassed by the theoretical \hh regions boundaries. The line ratios of the
radial averaged spectra are overlaid on each plot as filled diamonds. The
colour-coding of both samples is related to the spatial
position of a given fibre/annulus. Lighter tones correspond to the inner regions
of the galaxy, while darker colours correspond to positions with increasing
galactocentric radius. Clear trends can be noticed in each of the diagrams, in
the case of Panel a) the spectra corresponding to the inner regions, both for the 
\fxf\ and radial average sample, tend to have lower line ratios
for both indices; for regions at the outer part of the galaxy, the ratios
increase approaching the theoretical boundary. The reason for this behaviour
can be understood from the emission line maps presented in \citetalias{Sanchez:2011p3844},
the inner parts of the galaxy lack emission in \oii\ and \oiii, while towards the
outer parts, the emission from these species is prominent, increasing the two
line ratios involved in this diagram. A slightly different trend is shown by the
radial average spectra, which stay with a nearly constant \oiii/\hb\ value and
increasing \oii/\hb\ ratio, with increasing galactocentric distance up to
\oii/\hb\ $\sim$ 0.2, where the \oiii/\hb\ ratio increases considerably.
In the case of Panels b) and c), the behaviour of both samples is quite
similar. The \oiii/\hb\ ratio increases with galactocentric distance, but the
\nii/\ha\ and \sii/\ha\ ratios do not vary much (except for some outlying blue
fibres) and are concentrated along a vertical pattern centered at log
\nii/\ha\ $\sim$ --0.6 and log \sii/\ha\ $\sim$ --0.7, with the \sii/\ha\
showing a slightly higher scatter. The radial average values follow the same
trend in both cases, i.e. the azimuthally-average values of the
\nii/\ha\ and \sii/\ha\ ratios do not change appreciably with increasing
galactocentric distance.
The locus of the \fxf\ sample in all cases is consistent with
regions in which the dominant ionization mechanism giving rise to the line
emission of these spectra is a thermal continuum (i.e. hot OB
stars). Therefore, the selection criteria applied in order to obtain the
\fxf\ sample did in fact extract those regions with spectra showing features of
real \hh regions.

\begin{figure*}
  \centering
    \includegraphics[width=0.85\textwidth]{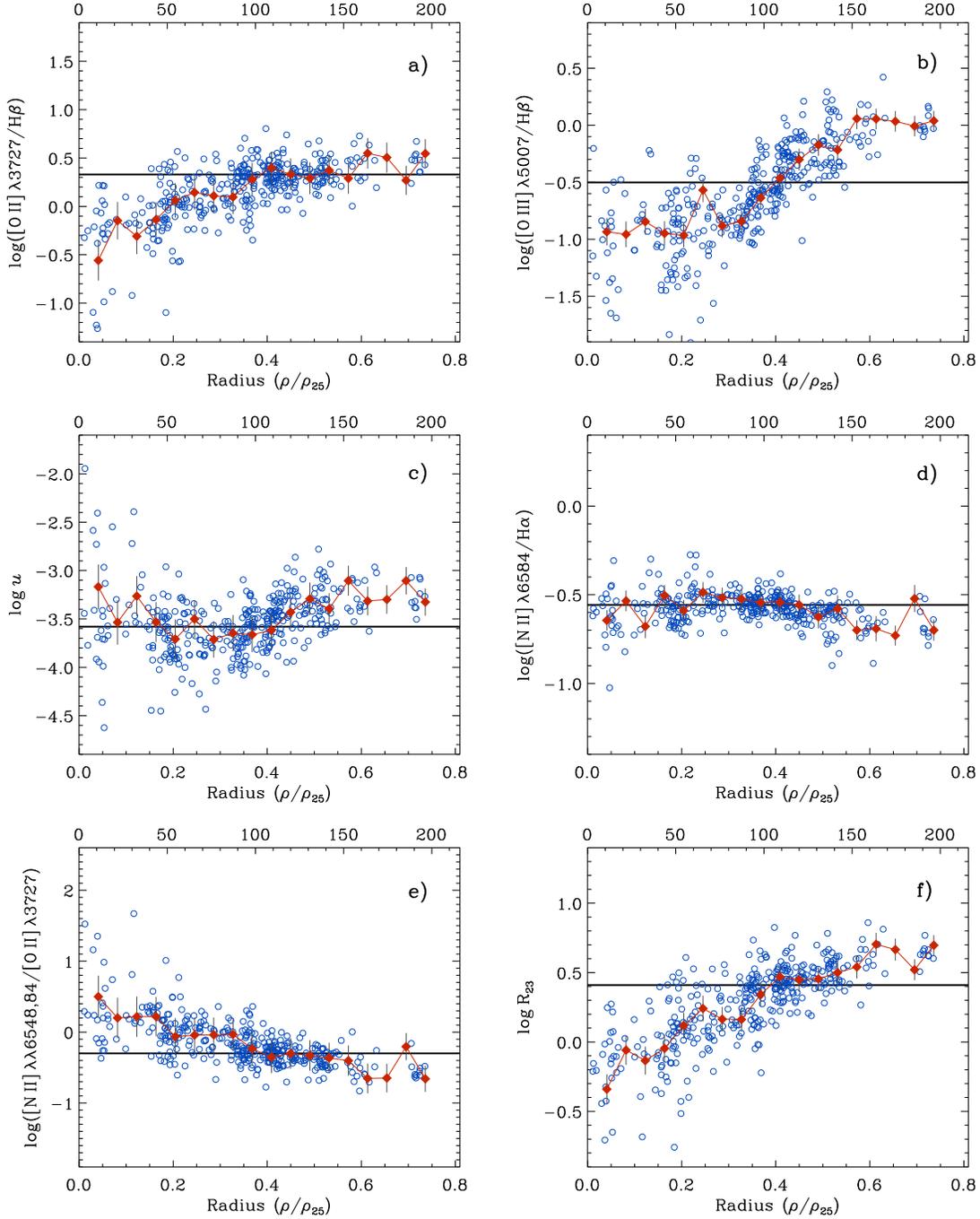}
    \caption[Method I: radial variation of selected physical properties for NGC\,628]
    {
      Radial variation of different line
      ratios and physical properties of \ngc\ for the \fxf
      sample (blue circles) and the azimuthally-averaged spectra (red
      diamonds). The deprojected radial position for the blue symbols
      has been normalised to the size of the optical disk at the 25 mag
      arcsec$^{-2}$ isophote. The radial position of the red diamonds
      correspond to the projected outer radius of the corresponding
      annulus. The top values on the $X$-axes show the linear-projected
      galactocentric radius in arcsec. The horizontal lines in each diagram
      correspond to the values derived from the integrated spectrum of
      \ngc\ presented in Paper~I. Characteristic error bars are only drawn for
      the radial-averaged spectra for the sake of clarity.
      \label{fig:radial_fxf}
    }
\end{figure*}

The last panel of \autoref{fig:bpt_fxf} corresponds to the 
\nii\ \lam6584/\oii\ \lam3727 vs. (\oii\ \lam3727 + \oiii
\lam4959,\lam5007)/\hb\ (or $N_2O_2$ vs. $R_{23}$) diagnostic diagram. This
is usually used to differentiate between the two branches of the $R_{23}$
abundance calibrator \citep{Pagel:1979p71} as explained in
\citetalias{Sanchez:2011p3844}, and it is a strong function of metallicity for
log \nii/\oii\ $\gtrsim$ --1.2 \citep{Kewley:2002p311}. 
As in the previous diagrams, the {\em diffuse} sample is spread over most
regions of the plot, while the \fxf\ sample and the radial average values are
found along a well-defined pattern consistent with inner regions of the galaxy
having higher $N_2O_2$ values and outer regions showing lower $N_2O_2$ ratios,
which combined with the opposite behavior of $R_{23}$, create a correlation with
a negative slope. All $N_2O_2$ values for the \fxf\ sample correspond to the
upper branch of the $R_{23}$ calibration, as implied in \citetalias{Sanchez:2011p3844}
during the analysis of the emission line maps for this galaxy.

The radial trends of the line ratios inferred from \autoref{fig:bpt_fxf} 
can be seen clearly in \autoref{fig:radial_fxf}. Panel a) corresponds to
the variation of \oii\ \lam3727/\hb; Panel b) to \oiii\ \lam5007/\hb;
Panel c) to the {\em ionization parameter}\footnote{Note that, as explained in Sec. 5.1.2 of
  \citetalias{Sanchez:2011p3844}, this definition is just an approximation of
  the ionization parameter, since the \oii/\oiii\ line ratio from which is defined
  is dependent on the temperature and the metallicity of the
  ionizing source \citep{Balick:1976p3949}.
  This parameter might be understood as an alternate definition of the optical
  {\em excitation diagnostic}, defined as the \oiii/\oii\ line
  ratio \citep[e.g. see][]{Dors:2003p3948,Morisset:2004p3947}. Nevertheless, we use
  the definition $\log u$ for consistency with \citetalias{Sanchez:2011p3844}.}
$\log u \equiv -0.80$\,log(\oii/\oiii)\,$-3.02$, after \citet{Diaz:2000p3371};
Panel d) to \nii\
\lam6584/\oii\ \lam3727; Panel e) to \sii\ \lam6717,31/\ha; and Panel f) to
$R_{23}$. The blue circles correspond to the \fxf\ spectra, while the
red-connected diamonds correspond to the values
derived from the radial averaged spectra. The deprojected radial position for
the blue symbols has been normalised to the size of the optical disk at the 25
mag arcsec$^{-2}$ isophote. In each panel, the horizontal line
corresponds to the value derived from the integrated spectrum of the galaxy in
\citetalias{Sanchez:2011p3844}.

The \oii/\hb\ and \oiii/\hb\ ratios on Panels a) and b) increase as a function of
the radius, with a somewhat steeper increase for the \oiii/\hb\ ratio from a
normalised radius $\sim$ 0.3, for lower radii this latter ratio shows some
level of scatter, but consistent with a constant value log(\oiii/\hb) $\sim$
--1.0, as already noticed in Panel a) of the BPT diagrams shown in
\autoref{fig:bpt_fxf}. The ionization parameter shows quite a lot of
scatter for radii lower than 0.3$\rho_{25}$, due to the low strength of oxygen emission
(from which this parameter is derived) in these inner regions. From
$\rho/\rho_{25} > 0.3$, $\log u$ increases slightly with increasing radius. 
On the other hand, the \nii/\ha\ ratio shown in
Panel d) confirm the very small variation of these ratios over the
surface of the galaxy. However, the \nii/\oii\ ratio
shows a negative gradient towards larger radii. The $R_{23}$ shows the
increasing radial pattern inferred previously, extending for more than one order
of magnitude from the inner (log\,$R_{23}$ $\sim$ --0.5) to the outer regions
(log\,$R_{23}$ $\sim$ --0.7). Note that all the indices and parameters in which
the oxygen \oii\ and/or \oiii\ lines are involved show a larger dispersion for
normalised radii lower than 0.3.

The radial average values traced by the red diamonds follow the
trends shown by the blue symbols in all the radial plots. The higher
signal-to-noise of the average spectra allows a better determination of the
line ratios, specially for the regions in the inner part of the galaxy. Note
also that the \fxf\ spectra cover practically all radii values
up to $\rho/\rho_{25} \sim 0.75$.
Interestingly, the values of the ratios and parameters derived from the
integrated spectrum (horizontal lines) are consistent in all cases with the
radial values found at $\rho/\rho_{25} \sim 0.4$.

The analysis of the \fxf\ sample shows that, the assumption of a single
fibre containing enough signal-to-noise to be analysed in individual basis
is --to a first order-- correct. 
However, as mentioned before, despite the quality selection
criteria and the low number of final selected spectra in the \fxf\ sample,
many of the selected fibres do not show characteristic signatures of a
``physical'' \hh region, i.e. fibres with a \oiii\ \lam5007/\lam4959 ratio not
consistent with the constraints imposed by the theoretical value ($\sim$ 3),
as suggested by the large dispersion shown in the bottom panels of
\autoref{fig:n628_ratios}.
In order to explored this possibility, we performed as an exercise an
extraction from the original {\em clean}
mosaic considering only those regions for which the \oiii\ \lam5007/\lam4959
ratio was consistent with the theoretical value, within a very small range of
observed ratios (i.e. \oiii\ \lam5007/\lam4959 = 2.98\,$\pm$\,0.3), and above
a flux threshold in \hb\ (5 $\times$ 10$^{-16}$ \flux), i.e. a lower limit
than in the \fxf\ case, hoping that the restriction of the \oiii\ line ratio
would provide better quality spectra even for fibres with low observed
intensity. This extraction resulted in 152 selected fibres, i.e. only 2\% of
the total number of fibres in the {\em clean} mosaic and a factor of
$\sim$\,2.5 lower fibres than the \fxf\ sample.

\begin{figure}
  \centering
  \includegraphics[width=\hsize]{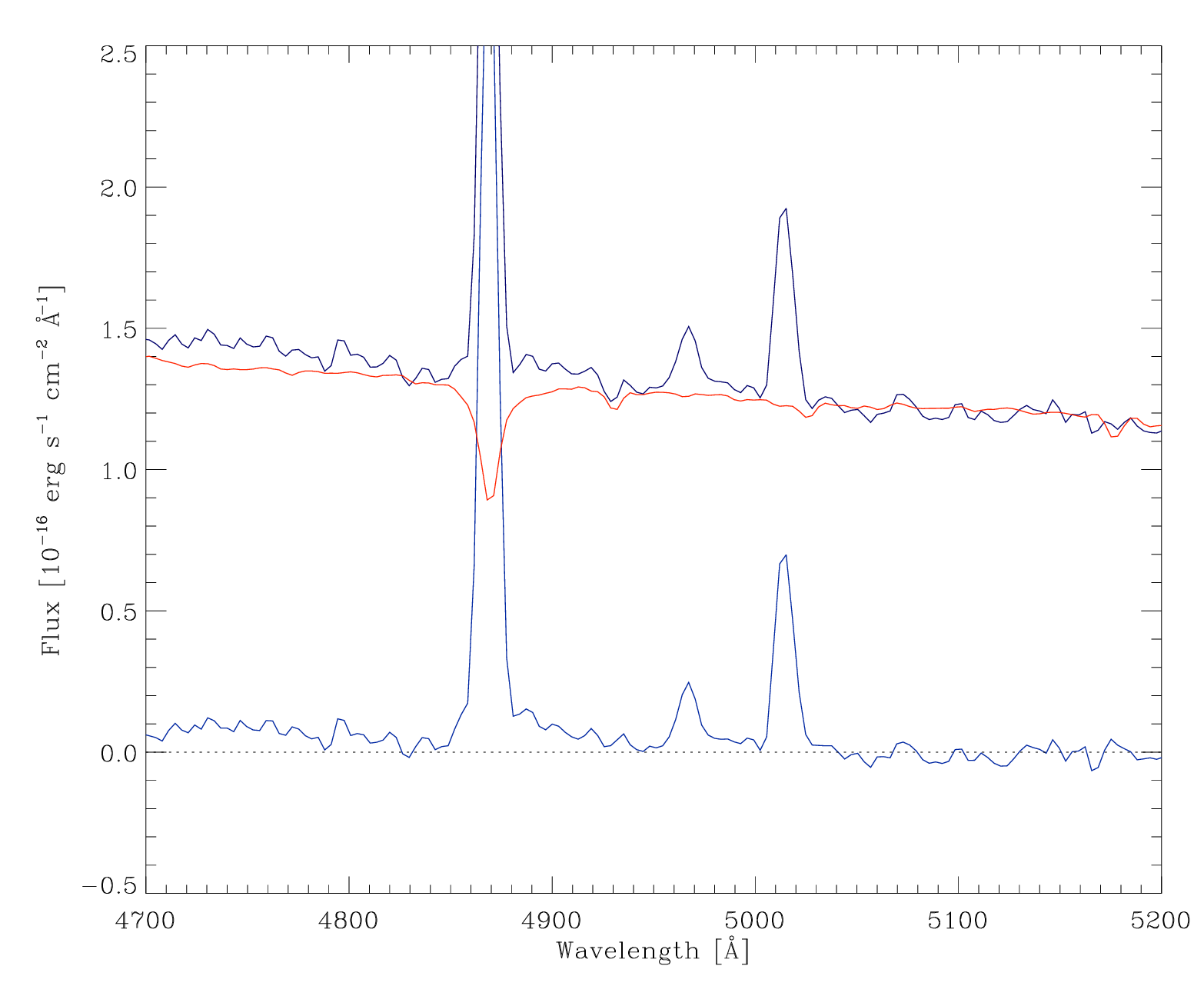}
  \caption[]
    {
      Example of a region with low emission in \oiii\ and a ``deficient'' SSP
      continuum subtraction (red line). Note that the SSP fit does not trace
      accurately the observed continuum (dark blue), leading to deviations from
      the theoretical \oiii\ \lam5007/\lam4959 ratio on the residual spectrum
      (light blue).
      \label{fig:n628_extra3}
    }
\end{figure}

The ``physically-selected'' sample was qualitatively compared with the \fxf\
sample. All the line ratios, spectral trends, level of dispersion
and ionization properties described by the \fxf\ sample were completely
followed by the new limited sample, with only one important difference: 
the new sample discarded regions with low emission in oxygen, e.g.
log(\oiii/\hb) below --1, points which are present in the case of the \fxf\
sample, as shown in \autoref{fig:bpt_fxf}.
Given that the selection criteria considered a lower flux threshold than in the
\fxf\ case, the flux limit cannot account for this effect. Therefore, the
reason for the lack of these regions is due to the selection criterion
based on the restricted range in the \oiii\ ratio.

Regions with low emission in oxygen correspond to the
inner parts of the galaxy, where the stellar population is more dominant in the
observed spectra, and therefore, the errors in the measurement of the residual
emission lines due to a deficient continuum subtraction during the SSP model
fitting are larger (e.g. see \autoref{fig:n628_extra3}). For the outer regions
of the galaxy, the contribution of the stellar population is lower, and
therefore, it is easier to recover the proper \oiii\ ratio from the derived
residual spectrum. 
However, in a region where the stellar population is more dominant and the
emission lines are weak, the derived \oiii\ ratio from the residual spectrum
might not be close to the theoretical ratio, but nevertheless, the total flux
of these lines and their line ratios are representative of the physical
conditions of the gas in that particular region. The fact that the inclusion
of spectra with ``non-physical'' \oiii\ line ratios in the \fxf\
sample produced well-defined trends (although with some level of scatter) in
those weak oxygen (inner) regions give support to this idea.

In summary, the \fxf\ method considered that each individual fibre samples a
large-enough physical region of the galaxy and contains enough
information for a complete spectroscopic analysis. Thanks to the different
spectra selection steps, we were able to identify areas within the galaxy
disk consistent with regions of nearly pure diffuse emission (in \ha\ and
\hb), and to differentiate those from areas with the characteristic emission
of well-defined \hh regions. The methodology introduced for the \fxf\ sample will
be implemented to the second extraction method considered for the IFS study of
\ngc, by performing similar quality/sanity checks and comparing the radial
trends in the line emission and diagnostic diagrams derived from the \fxf\
sample.


\subsection{Method II: {\sc HII region catalogue}}
\label{sec:hii}

Traditionally, spectroscopic studies of nearby galaxies have been performed by
targeting a number of (bright) \hh regions over the surface of a galaxy,
placing long-slits and/or fibres of different apertures on top the selected
regions, and integrating the flux over that aperture.
The classical chemical abundance diagnostics based on the
observation of strong emission lines ratios (e.g. $R_{23}$), were conceived as
empirical methods describing the physical properties of these large,
spatially-integrated, and individually defined \hh regions. The calibration of
these metallicity indicators were performed by using grids of photoionization
models for a given range of metallicities and ionization parameters, and
therefore, are not based on observational data alone. Given the large parameter
space under investigation, these calibrations have generally assumed
spherical or plane-parallel geometries without considering the effects of the
distribution of gas, dust, and multiple, non-centrally located ionizing
sources. These geometrical effects may affect the temperature and ionization
structure of the regions.

It has been argued that the geometrical distribution of ionization sources may
partially account for the large scatter in metallicities derived using
model-calibrated empirical methods \citep[][hereafter EBS07]{Ercolano:2007p3364}. 
According to recent results based on 3D photoionization models with various
spatial distributions of the ionizing sources, for intermediate to high
metallicities, models with fully distributed configurations of stars display
lower ionization parameters than their fully concentrated counterparts. The
implications of this effect varies depending on the sensitivity of the
metallicity indicator to the ionization parameter (EBS07).

Generally speaking, results derived from the use of the empirical metallicity
indicators should be considered within a statistical framework, as the error
due to intrinsic temperature fluctuations and chemical inhomogeneities on a
single region may be very large, even when the temperature of the region can
be directly determined
\citep[e.g.][]{Peimbert:1967p1928,GarciaRojas:2006p2832,Ercolano:2007p3364}.
The spectra extracted in the previous selection method were based on
assuming that the spectra of individual fibres would contain enough information
in order to derive the physical properties of the region sampled by the
individual fibre aperture.
However, as shown in \autoref{fig:n628_offset}, the selected fibres trace
morphologically complex regions, which do not resemble the classical picture of
well-defined spherical \hh regions. Some of the most prominent emitting regions
are embedded in giant \hh complexes without an established geometrical centre,
and most importantly, as discussed in Sec. 7 of
\citetalias{RosalesOrtega:2010p3836}, regions which would be considered as
individual \hh regions in classical terms, show fibre-to-fibre variations on
their emission line intensities.

One question that we might rise at this point is, in the case of the \fxf\
sample, whether we are observing real
point-to-point variations of the physical properties within a region, i.e. if
the different measured line ratios are reflecting a real distribution of the
ionizing sources, gas content, dust extinction and ionization structure within
these regions, or the line intensity variations are just spurious effects due to
the relatively low signal-to-noise of those emitting regions.
In that respect, one of the main issues that IFS observations of emission line
regions should aim to assess is: how valid are the results derived from the use
of strong line calibrators applied on a point-to-point (fibre-to-fibre) basis?
compared to the co-added spectrum of a larger, classically well-defined \hh
region. 
In order to try to answer this question, and at the same time perform a robust
2D spectroscopic analysis of \ngc, an analysis method was envisaged
based on considering ``classical'' \hh regions as the source of analysable
spectra.  For doing so, a number of \hh regions in \ngc\ was identified and
classified by hand, based on the \ha\ emission line map of the galaxy
and on the {\em diffuse} plus \fxf\ spatial distribution of
fibres, as shown in \autoref{fig:n628_offset}.
The selection of the fibres considered to belong to an individual \hh region
was performed following a purely geometrical principle, i.e. fibres located
within the same region, which seemed to be geometrically connected, were
considered as a single \hh region. This criterion might be relatively
subjective, but ``classical'' \hh regions in other spectroscopic studies were
chosen following the same principle, e.g. by selecting the more prominent
(high surface brightness) regions in \ha\ narrow band images.

\begin{figure*}
  \centering
    \includegraphics[width=0.95\textwidth]{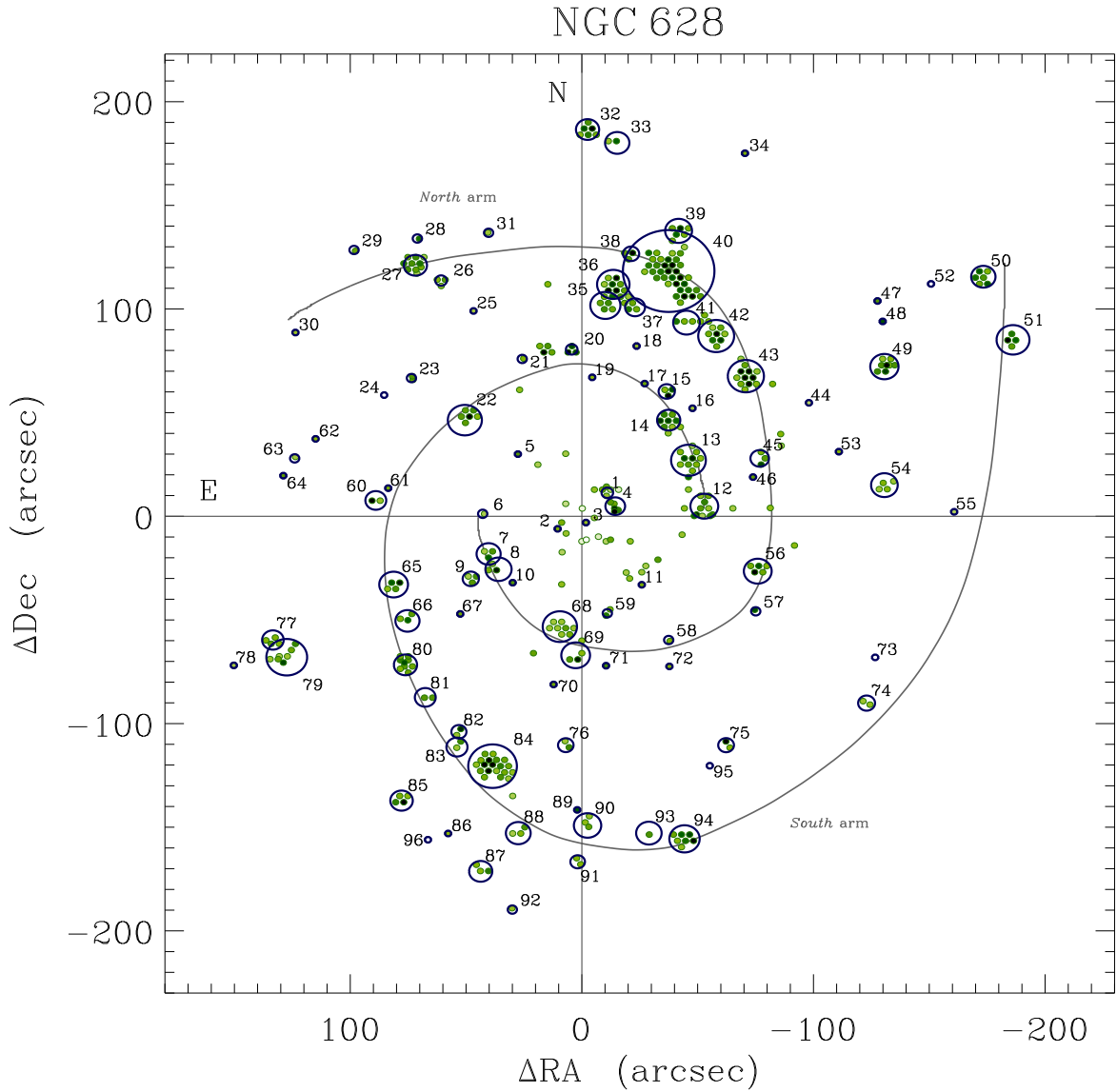}
    \caption[Method III: map of the HII regions catalogue]
    {
      Spatial location and identification of the selected \hh regions for
      \ngc. The background fibres correspond to the \fxf\
      sample, with colour intensities scaled to the flux of the H$\alpha$
      emission line. The circles correspond to the real or equivalent aperture
      diameter, as explained in the text. 
      The grey-thick lines define the North and South {\em operational}
      spiral arms, as in Fig. 1 of \citet{Kennicutt:1980p3756}.
      \label{fig:n628_id}
    }
\end{figure*}

The actual mechanism in order to generate the \hh region catalogue for \ngc\
was the following:

\begin{enumerate}

\item A group of fibres is identified by eye as an individual \hh
  region from the {\em sub-mosaic} extracted from the {\em clean}--residual
  mosaic, corresponding to the {\em diffuse} plus the \fxf\
  samples, as shown in \autoref{fig:n628_offset}. The positions and IDs of the
  selected fibres are stored and
  associated to the corresponding \hh region. The fibre selection mechanism
  was based on two different methods: 1) by choosing the fibres individually by hand,
  following the morphological structure of the selected \hh region; and 2) by
  considering all the fibres within a pre-established circular aperture
  centered at an arbitrary position, which might not coincide with
  the centre of any specific fibre. In the first case, the associated
  ``location'' of the \hh region corresponds to centre of the first selected
  fibre, which was chosen to coincide nearly with the geometrical centre of
  the group of fibres considered as a \hh region. In the second case, the
  location of the \hh region corresponds to centre of the circular
  aperture, which was also chosen to coincide with the geometrical centre of
  the \hh region. In the case of the circular aperture, different diameters
  were tested until the encompassed fibres would correspond to the visually
  selected region.

\item Once the positions and IDs of the fibres corresponding to a given \hh
  region are identified in the {\em sub-mosaic} described above, the fibres
  corresponding to the same positions and IDs are recovered from the {\em
  clean}--observed mosaic, i.e. the RSS file before performing the SSP
  model subtraction. The spectra belonging to those fibres are co-added,
  obtaining a single spectrum corresponding to the selected \hh region.

 \item The integrated \hh region spectrum is fitted by a linear combination of
 SSP templates by exactly the same procedure as described before, in order to
 decouple the contribution of the stellar population. Once the model of the
 underlying stellar population was derived, this is subtracted from the
 original spectrum, obtaining a {\em residual} \hh region spectrum.

\item Individual emission line fluxes are measured from the residual spectrum
  by fitting single Gaussian functions as explained previously, obtaining a
  set of emission line intensities for each \hh region.

\item The process is repeated for each group of fibres identified as a single
  \hh region, until the whole surface of the galaxy mosaic is covered.

\end{enumerate}

\begin{figure*}
  \centering
  \includegraphics[width=\textwidth]{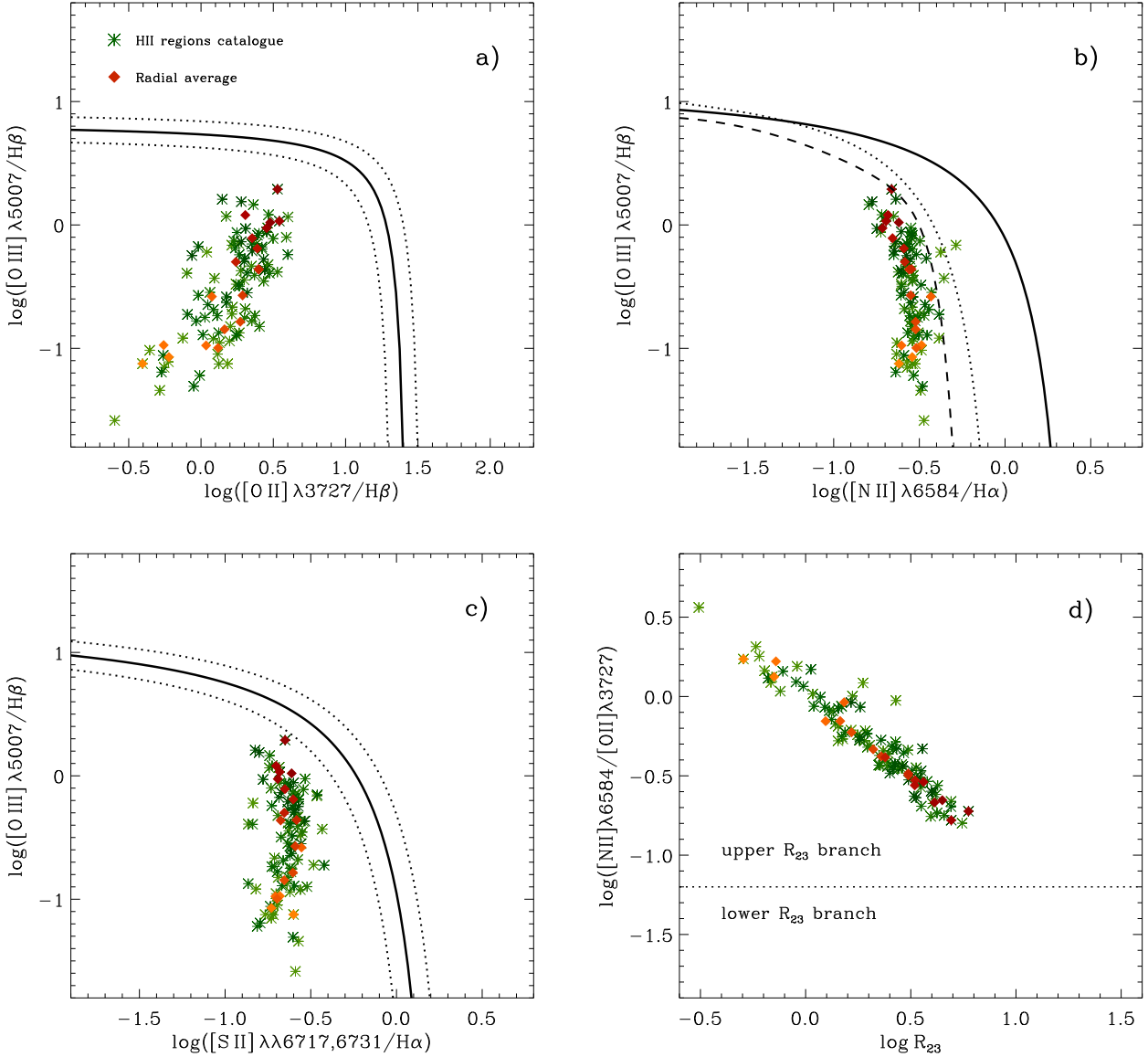}
  \caption[Method III: BPT diagrams]
  {
    Diagnostic diagrams for \ngc\ corresponding to the selected \hh regions
    catalogue (green symbols), the (reddish) diamonds correspond to the
    azimuthally-averaged radial values. Lighter tones correspond to inner
    regions of the galaxy, darker colours to increasing galactocentric
    distance. 
    The different lines correspond to the theoretical boundaries dividing
    star-forming regions from other types of ionization.
    See the caption of the similar \autoref{fig:bpt_fxf} for references and a
    more complete explanation.
    \label{fig:bpt_HII}
    }
\end{figure*}

A total of 108 \hh regions were selected following the procedure described
above. The residual spectra of the catalogue was tested against the quality
criteria outlined in the case of the \fxf\ analysis (i.e. the presence of
\oii, \oiii\ and finite floating numbers for the derived value of
c(H$\beta$)). Twelve spectra did not satisfy all the criteria and were
discarded from the final catalogue.
\autoref{fig:n628_id} shows the final sample of 96 selected \hh regions for
\ngc. The fibres displayed in this figure correspond to the \fxf\ sample
shown in \autoref{fig:n628_offset} (the {\em diffuse} sample was not included
for the sake of clarity). The circles define the selected \hh regions, the
numbers next to the circles correspond to the \hh region ID used in this paper.
The diameter of the circles correspond to: 1) an ``equivalent aperture'' in the
case where the \hh region was selected by choosing individual fibres by hand,
e.g. region N628--40, at ($\Delta\alpha$,\,$\Delta\delta$) $\sim$ (--40,130);
2) to the real diameter of the circular aperture when the selected \hh region
was chosen on this basis. Note that some fibres in \autoref{fig:n628_id} are
not associated with any \hh region (especially in the central region of the
galaxy), these fibres correspond to the \hh regions discarded due to the reasons
explained in the previous paragraph.
Note also that many \hh regions are consistent with a single fibre. In those
cases, the area surrounding the individual fibre did not show spectra with
significant signal; therefore, they were not considered as their inclusion
would only add noise to the integrated spectrum.


\begin{figure*}
  \centering
    \includegraphics[width=0.83\textwidth]{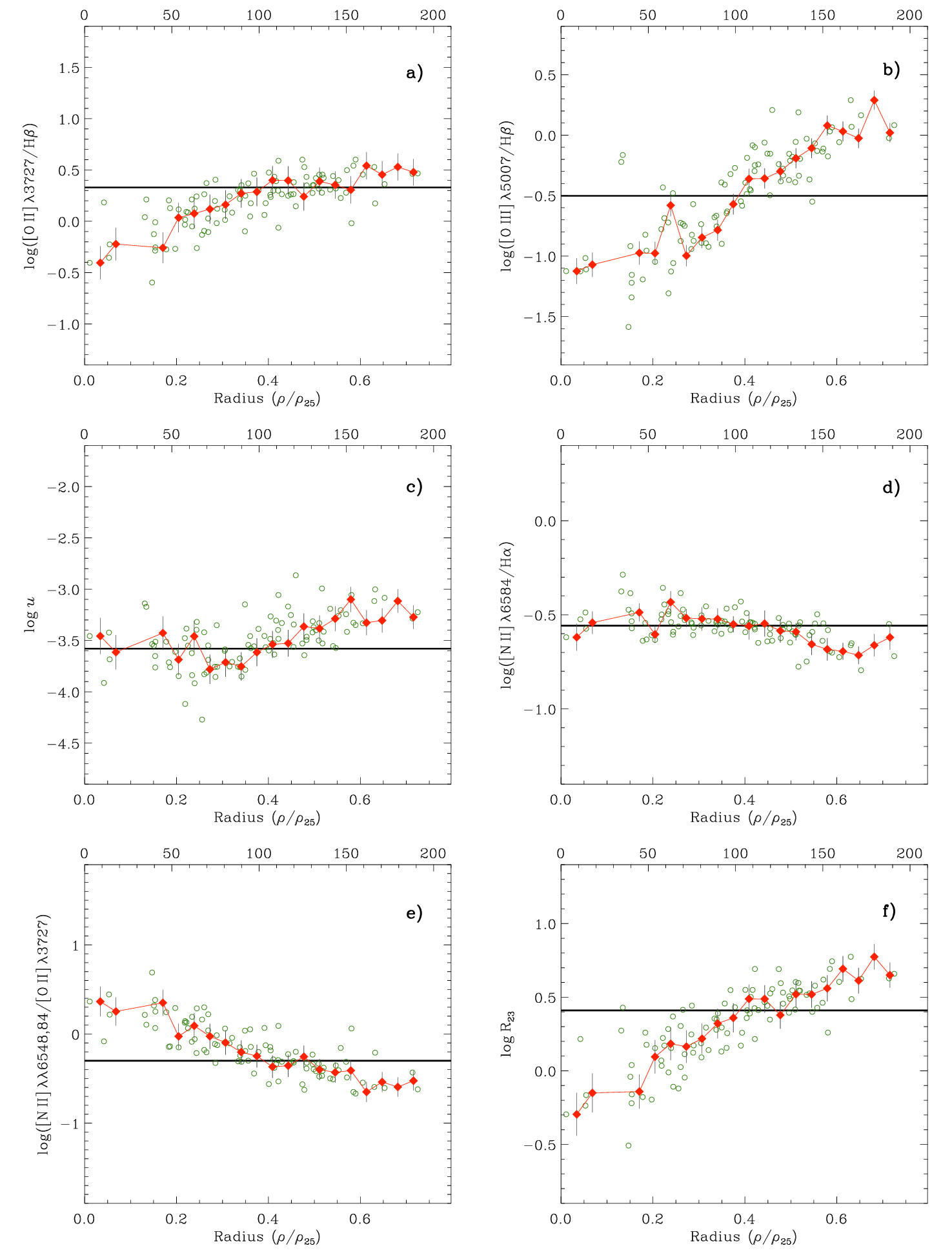}
    \caption[Method III: radial variation of selected physical properties]
    {
      Radial variation of different line
      ratios and physical properties of \ngc\ for the \hh regions catalogue
      (green circles) and the azimuthally-averaged spectra (red
      diamonds). Characteristic error bars are only drawn for
      the radial-averaged spectra for the sake of clarity.
      See the caption of the similar \autoref{fig:radial_fxf} for a full
      explanation.
      \label{fig:radial_HII}
    }
\end{figure*}

After the quality of the derived spectra was confirmed by applying the same
sanity-checks as in the \fxf\ case,
we proceeded to correct the measured emission line intensities by extinction
in order to derive the physical properties from this new sample. Given the higher
signal-to-noise of the \hh regions compared with the previous samples, the
logarithmic extinction coefficient c(\hb) was obtained using both the
\ha/\hb\ and \hg/\hb\ line ratios. The extinction law of
\citet{Cardelli:1989p136} with a total to selective extinction ratio $R_V =
3.1$ was adopted for the interstellar reddening correction. Formal errors were
derived by propagating in quadrature the uncertainty in the flux calibration,
the statistical error of the line emission fluxes and the error in the c(\hb)
term. Additional information of the \hh regions catalogue, including ID,
coordinates, offsets, extraction method, equivalent/real aperture diameter (in
arcsec and pc), and the number of extraction fibres for each selected \hh
region are included in Appendix~\ref{app:online}.

\autoref{fig:bpt_HII} and \autoref{fig:radial_HII} show the
diagnostic diagrams and radial trends derived from the \hh region catalogue of
\ngc (green symbols), and a radial average spectra sample (reddish diamonds),
obtained after co-adding successive annulus of 10 arcsec in an azimuthally
radial way, as in the previous cases. Comparison of these plots with the
corresponding diagrams of the \fxf\ sample show that, in general terms, all
trends are exactly reproduced, with the difference being the lower level of
scatter in the case of the \hh region catalogue diagrams. As in the previous
similar plots, lighter colours in \autoref{fig:bpt_HII} correspond to inner
regions in the galaxy, darker tones to outer parts. In particular, Panel a) of
\autoref{fig:bpt_HII} shows a clear trend of increasing oxygen
intensity for both \oii\ and \oiii\ species as a function of radius. The trends shown
in the rest of the panels show much narrower correlations than in the case of the
\fxf\ analysis, extending to relatively low ratios of \oii\ \lam5007/\hb\ and
$\log R_{23}$, i.e. corresponding to the innermost regions of the galaxy. The
\nii/\oii\ ratio obtained from the \hh regions sample confirms
that all the spectra are consistent with $R_{23}$ values corresponding to the
upper branch of the O/H vs. $R_{23}$ relation.

The radial trends shown in \autoref{fig:radial_HII} are consistent with
those derived for the \fxf\ sample, where spectra of the inner regions of the
galaxy ($\rho \lesssim 0.3\rho_{25}$) were included by the selection
criteria. In the case of the \hh region catalogue, the number of regions
sampling this zone is low, but they are enough to indicate the trends of all the
line ratios and to produce reliable radial-averaged spectra. Furthermore, the
scatter in the inner regions has been reduced compared
to the similar  \fxf\ plots. For radii $\rho > 0.4\rho_{25}$, the trends for
all three different methods are practically identical, with a lower level of
scatter for the \hh regions sample. 
The position of the \hh regions cover practically all radii from the inner
regions of the galaxy (where the closest \hh region to the centre is located
at  $\rho/\rho_{25} \sim$ 0.01), to the outer parts (where the \hh region with
the largest radius is $\rho/\rho_{25} \sim$ 0.72). However there is a gap
between $\rho/\rho_{25} \sim$ 0.05 and 0.15 where no \hh regions are found.

\subsection{Comparison with H\,II regions from the literature}
\label{sec:lit}

\begin{table*}
  \begin{center}
    \caption[]
    {
      \hh regions within the IFS mosaic of \ngc\ observed by previous
      long-slit spectroscopic studies. 
      Columns. (1-2): \hh region identification in this work and the
      corresponding one found in the literature. 
      (3-4): Offsets of the coincident \hh regions in arcsec; in the case of
      PINGS, the offsets are reported with respect to the centre of the IFS
      mosaic ($\alpha$: 24.17410, $\delta$: 15.78392, 2000 equinox).
      (5): Galactocentric radius normalised to the size of the optical disc at the
      $B_{25}$ mag arcsec$^{-2}$ isophote.
      (6-7): Value of the $R_{23}$ index, defined as the ratio of the emission line
      intensities: (\oii\ \lam3727 + \oiii\ \lam\lam4959, 5007)/\hb.
      (8): References. 1: \citet{Talent:1983p3838}; 2: \citet{McCall:1985p1243};
      3: \citet{Bresolin:1999p3837}; 4: \citet{Ferguson:1998p224}; 5: \citet{vanZee:1998p3468}.
      \label{tab:lit}
    }
    \begin{tabular}{@{\extracolsep{\fill}} rlllccccccrrc }\hline
      \\[-5pt]
      &\multicolumn{2}{c}{Region ID} && \multicolumn{5}{c}{Region offset ($\Delta\alpha$,$\Delta\delta$)} &&
      \multicolumn{2}{c}{$R_{23}$}\\[2pt]
      \cline{2-3} \cline{5-9} \cline{11-12}\\[-5pt]
      &\multicolumn{1}{c}{PINGS} & \multicolumn{1}{c}{Literature} &&
      \multicolumn{2}{c}{PINGS} && \multicolumn{2}{c}{Literature} &
      $\rho/\rho_{25}$  & \multicolumn{1}{c}{PINGS} &\multicolumn{1}{c}{Literature} & Reference \\[2pt]
      & \multicolumn{1}{c}{\scriptsize (1)} &
      \multicolumn{1}{c}{\scriptsize (2)} & &
      \multicolumn{2}{c}{\scriptsize (3)} &&
      \multicolumn{2}{c}{\scriptsize (4)} & {\scriptsize (5)} &
      \multicolumn{1}{c}{\scriptsize (6)} &
      \multicolumn{1}{c}{\scriptsize(7)} &
      {\scriptsize (8)} \\[3pt]\hline
      \\[-4pt]
      1 & N628--22 & H292      &&   50   &   46   &&   49 &   52  & 0.23  & 1.67~(0.20) & 2.59~(0.39) & 1 \\[2pt]
        &          &           &&        &        &&      &       &       &             & 3.04~(0.34) & 2 \\[2pt]
        &          &           &&        &        &&      &       &       &             & 2.44~(0.17) & 3 \\[2pt]

      2 & N628--51 & H154-155  && -186   &   85   && -186 &   86  & 0.72  & 4.24~(0.19) & 4.84~(0.46) & 2 \\[2pt]
        &          &           &&        &        &&      &       &       &             & 5.29~(0.24) & 3 \\[2pt]

      3 & N628--56 & FGW628A   &&  -75   & -26    &&  -73 &  -29  & 0.26  & 1.30~(0.17) & 2.28~(0.08) & 4 \\[2pt]
        &          & H451      &&        &        &&  -74 &  -22  &       &             & 1.81~(0.27) & 1 \\[2pt]
        &          &           &&        &        &&      &       &       &             & 1.65~(0.18) & 2 \\[2pt]
        &          &           &&        &        &&      &       &       &             & 1.75~(0.18) & 3 \\[2pt]

      4 & N628--75 & H572      &&  -62   & -110   &&  -60 & -107  & 0.39  & 2.28~(0.23) & 2.97~(0.41) & 2 \\[2pt]
        &          &           &&        &        &&      &       &       &             & 2.82~(0.18) & 3 \\[2pt]

      5 & N628--84 & H598      &&   38   & -120   &&   42 & -116  & 0.41  & 2.66~(0.21) & 3.49~(0.52) & 1 \\[2pt]
        &          &           &&        &        &&      &       &       &             & 3.55~(0.41) & 2 \\[2pt]
        &          &           &&        &        &&      &       &       &             & 3.13~(0.25) & 3 \\[2pt]

      6 & N628--85 & +081-140  &&   77   & -137   &&   81 & -140  & 0.55  & 3.28~(0.15) & 3.68~(0.08) & 4 \\[2pt]

      7 & N628--87 & +044-175  &&   43   & -171   &&   44 & -175  & 0.60  & 4.92~(0.18) & 4.55~(0.11) & 4 \\[2pt]

      8 & N628--94 & H627      &&  -44   & -155   &&  -42 & -154  & 0.51  & 3.95~(0.19) & 4.80~(0.52) & 1 \\[2pt]
        &          &           &&        &        &&      &       &       &             & 5.02~(0.18) & 2 \\[2pt]
        &          &           &&        &        &&      &       &       &             & 4.24~(0.18) & 3 \\[2pt]

      9 & N628--96 & +062-158  &&   66   & -156   &&   62 & -158  & 0.57  & 2.89~(0.18) & 2.67~(0.16) & 4 \\[3pt]

      \hline
    \end{tabular}
  \end{center}
\end{table*}

The PINGS IFS mosaic of \ngc\ covers a substantial fraction of the disc of the
galaxy, therefore we are in the position to compare the emission line
ratios and metallicity abundances derived in this work, with the coincident
\hh regions analysed by previous long-slit spectroscopic studies. A total of 47
independent observations of \hh regions are reported in the literature for this
galaxy, the first observations were performed by \citet{Talent:1983p3838}
(hereafter Tal83) who observed 5 \hh regions from the catalogue of \citet{Hodge:1976p3867},
7 correspond to \citet{McCall:1985p1243} (hereafter MRS85), 18 to
\citet{vanZee:1998p3468} (hereafter vZ98), 6 to \citet{Ferguson:1998p224}
(hereafter FGW98), 7 to \citet{Bresolin:1999p3837} (hereafter BKG99) and 4 to
\citet{Castellanos:2002p3372} (hereafter CDT02); the spectroscopic analyses by 
\citet{Zaritsky:1994p333} and \citet{Moustakas:2010p3856} made use of the line
ratios provided by the previous references. The 5 \hh regions observed by Ta83
were included in the sample observed by MRS85, the 7 regions observed
by BKG99 are the same as MRS85 (with different wavelength coverage), region 628A
from FGW98 is the same as region H451 from MRS85 and BKG99, and region H13
present in Tal83, MRS85 and BKG99 was also observed by CDT02. Therefore, the total
number of non-duplicated \hh regions found in the literature for \ngc\ is
33. 

Only 9 of these regions fall within the FOV of the IFS mosaic observed by
PINGS, they were identified with the corresponding \hh regions defined in the
previous section by comparing the offset reported in the literature and by
visual inspection with \autoref{fig:n628_id}. \autoref{tab:lit} lists the
coincident \hh regions observed by PINGS and previous long-slit spectroscopic
studies, 6 of them have multiple observations, with N628--56 having the maximum
number of references as it was observed by Tal83, MRS85, FGW98 and BKG99. 
Note that direct comparison between the \hh regions reported in
\autoref{tab:lit} has to be taken with caution, as the
identification is somewhat ambiguous given the differences in the offsets
between PINGS and the literature (due to the relatively arbitrary choice of the
reference point in each study), but most importantly, to the different
extraction apertures between long-slit observations and the fibre-defined \hh
regions of PINGS. \autoref{tab:lit} shows also the value of the $R_{23}$ index
for each of the coincident regions compared with those reported in the
literature. 
While in some cases we find a relative agreement (e.g. N628--51, 75,
85, 87, 96), the PINGS values are lower that the literature values in 7 of the
9 cases. However, even if we just consider the values found in the literature
we find some important deviations in the reported $R_{23}$ index
(e.g. N628--22, 51, 94). A thorough discussion of
the comparison between the emission line ratios and the derived properties of
these coincident \hh regions will be presented in the next section.

\subsection{On the spectra selection criteria}


In this section we have explored two methods in order to 
extract spectra from an IFS mosaic in order to perform a 2D spectroscopic
study. The first one considered the spectra contained in single fibres to
be representative of the physical conditions of those regions sampled by the
fibre aperture, while the second consisted in creating a catalogue of
``classical'' \hh regions by co-adding fibres corresponding to the same
morphological regions. The analysis of both samples resulted in similar trends
but with a much reduced scatter in the case of the \hh region spectra sample.
From this exercise, we might be tempted to conclude that the best, or
optimal selection method consists in obtaining a \hh region catalogue in the
``classical'' sense, i.e. integrating the emission from a group of fibres
associated to a given emitting region. However, we may leave open the
possibility that --to some level-- the scatter seen in the \fxf\ sample might be
actually due to intrinsic point-to-point variations of the emitting
regions. It is important to note that this information is lost in typical
long-slit spectroscopy, where the spectra is obtained very similarly as in the
\hh region catalogue method. The capability to detect these point-to-point
variations, if real, might be one of the power of IFS observations. Some
implications of the above findings are discussed in the following sections.


\section{The IFS-derived abundance gradient of NGC\,628}
\label{sec:grad}


The gas-phase chemical content of \ngc\ has been previously analysed in a
number of long-slit spectroscopic studies
\citep[e.g.][]{Talent:1983p3838,McCall:1985p1243,Zaritsky:1994p333,vanZee:1998p3468,Ferguson:1998p224,Bresolin:1999p3837,Castellanos:2002p3372,Moustakas:2010p3856}.
These works have derived the abundance gradient of \ngc\ up to
relatively large galactocentric radii ($\rho \sim 1.7R_{25}$), using mainly empirical
metallicity indicators based on the ratios of strong emission lines. 
They have found a higher metallicity content in the inner part of the
galaxy, that the slope of the gradient is {\em constant} across the range of
galactocentric distances sampled by the different studies, that the oxygen
abundance decrease is relatively small, and that the average metallicity content
is relatively high, consistent with solar and super-solar values. However, these
results have been drawn from relatively few spectroscopically observed \hh
regions, and none of these within a radius of $0.2R_{25}$.

Apart of the previous spectroscopic studies mentioned in \autoref{sec:grad},
\citet{Belley:1992p3779} made use of imaging spectrophotometry, i.e. using
narrow-band interference filters with the bandpass centered on several key
nebular lines, in order to derive reddenings, \hb\ equivalent widths,
diagnostic line ratios and metallicities for 130 \hh regions across a large area
of the galaxy disk. They found that the excitation, and some diagnostic line
ratios are strongly correlated with galactocentric radius. They were able also
to derive an oxygen abundance gradient of \ngc, based on the \oiii/\hb\
ratio. Although strictly speaking, their results were not based on
spectroscopic observations, this work represented an early and successful
attempt to obtain the 2D distribution of the emission line properties of \ngc.


The PINGS observations of \ngc\ allow to perform for the first time a full 2D
spectroscopic abundance analysis based on the spectra samples obtained in the
previous section, with an unprecedented number of spectroscopic data points.
The reddening corrected line ratios for both the \fxf\ and \hh region catalogues
were used to derived the oxygen abundance for each individual spectrum, using
a subset of the abundance diagnostics employed in
\citetalias{Sanchez:2011p3844}. Different abundance estimators were used in
order explore the effects of a particular calibration depending on the
physical properties of the galaxy
\footnote{For a review of the most common empirical calibrations used to
  estimate the nebular oxygen abundances see \citet{Kewley:2008p1394}, and 
  for a discussion of the differences between those indicators (within the
  context of 2D spectroscopic data), see 
  \citet{LopezSanchez:2010p3920,LopezSanchez:2011p3919}.}, they correspond
to a $R_{23}$-based calibrator \citep[][hereafter KK04]{Kobulnicky:2004p1700}, 
an ``index-empirical'' method after \citet[][]{Pettini:2004p315} (hereafter
PP04), and two additional strong-line empirical methods proposed by
\citet{Pilyugin:2005p3401} and \citet{Pilyugin:2007p3381}.

The \citetalias{Kobulnicky:2004p1700} calibrator is based on the stellar
evolution and photoionization grids from \citet{Kewley:2002p311}, and
takes into account the effects of the ionization parameter, providing
parametrizations for both branches of the $R_{23}$ relation.
A guess value of the metallicity has to be first inferred
depending on the R$_{23}$ branch (previously determined from the
\nii/\oii\ ratio, these nominal values (12+log(O/H)$_{lower}$ = 8.2 and
12+log(O/H)$_{upper}$ = 8.7) are used to calculate an ionization parameter $q$,
i.e.

\begin{eqnarray}
  \label{eq:kk1}
  \log q &=&  \{ 32.81 - 1.153y^2  \nonumber\\
         & &  + z( - 3.396 - 0.025y + 0.1444y^2 ) \} \nonumber\\
         & &  \times \{ 4.603 - 0.3119y - 0.163y^2 \nonumber\\
         & &  + z( - 0.48 + 0.0271y +0.02037y^2  ) \}^{ - 1},
\end{eqnarray}

\noindent where $z \equiv {12 + \log ({\rm{O/H}})}$, $ y \equiv \log O_{32}$ and

\begin{equation}
  O_{32} =   {\frac{{\left[ {{\rm{O\,III}}  } \right]\lambda\lambda 4959,5007}}{{\left[ {{\rm{O\,II}}} \right]\lambda 3727}}}.
\end{equation}\vspace{5pt}

\noindent The initial resulting ionization parameter is used to derive an
initial metallicity estimate depending on the R$_{23}$ branch

\begin{eqnarray}
\label{eq:kk2}
12 + \log ({\rm{O/H}})_{{\rm{lower}}} & = & 9.40 + 4.65x - 3.17x^2  \\ 
                                   &  & - \log q( {0.272 + 0.547x - 0.513x^2 }), \nonumber\\   
\nonumber\\
\label{eq:kk3}
12 + \log ({\rm{O/H}})_{{\rm{upper}}} & = & 9.72 - 0.777x - 0.951x^2  \\
                                   &  &- 0.072x^3  - 0.811x^4 \nonumber\\
                                   &  & - \log q( 0.0737 - 0.0713x \nonumber\\
                                   &  & - 0.141x^2 + 0.0373x^3  - 0.058x^4 ), \nonumber
\end{eqnarray}

\noindent where $x = \log R_{23}$. Equations \ref{eq:kk1} and \ref{eq:kk2} (or
\ref{eq:kk3}) are iterated until 12+log(O/H) converges.


The $O3N2$ index was first introduced by \citet{Alloin:1979p2878}, a
slightly different definition was proposed by \citetalias{Pettini:2004p315}

\begin{equation}
O3N2 = \log \left( {\frac{{\left[ {{\rm{O\,III}}} \right]\lambda 5007/{\rm{H}}\beta }}{{\left[ {{\rm{N\,II}}} \right]\lambda 6584/{\rm{H}}\alpha }}} \right) ,
\label{eq:o3n2}
\end{equation}\vspace{1pt}

\noindent considering only the \oiii\ \lam5007 line in the numerator. 
This ratio is sensitive to the metallicity as measured by the oxygen
abundance through a combination of two effects. As O/H decreases below solar,
there is a tendency for the ionization to increase (either from the hardness
of the ionizing spectrum or from the ionization parameter, or both),
decreasing the ratio [N\,{\footnotesize II}]/[N\,{\footnotesize III}]; on the
other hand, the N/O ratio decreases at the high-abundance end, due to the
secondary nature of nitrogen. The inclusion of \oiii\ could be useful in the
high metallicity regime where \nii\ saturates but the strength of
\oiii\ continues to decrease with increasing metallicity. This
index is almost independent of reddening correction or flux calibration.
\citetalias{Pettini:2004p315} fitted the observed relationship
between this ratio and $T_e$ based metallicities, obtaining monotonic
relationship given by

\begin{equation}
12 + \log ({\rm{O/H}}) = 8.73 - 0.32 \cdot O3N2 ,
\end{equation}\vspace{1pt}

\noindent valid for $O3N2 < 2$, with an estimated accuracy of $\sim$ $\pm$0.25
dex.

The third considered calibrator is the ff-$T_e$ method, i.e. the combination of
the flux-flux (or $ff$--relation) proposed by \citet{Pilyugin:2005p3401}, and an
updated version of the $T_e$-based method for metallicity determination
\citep{Izotov:2006p3429}. The ff-relation links the flux of the auroral line
\oiii\ \lam4363 to the total flux in the strong nebular lines \oii\ \lam3727 and
\oiii\ \lam4959, \lam5007. This relation is metallicity-dependent at low
metallicities, but becomes independent at metallicities higher than 12+log(O/H)
$\sim$ 8.25, i.e. the regime of high-metallicity \hh regions. Using this
relation, an inferred value of the \oiii\ \lam4363 line can be derived, which
translates to an electronic temperature of the high-ionisation zone $t_3 \equiv$
$t$(\oiii). Defining the following notations:

\begin{equation}
   R  =  \frac{{\left[ {{\rm{O\,III}}} \right]\lambda 4363}}{{{\rm{H}}\beta }},
\end{equation}

\begin{equation}
 R_2  =  \frac{{\left[ {{\rm{O\,II}}} \right]\lambda 3727}}{{{\rm{H}}\beta }},\\
\end{equation}

\begin{equation}
 R_3  =  \frac{{\left[ {{\rm{O\,III}}} \right]\lambda 4959 + \left[ {{\rm{O\,III}}} \right]\lambda 5007}}{{{\rm{H}}\beta }},
\end{equation}

\begin{equation}
 R_{23}   =  R_2  + R_3 .
\end{equation}

\noindent we can express the excitation parameter $P$, as

\begin{equation}
 P = \frac{{R_3 }}{{R_2  + R_3 }} .
\end{equation}

\noindent The $ff$ relation is defined as the relationship between the flux $R$
in the auroral line and the total flux $R_{23}$ in the strong nebular lines
through a relation of the type  $\log R  = a + b \log R_{23}$, but since $R_{23}
= R_3/P$, the $ff$ relation can be also expressed in the form $R=f(R_3,P)$. This
last relation was parametrised by \citet{Pilyugin:2006p3376} in the following
way

\begin{eqnarray}
\log R &=& -4.151 - 3.118\log P \nonumber\\
       & & + 2.958\log R_{23} - 0.680(\log P)^2.
\end{eqnarray}

\noindent From this equation, a ratio of the nebular \oiii\ \lam4363 line to
H$\beta$ is obtained, and therefore, a value of the electron temperature $T_e$
can be derived. This temperature, coupled with the observed strong-line intensity
ratios are used in order to derive the chemical abundance using the revised
direct method by \citep{Izotov:2006p3429}. The abundances derived through this
method will be referred as the ff--$T_e$ abundances.

The last of the considered empirical calibrators is based on the prediction of
the ratio Q$_{\rm \nii}$ = \nii\ \lam\lam6548,6584/\nii\ \lam5755 from $R_2$ and
$P$, using a calibration of high metallicity \hh regions after
\citet{Pilyugin:2007p3381} 

\begin{eqnarray}
\log {\rm Q_{ \nii}} &=& 2.619 - 0.609 \log R_2 - 0.010 [\log R_2]^2 \\
               & & + 1.085 \log(1-P) + 0.382[\log(1-P)]^2.\nonumber
\end{eqnarray}

\noindent From this ratio, the value of the $t_e$(\nii) temperature is determined, which is
taken as the characteristic temperature of the O$^+$ low ionization region. Assuming 
$t_e$(\nii) $\approx$ $t_e$(\oii) $\approx$ $t_e$(\sii) $\equiv$ $t_2$, the $t_2$
temperature is determined via

\begin{equation}
 t_2 = \frac{1.111}{\log {\rm Q_{ \nii}} - 0.892 - 0.144\log t_2 + 0.023 t_2}.
\end{equation}

\noindent The $t_2-t_3$ relation between the O$^+$ and O$^{++}$ zones electron
temperatures proposed by \cite{Pilyugin:2007p3381} is used to derive the
$t_3$(\oiii) temperature, which is characteristic of the zone of high ionization:

\begin{equation}
 t_2 = 0.41\frac{1}{t_3} - 0.34P +0.81.
\end{equation}

\begin{figure*}
  \centering
    \includegraphics[width=\textwidth]{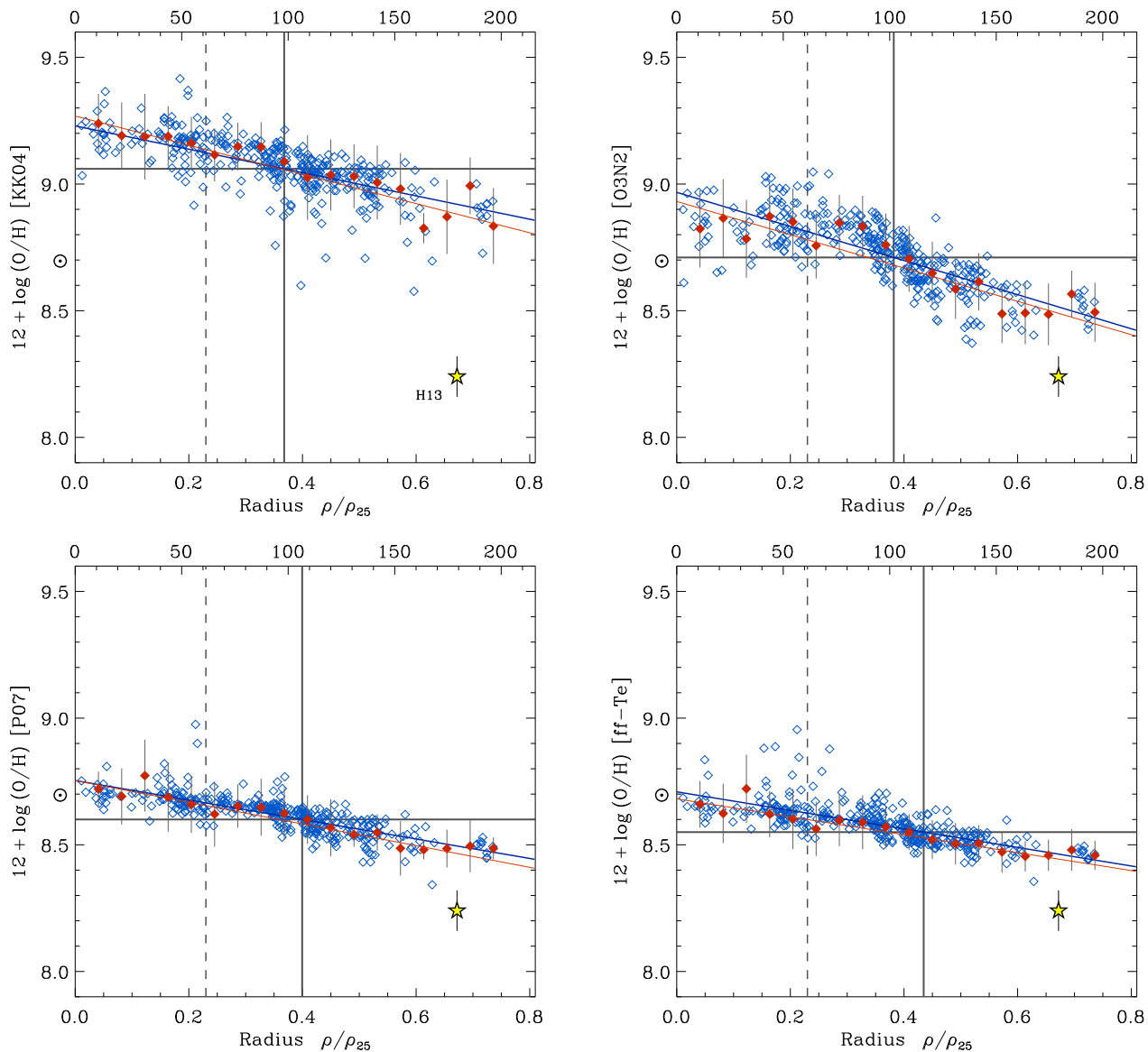}
    \caption[Method I: radial abundance gradients for NGC\,628]
    {
      Radial abundance gradients
      derived for \ngc\ based on the \fxf\ analysis (blue symbols) for
      different abundance diagnostics. The red diamonds
      correspond to the azimuthally-averaged radial spectra, 
      for which characteristic error bars are drawn.
      The blue and red thick lines correspond to the linear least-squares fit
      to the \fxf\ and radial average data points, respectively. 
      The dashed vertical line marks the minimum radius of a \hh region
      previously reported by the literature.
      The horizontal grey lines correspond to the abundance derived for a
      given calibrator using the integrated spectrum as reported in Paper~I.
      The top $X$-axis values correspond to the projected radii in
      arcsec for the radial average data. 
      The yellow star corresponds to the {\em direct} oxygen abundance of the
      \hh region H13 derived by \citet{Castellanos:2002p3372}.
      \label{fig:grad_fxf}
  }
\end{figure*}

\noindent This relation takes into account the effects of the excitation
parameter $P$. Using these temperatures, the electronic density and the emission
line intensities, the chemical abundances are derived using the equations of the
direct method \citep{Izotov:2006p3429}. The abundances derived through this
method will be referred as the P07 abundances.

One-dimensional radial abundance gradients for \ngc\ were derived based on the 
\fxf\ and \hh region catalogue samples for the different abundance diagnostics
mentioned above. The radial gradients for the \fxf\ sample 
are shown in \autoref{fig:grad_fxf}. In each plot, the blue symbols correspond
to the spectra of the \fxf\ sample, the red diamonds
correspond to the azimuthally-averaged radial spectra, as in previous
figures. The radial average spectra show 1$\sigma$ error bars derived through
a Monte Carlo simulation by propagating Gaussian distributions with a width
equal to the errors of the emission line intensities, modulated by recomputing
the distribution over 500 times. The errors of the \fxf\ sample are slightly
larger (due to the lower signal-to-noise of the individual spectra), but
comparable in average with those calculated for the radial spectra (not drawn
for the sake of clarity).
The blue and red thick lines correspond to the linear least-squares fit
to the \fxf\ and radial average data points, respectively. 
In each panel, the values on the top $X$-axes correspond to the galactocentric
distance in arcsec, the horizontal line correspond to the oxygen abundance
value obtained from the integrated spectrum of the galaxy for that particular
calibrator \citepalias[see][]{Sanchez:2011p3844}. The vertical solid line
corresponds to the radius at which the \fxf\ linear fit (blue line) equals the
integrated abundance value. 
The solar abundance (12~+~log(O/H)$_{\odot}$ = 8.70,
\citealt{Scott:2009p3377}) is shown with the $\odot$ symbol in the $Y$-axes.
The vertical dashed line marks the minimum
galactocentric radius with reported observations of a \hh region in the
literature (H292 or FGW628A at 0.23$\rho_{25}$). Therefore, this work presents
for the first time observations and chemical abundances of the innermost
regions of \ngc.

The yellow star in all panels of \autoref{fig:grad_fxf} corresponds to the
oxygen abundance of the \hh region H13 derived by
\citetalias{Castellanos:2002p3372}, who were able to
determine observationally the electronic temperature from optical
forbidden auroral to nebular line ratios and to perform a {\em direct} oxygen
abundance determination. This region represents the unique {\em direct} abundance
measurement for \ngc\ reported in the literature, and it is included as an
{\em anchor} abundance reference in order to compare the scales and offsets of
the different abundance calibrators (the linear fits do not include this point
in the calculation). No other points from the literature were
considered when building the \fxf\ abundance gradient, as the ``source'' of
the spectra is conceptually different from those found in
the literature (which are based on spectra integrated within apertures
centered in classical \hh regions); consequently,  the baseline of the \fxf\
abundance gradient spans just up to $\sim 0.75\rho_{25}$.

\begin{figure*}
  \centering
    \includegraphics[width=\textwidth]{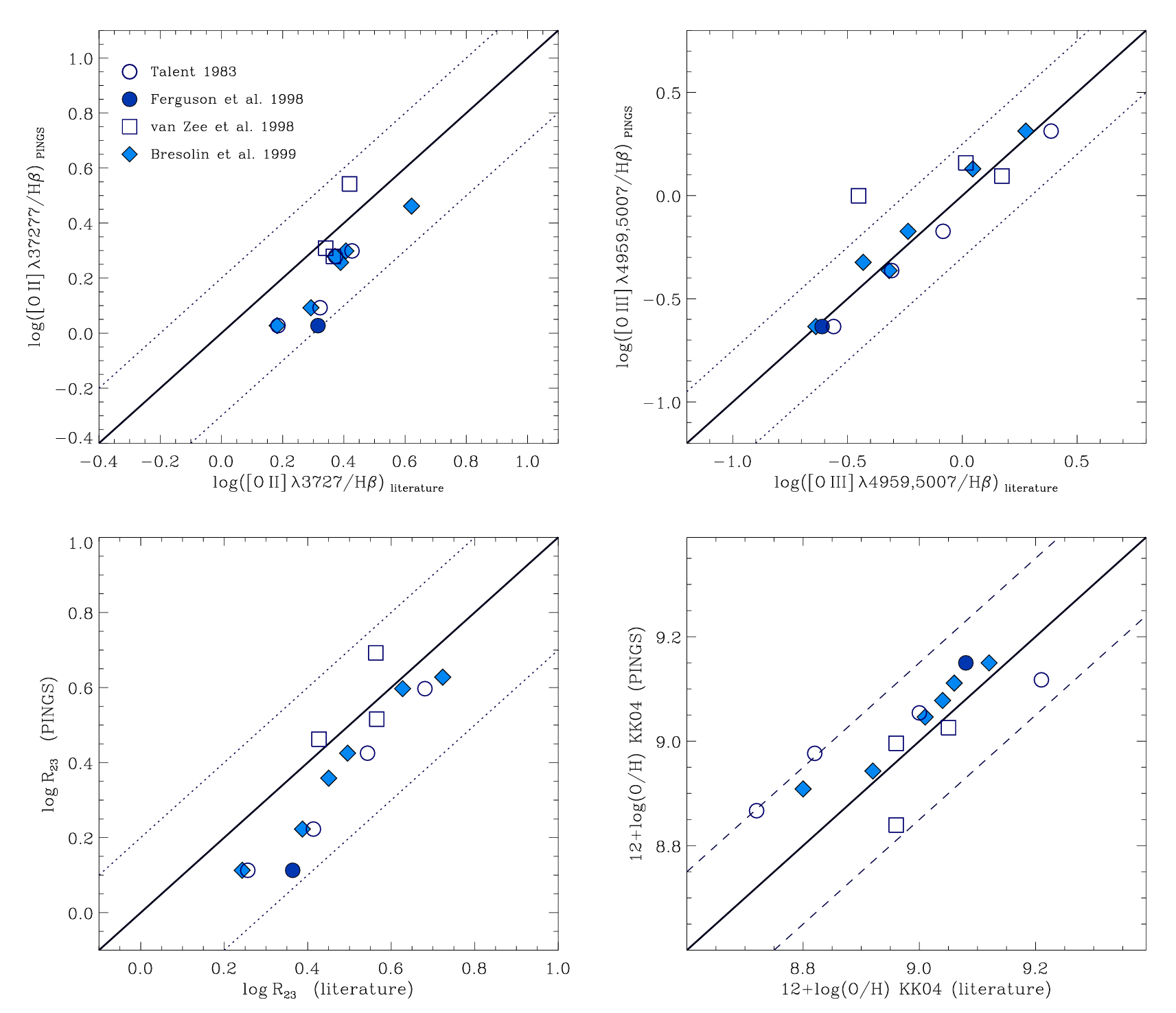}
    \caption[]
    {
      Comparison of the emission line intensities and oxygen abundance derived
      from this work with the coincident \hh regions found in the literature
      within the FOV of the IFS mosaic of \ngc. Top-left panel: \oii\
      \lam3727/\hb; top-right: \oiii\ \lam\lam4959,5007/\hb; bottom-left:
      $R_{23}$ index; bottom-right: 12~+~log(O/H) derived after the
      \citet{Kobulnicky:2004p1700} $R_{23}$-based calibrator. See the text for
      a full explanation.
      \label{fig:lit}
    }
\end{figure*}

The oxygen abundances derived using the $R_{23}$ KK04 method were calculated
using the corresponding branch parametrization, based on the values of the 
\nii/\oii\ ratio as shown in \autoref{fig:bpt_fxf}. The $R_{23}$-based KK04
method shows a noticeable higher mean oxygen abundance, a higher level of
dispersion for the same galactocentric radius, and a steeper slope than the
rest of the other diagnostic methods. The gradient derived from the
KK04 method is --0.46\,$\pm$\,0.04 dex $\rho^{-1}_{25}$. The maximum oxygen
values at $\rho = 0$ inferred from this gradient is: 12~+~log(O/H) = 9.23. The
oxygen abundance derived from the integrated spectrum matches the linear fit
for a radius $\rho/\rho_{25}$ very close to 0.4 ($\sim$ 100 arcsec).
The scatter of the data points is somewhat larger than the intrinsic errors of
the derived abundances, specially for regions located between 0.2 and 0.6 in
normalised radius units. On the other hand, the radial average spectra shows a
very linear relationship, with a low level of scatter. The gradient derived
from the linear fit of this sample corresponds to --0.58\,$\pm$\,0.06 dex
$\rho^{-1}_{25}$, with corresponding central oxygen abundances of 9.27 in
12~+~log(O/H) units.

The $O3N2$ derived gradient (top-right panel in \autoref{fig:grad_fxf})
presents a similar trend to the $R_{23}$ KK04 calibrator, although it does not
show a clear log-linear relationship as in the previous case. The pattern is
more consistent with a semi-sinusoidal trend, with a steep gradient between
0.25 and 0.55$\rho_{25}$, a flattening of the gradient for the innermost
and outermost regions of the galaxy (within the considered radial baseline),
and a lower mean oxygen content with respect to KK04.
The $O3N2$ abundances for those regions within 0.3$\rho_{25}$ show a large
level of scatter, which is due to the combined effect of the high dispersion of the
\nii/\ha\ and \oiii/\hb\ ratios observed for those zones (as shown in
\autoref{fig:radial_fxf}), but the scatter of the data is somewhat
within the calculated error bars. The abundance gradients derived from the
$O3N2$ calibrator are --0.67\,$\pm$\,0.03 and --0.66\,$\pm$\,0.07 for the
\fxf\ and radial average samples respectively, with central oxygen abundances of
8.97 and 8.93 in each case. Similarly to the $R_{23}$ method, the oxygen
abundance obtained from the integrated spectrum matches the $O3N2$ linear fit at
normalised radius $\sim$ 0.4.

The \fxf\ oxygen abundance gradient based on the P07 and ff--$T_e$ method are
shown in the bottom panels of \autoref{fig:grad_fxf}. Both show very similar
trends in terms of the mean abundance scale and steepness of the gradient.
The scatter of the data points along all galactocentric radii is of the same
level than the corresponding error bars. However, for regions of the galaxy at
$\rho \sim 0.2\rho_{25}$, the dispersion increases considerably with a number
of outlying data points approaching oxygen values of 12~+~log(O/H) $\sim$ 9.0
especially for the ff--$T_e$ calibrator. Considering that the signal-to-noise
of the \oii\ and \oiii\ emission lines for these inner regions of the galaxy
is relatively low, this behaviour suggests that this method is sensitive to
the signal-to-noise of the strong oxygen lines used in the determination of
the oxygen abundance. The abundance gradient derived from the
\fxf\ sample based on the P07 calibrator is --0.38\,$\pm$\,0.02 dex
$\rho^{-1}_{25}$ with a central abundance of 8.75, the same value as the
radial average fit, but with a steeper gradient consistent with
--0.43\,$\pm$\,0.03 dex $\rho^{-1}_{25}$. In the case of the ff--$T_e$ method,
the \fxf\ gradient is --0.37\,$\pm$\,0.02 dex $\rho^{-1}_{25}$, while
the slope obtained from the radial average is --0.35\,$\pm$\,0.03 dex
$\rho^{-1}_{25}$, with central abundances of 8.71 and 8.68 respectively,
i.e. the flatter gradients and lower central oxygen abundances among all the
considered calibrators. As in the previous cases, the oxygen
abundance determined from the integrated spectrum matches the ff--$T_e$ and
P07 radial trends for a normalised radius $\sim$ 0.4.
\autoref{tab:n628_grad1} summarises the results of the abundance gradients
analysis of \ngc\ for the \fxf\ sample. The first two columns for each of the
metallicity indicators correspond to the \fxf\ and radial average sample
respectively. For each sample, the values of the extrapolated central
abundance at galactocentric radius $\rho = 0$, of the {\em characteristic}
oxygen abundance at  $\rho = 0.4\rho_{25}$ (in 12\,+\,log(O/H) units), and the
slope of the abundance gradients in dex/$\rho_{25}$ and dex/kpc are presented.

Despite the evident scale offsets and small pattern variations among the
different oxygen calibrators, all the metallicity gradients are quite similar
in qualitative terms. 
They show a somewhat flat-distribution of the metallicity for the innermost
region ($\rho < 0.3\rho_{25}$) and a log-linear correlation between the oxygen
abundance and galactocentric distance for the outer radii ($0.3 < \rho/\rho_{25} < 0.8$),
the slopes derived from the \fxf\ and radial average samples are equivalent
within the errors, and as a common
interesting feature, the integrated abundance for all the indicators coincides
with the \hh region abundance gradient for a normalised radius $\rho/\rho_{25}
\sim 0.4$, i.e. that the integrated abundance correlates with the spatially
resolved abundances, independent of the abundance calibration used. This
effect was previously indicated by \citet{Moustakas:2006p313}, who observed
--to a first order-- that the abundance inferred from the integrated spectrum of
a galaxy is representative of the gas-phase oxygen abundance at  $\rho =
0.4\rho_{25}$, i.e. the {\em characteristic} abundance of a galaxy, even in
the presence of an abundance gradient, emission from the diffuse medium or
variations in dust reddening. They reached this conclusion based on the
($R_{23}$ based) \citet{McGaugh:1991p314} calibrator (hereafter M91) and the
\citet{Pilyugin:2005p308} (empirical-index) calibrator. The latter showed a
similar behaviour in their sample, but with a much higher dispersion. Here we
confirm this result for the case of \ngc, extending the number of calibrators 
also showing this trend.

Furthermore, the comparison between different calibrators helps us to infer
some trends that are not noticeable when employing only one type of
metallicity estimator. For example, the $O3N2$ and P07 methods somewhat
suggest a flattening of the gradient for the innermost of the galaxy (for
radii $< 0.3\rho_{25}$), and a large dispersion in the log(O/H) values, with
differences up to $\pm 0.5$ dex.
Obviously, the central ($\rho = 0$) and mean oxygen abundances are different
considering the well-known offsets between the empirical calibrators, 
but even taking into account these discrepancies, it is interesting that the
H13 \hh region with the {\em direct} oxygen abundance measurement (12~+~log(O/H)
= 8.24 at 0.67$\rho_{25}$), lies significantly below the metallicity content
inferred by any of the strong-line abundance calibrators at the same
galactocentric distance, especially with respect to the KK04 calibrator, with
an offset of $\sim$ 0.7 dex.
If the oxygen abundance of the H13 \hh region is considered reliable, these
offsets indicate that the empirical calibrations overestimate the metallicity
content of the \ngc\ \hh regions by a factor between 2 ($O3N2$, P07, ff--$T_e$)
and 5 (KK04), in accordance with recent studies comparing the abundances
provided by the direct method with those obtained through empirical
calibrations for star-forming regions
\citep{LopezSanchez:2010p3920,LopezSanchez:2011p3919}.

\begin{figure*}
  \centering
    \includegraphics[width=\textwidth]{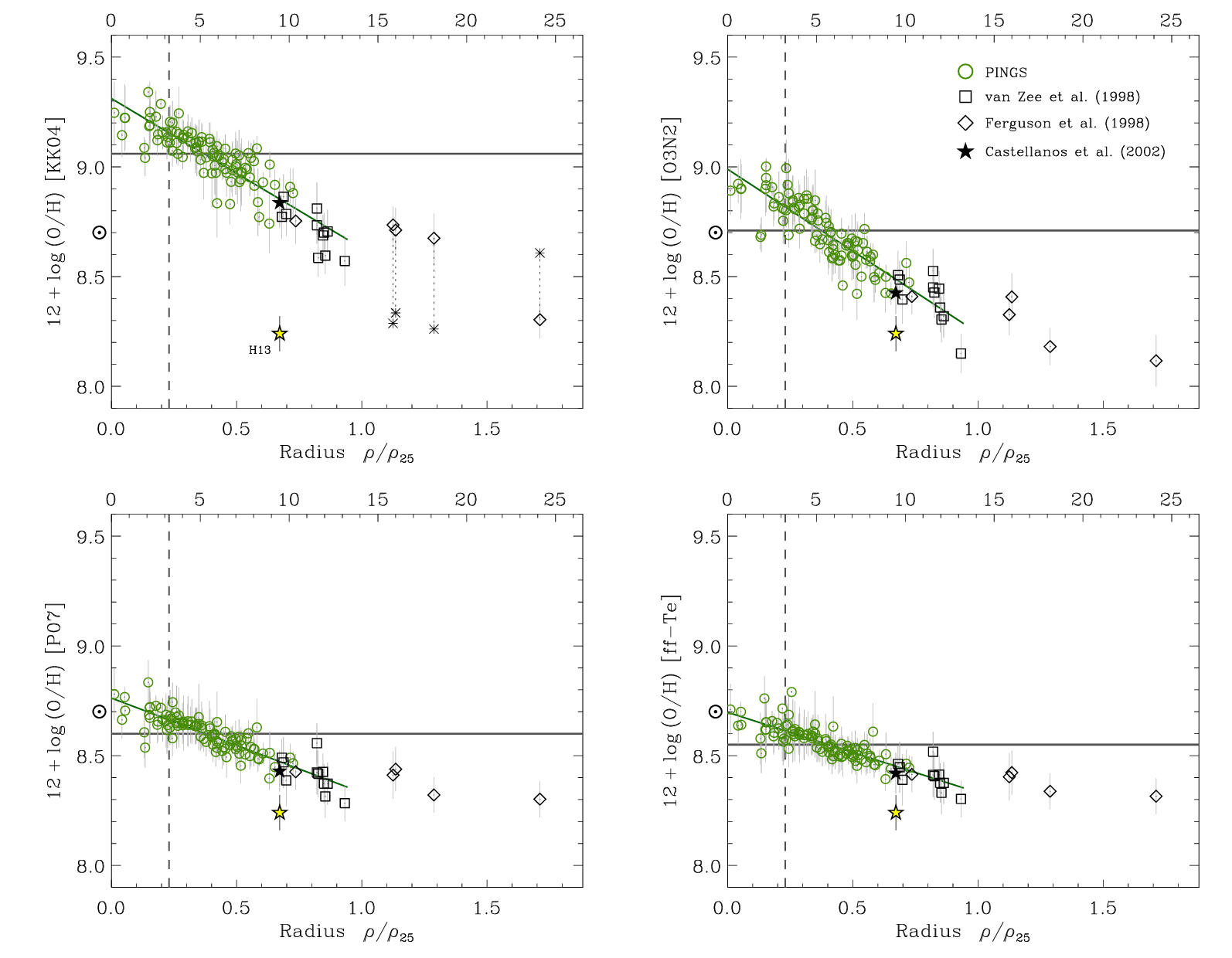}
    \caption[Method III: radial abundance gradients]
    {
      Radial abundance gradients derived for \ngc\ based on the PINGS \hh
      region catalogue (green symbols), and \hh regions from the literature
      (black symbols). 
      For the KK04 estimator, we show two dotted-connected
      abundance values for those \hh regions with an ambiguous choice on the
      $R_{23}$-branch.
      The yellow star corresponds to the $T_e$-derived abundance of the
      \hh region H13 by \citet{Castellanos:2002p3372}.
      Note the flattening of the gradient for radii $\gtrsim \rho_{25}$.
      See the text and the caption of the similar \autoref{fig:grad_fxf} for a
      full explanation.
      \label{fig:grad_HII}
    }
\end{figure*}


The large number of spectroscopic points provided by this work 
has allowed us to build the most detailed metallicity gradient
of \ngc\ within the optical radius of the galaxy. However, the radial
baseline of the \fxf\ sample is relatively short in order to account for the
true behaviour of the abundance gradient of \ngc. Nevertheless, we can made use
of emission-line measurements of \hh regions found in the literature (which span
up to $1.7\rho_{25}$), in order to cover a larger galactocentric baseline of the
galaxy. Oxygen abundances for the 47 reported \hh regions found in the literature
were calculated using published emission line ratios and by applying the four
strong-line calibrators introduced previously. These values were incorporated to
the oxygen abundances derived from the \hh region catalogue sample observed by
PINGS. Seventeen \hh regions from the literature are located at $\rho \gtrsim
0.7\rho_{25}$, i.e. the maximum baseline of the PINGS \hh region sample,
permitting a much better characterisation of the radial abundance gradient.

\begin{table*}
\centering
\caption[Abundance gradients results]{
  Radial abundance gradients for \ngc. For each of the metallicity calibrators
  employed, the results from the \fxf\ sample, radial average and \hh region
  catalogue are presented. 
  The rows correspond to: the central abundance at galactocentric
  radius $\rho = 0$; the {\em characteristic} abundance at  $\rho =
  0.4\rho_{25}$; and the slope of the abundance gradients in
  dex~$\rho^{-1}_{25}$ and dex~kpc$^{-1}$, respectively.
  Given the flattening of the abundance gradient for the outer parts of the
  galaxy, the values obtained from the linear fite were derived only for those
  regions with $\rho < \rho_{25}$, as explained in the text.
  \label{tab:n628_grad1}
}

{\small

\begin{tabular*}{\textwidth}{@{\extracolsep{\fill}} lrrrrrr }\hline\\[-5pt]
&
\multicolumn{3}{c}{KK04}  & \multicolumn{3}{c}{O3N2} 
\\[2pt]
\cline{2-4}
\cline{5-7} 
\\[-5pt]

& 
\multicolumn{1}{c}{\em Fibre-by-fibre} & \multicolumn{1}{c}{\em Radial} & \multicolumn{1}{c}{\em H~{\tiny II} regions} &
\multicolumn{1}{c}{\em Fibre-by-fibre} & \multicolumn{1}{c}{\em Radial} & \multicolumn{1}{c}{\em H~{\tiny II} regions} 
\\[2pt]
\hline\\[-5pt]

12\,+\,log(O/H)$_{\rho=0}$ &
   9.23~$\pm$~0.02  &  9.27~$\pm$~0.02  & 9.31~$\pm$~0.02 & 8.97~$\pm$~0.01  &  8.93~$\pm$~0.03 & 8.99~$\pm$~0.02 \\[5pt]

12\,+\,log(O/H)$_{\rho=0.4\rho_{25}}$ &
   9.05~$\pm$~0.02  &  9.04~$\pm$~0.02  & 9.04~$\pm$~0.02 & 8.70~$\pm$~0.01  &  8.67~$\pm$~0.03  & 8.69~$\pm$~0.02 \\[5pt]

log(O/H)$_{\rm (dex~\rho^{-1}_{25})}$ &
   --0.46~$\pm$~0.04  &  --0.58~$\pm$~0.06 & --0.68~$\pm$~0.03 & --0.67~$\pm$~0.03  &  --0.66~$\pm$~0.07  & --0.75~$\pm$~0.04\\[5pt]

log(O/H)$_{\rm (dex~kpc^{-1})}$ &
   --0.033~$\pm$~0.003  &  --0.041~$\pm$~0.004  & --0.048~$\pm$~0.002 & --0.048~$\pm$~0.002  &  --0.047~$\pm$~0.005  & --0.053~$\pm$~0.003\\[5pt]

\hline
\end{tabular*}

\vspace{10pt}



\begin{tabular*}{\textwidth}{@{\extracolsep{\fill}} lrrrrrr }
&
\multicolumn{3}{c}{P07} &  \multicolumn{3}{c}{ff--$T_e$}  
\\[2pt]
\cline{2-4}
\cline{5-7} 
\\[-5pt]

& 
\multicolumn{1}{c}{\em Fibre-by-fibre} & \multicolumn{1}{c}{\em Radial} & \multicolumn{1}{c}{\em H~{\tiny II} regions} &
\multicolumn{1}{c}{\em Fibre-by-fibre} & \multicolumn{1}{c}{\em Radial} & \multicolumn{1}{c}{\em H~{\tiny II} regions} 
\\[2pt]
\hline\\[-5pt]

12\,+\,log(O/H)$_{\rho=0}$ &
  8.75~$\pm$~0.01  &  8.75~$\pm$~0.01  & 8.76~$\pm$~0.01  & 8.71~$\pm$~0.01  &  8.68~$\pm$~0.01  & 8.70~$\pm$~0.01 \\[5pt]  

12\,+\,log(O/H)$_{\rho=0.4\rho_{25}}$ &
  8.60~$\pm$~0.01  &  8.58~$\pm$~0.01  & 8.59~$\pm$~0.01  & 8.56~$\pm$~0.01  &  8.54~$\pm$~0.01  & 8.55~$\pm$~0.01 \\[5pt]  

log(O/H)$_{\rm (dex~\rho^{-1}_{25})}$ &
  --0.38~$\pm$~0.02  &  --0.43~$\pm$~0.03  & --0.43~$\pm$~0.02 & --0.37~$\pm$~0.02  &  --0.35~$\pm$~0.03  & --0.37~$\pm$~0.02 \\[5pt]  

log(O/H)$_{\rm (dex~kpc^{-1})}$ &
  --0.027~$\pm$~0.001  &  --0.030~$\pm$~0.002  & --0.031~$\pm$~0.002 & --0.026~$\pm$~0.001  &  --0.025~$\pm$~0.002  & --0.026~$\pm$~0.001 \\[5pt]  

\hline 
\end{tabular*}
}
\vspace{5pt}

\end{table*}

%
%
%






The determination of the oxygen abundance for the \hh regions found in the
literature allow us to make a direct comparison of the metallicity content for
the coincident \hh regions within the FoV of PINGS, based on the values of the
published (literature) and observed (PINGS) emission line
intensities. \autoref{fig:lit} shows this comparison, the top-panels correspond
to the comparison between \oii\ \lam3727/\hb (left), and
\oiii\ \lam\lam4959,5007/\hb (right), while in the bottom we compare the
$R_{23}$ index (left), and the 12~+~log(O/H) derived after the
\citet{Kobulnicky:2004p1700} $R_{23}$-based calibrator, i.e. the method in which
we find the larger disagreement between PINGS and the literature abundance.
For the 6 common regions between \citetalias{McCall:1985p1243} and
\citetalias{Bresolin:1999p3837}, only those by \citetalias{Bresolin:1999p3837}
are drawn (blue-filled diamonds). N628--56 is also common with
\citetalias{Ferguson:1998p224}, shown as a blue-filled circle. The
regions common to \citetalias{vanZee:1998p3468} are drawn as open squares. In
each panel, the solid line represents equality between the values, the dashed
lines in the bottom-right (abundance) panel stand for a typical $\pm0.2$ dex
error associated with the determination of an oxygen abundance through the KK04
metallicity calibrator (and generally speaking, to any strong-line method). 
The dotted lines in the other panels stand for the error that the
corresponding emission line ratio (or index) should posses in order to account for
the $\pm 0.2$ dex error of the abundance determination, i.e. if we propagate
back the uncertainty in the metallicity determination as due to a difference in
a particular emission line ratio or index. Note that these errors are not
symmetric due to the definition of the KK04 calibrator (based on the $R_{23}$
index), as defined in \autoref{eq:kk1} to \autoref{eq:kk3}.

The values of the emission line ratios and the derived oxygen abundance show in
general a good agreement between both samples. Note that any difference between
the observed (PINGS) and published (literature) data fall within the errors, as
defined by the $\pm 0.2$ dex intrinsic uncertainty of the KK04 strong-line
calibrator, with the the exception of one \hh region from
\citetalias{vanZee:1998p3468} in the \oiii/\hb\ ratio. 
Any small deviation between the published and observed values is
somewhat expected, considering that: i) the spatial identification of the \hh
regions is ambiguous; ii) the extraction apertures differ importantly between
the previous long-slit and the fibre-based PINGS observations; iii) the emission
line intensities found in the literature were obtained directly from the
observed spectra, while in the case of this work, we measured the line
intensities from the residual spectra which took into account the effects of the
underlying stellar population; and iv) the reddening correction was performed with
different extinction laws. Considering all these facts, the good agreement
between the literature and PINGS values give us confidence regarding the
high-quality of the spectra, and in general to the abundance analysis method for
the IFS mosaic of \ngc.

The derived abundance gradients for the PINGS \hh region catalogue, plus the \hh
regions from the literature of \ngc\ are shown in \autoref{fig:grad_HII}, using
the same abundance estimators as in the \fxf\ case. In each panel, the green
open-circles correspond to the PINGS \hh region sample, while the black symbols
correspond to \hh regions from the literature. One-sigma error bars are display
for each data point. 
For the sake of clarity, only those regions from the literature with $\rho
\gtrsim 0.7\rho_{25}$ are drawn, since those with lower radii fall within the
cloud of the PINGS data points, following the same trend marked by the inner
\hh regions. 
The $T_e$-derived oxygen abundance of the H13 \hh region from
\citet{Castellanos:2002p3372} is shown as a yellow star. The values on the
top-horizontal axis correspond to the deprojected galactocentric distance in kpc.
The dashed vertical line marks the minimum galactocentric distance of a \hh
region found in the literature ($\rho \approx 0.2\rho_{25} \approx 3$ kpc).
At first glance, there is one noticeable
feature in all four panels of \autoref{fig:grad_HII}, i.e. that the negative
slope of the metallicity with increasing galactocentric radius changes
to a nearly constant abundance for regions with $\rho \gtrsim \rho_{25}$,
implying that there is flattening of the abundance gradient of \ngc\ for the
outermost regions of the galaxy, feature that has not been previously reported
for this galaxy.

All previous determinations of the abundances gradient for \ngc\ considered only
regions within $\rho \lesssim \rho_{25}$, (e.g. \citetalias{Talent:1983p3838},
\cite{Zaritsky:1994p333}, \citetalias{vanZee:1998p3468},
\cite{Moustakas:2010p3856}), with the exception of
\citetalias{Ferguson:1998p224}, who obtained an abundance gradient extending to
$\rho \sim 1.8\rho_{25}$ based on 6 \hh regions observed by these
authors (two of them within the radii of the PINGS mosaic), and by adding
7 \hh regions analysed previously by \citet{McCall:1985p1243} (all of them
within the range of observed \hh regions of this work), for a total of
13 regions. They made use of the $R_{23}$-based M91 calibrator, deriving a
steep gradient 12~+~log(O/H) = 9.08 $-(0.73\,\pm\,0.12)\rho/\rho_{25}$. The
flattening of the oxygen abundance is somewhat apparent in their work (see
Fig. 10 and 11 of \citetalias{vanZee:1998p3468}). However, the lack of spectroscopic
points between the inner and outer regions of \ngc, and the choice of the $R_{23}$
branch in the calibration for the outer \hh regions, made
\citetalias{Ferguson:1998p224} infer that the abundance gradient of the galaxy
was steep along all galactocentric radii.

The top-left panel of \autoref{fig:grad_HII} shows the abundance gradient of
\ngc\ derived after the KK04 $R_{23}$-based calibrator. For regions within
$\rho_{25}$, the galaxy presents a clear steep negative gradient. However, for
regions beyond that radius, the $R_{23}$ calibrator enters into the ill-defined
region between the two $R_{23}$ branches. For each of the outer \hh regions, the
asterisk mark the value of the oxygen abundance derived from the opposite branch
as defined by \citetalias{Kobulnicky:2004p1700}, a dotted line connects the values
of both branches. For those regions between $1 \rho/\rho_{25} 1.5$,
\citetalias{Ferguson:1998p224} took the lower-branch O/H value, while according
to the prescriptions of \citetalias{Kobulnicky:2004p1700} based on the
\nii/\oii\ ratio, we determine the upper-branch values. From this example, it is
evident that, when the metallicity is derived through a $R_{23}$-based
calibrator, the choice of the $R_{23}$ branch plays an important role in
defining the nature of the abundance gradient at larger radii, as it could lead
to different interpretations, e.g. that the slope of the gradient is constant
across all the galaxy, that there is a flattening of the oxygen distribution
after a certain radius, or even that there is a break of the metallicity content
of $\sim 0.3$ dex with respect to the edge of the optical disk.

In the top-right panel of \autoref{fig:grad_HII} we use the $O3N2$ empirical-index
calibrator \citepalias{Pettini:2004p315} to construct the abundance gradient of
\ngc. Apart from the well-known shift in the metallicity scale with respect to
\citetalias{Kobulnicky:2004p1700}, we note in the outer zones a feature that can
be either a flattening of the gradient to a constant metallicity, or a change in
the slope to a nearly flat gradient. This feature is also present in the other
two calibrators employed, P07 and ff$-T_e$, shown in the bottom panels of
\autoref{fig:grad_HII}. These three --conceptually different-- metallicity
estimators suggest that this effect is indeed real, that the oxygen abundance
decreases with increasing radius up to a point in which the variation in
metallicity is almost negligible, and that there is no discontinuity
in the radial abundance distribution, as one choice of the $R_{23}$ branch for the
\citetalias{Kobulnicky:2004p1700} calibrator might incorrectly suggest.

Recently, \citep{Bresolin:2009p3470} observed a flat oxygen abundance
gradient and a drop in abundance beyond the $\rho_{25}$ isophotal radius of
M83. The result was confirmed by a  wide range of abundance indicators and tests
considering the impact of different ionizations at larger radii. Likewise, 
\citet{Goddard:2010p3912} produced an accurate abundance gradient of the
XUV spiral galaxy NGC\,4625 out to $2.5\rho_{25}$, finding a flattering of the
radial gradient in the outer disc and an apparent discontinuity in the abundance
close to the optical edge of the disc. A possible interpretation of this
feature may be an intrinsic difference in the star formation rate between the
inner and outer disc linked to the 2D gas surface density. In this scenario,
the formation of massive stars in the outer regions is rare, and therefore
there is a lack of enrichment from supernovae in the outer disc compared with
the inner regions. The results from this work give support to the existence of
these features in nearby spiral galaxies, although the definite existence (and
interpretation) of these attributes will require further observations and a
larger statistical sample.

In all panels of \autoref{fig:grad_HII}, the green thick line corresponds to a
linear least-squares fit applied to those regions within $\rho/\rho_{25} < 1$. 
\autoref{tab:n628_grad1} shows the values derived from the abundance
gradient fitting of the \fxf, radial average and \hh regions samples.
Direct comparison among the different samples shows in general a good agreement
in terms of the derived slopes and central abundances for each calibrator, with
the exception of KK04, for which the central oxygen abundance is higher ($\sim
0.8$ dex) and the slope is much steeper ($\sim 0.2$ dex $\rho_{25}^{-1}$) for
the \hh region sample. The central oxygen values at $\rho = 0$ are in agreement
within 0.1\,dex with the values obtained previously for the \fxf\ sample. 
Likewise, the oxygen abundance derived from the integrated spectrum (horizontal
gray-line) matches the \hh region abundance of each calibrator for a radius
$\rho \sim 0.4\rho_{25}$ for all methods. The slope of the gradient varies also
considerably among different calibrators, with $O3N2$ showing the steepest
value ($-0.75 \pm 0.04$ dex $\rho_{25}^{-1}$), compared to the flatter slope of
the ff$-Te$ method ($-0.37 \pm 0.02$ dex $\rho_{25}^{-1}$), i.e. a difference of
$\sim 0.38$ dex. The reason of this behavior might reside in the fact that, the
metallicity determination with empirical indices based on the \nii\ emission lines
(e.g. $N2$, $O3N2$), overestimate the abundance at high N/O ratios and vice
versa \citep{PerezMontero:2009p3904}, causing a larger difference in metallicity
between the inner and outer \hh regions of the galaxy.

Another interesting feature present in all panels of
\autoref{fig:grad_HII}, is the high level of scatter in the oxygen abundance for
a given radius within the innermost regions of \ngc. In all cases, differences
of $\sim 0.3$ dex in log(O/H) can be found for a given $\rho$. Furthermore, in
all instances, the \hh region with the maximum oxygen abundance is not located
at $\rho/\rho_{25} \sim 0$, but at a distance of $\rho/\rho_{25} \sim 0.15$
($\sim$~2~kpc in physical scale), radius at which the higher dispersion in the oxygen
abundance is found.
As discussed in \citetalias{Sanchez:2011p3844}, previous IFS observations of the
stellar component in the central core of \ngc\ showed a ring-like structure and
an inversion of the metallicity gradient at $\sim$ 2~kpc
\citep{Wakker:1995p3788,Ganda:2007p3763}, somewhat suggesting a dynamical cold
inner disc.
Based on the results of this work, the metallicity of the gas seems to follow
the trend mark by the stellar populations at the centre of the galaxy.
This result, together with the confirmed presence of circumnuclear star-forming
regions in \ngc, might indicate a scenario in which the gas is being radially
transferred, inhibiting the enhancement of the gas-phase metallicity at the
innermost regions of the galaxy, where a moderate drop or flattening of the
abundance gradient can also be seen.
The combination of the inner and outer structure of the metallicity distribution of
\ngc\ suggests a {\em multi-modality} of the abundance gradient of the galaxy,
consistent with a nearly flat-distribution in the innermost regions of the
galaxy ($\rho/\rho_{25} < 0.2$), a steep negative gradient for $0.2 \lesssim
\rho/\rho_{25} < 1$, and a shallow or nearly-constant gradient beyond the
optical edge of the galaxy.

\subsection{The distribution of the N/O ratio}

\begin{figure}
  \centering
  \includegraphics[width=\hsize]{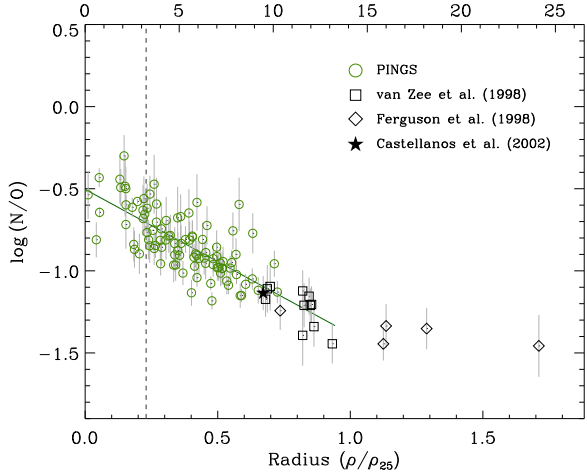}
  \caption[]
  {
    Radial distribution of the N/O ratio for the \hh region catalogue
    of \ngc, based on the results from the P07 calibrator. Green symbols
    correspond to the PINGS data, while black symbols stand for the \hh
    regions drawn from the literature. The linear fit, shown as a thick green
    line, was only performed to those regions with $\rho < \rho_{25}$.
    \label{fig:no}
  }
\end{figure}

The nitrogen-to-oxygen radial distribution is shown in \autoref{fig:no} with
similar symbol-colour coding that the previous figure. The N/O ratio was calculated
following the prescription by \citet{Hagele:2008p3480}:

\begin{eqnarray}
\log ({\rm N^+/O^+}) &=& \log \frac{\rm [N~II]\,\,\,I(6548+6584)}{\rm [O~II]\,\,\,I(3727)} + 0.281 \\\nonumber
                 & & - \frac{0.689}{t_e} + 0.089\log t_e
\end{eqnarray}

\noindent assuming N/O = N$^+$/O$^+$ and $t_e \equiv t({\rm \nii}) = t({\rm
  \oii})$, using the electronic temperature derived from the P07 calibrator.

The dispersion at a given radius is larger compared to the oxygen abundance
gradient, the reason might reside in local effects related to the time delays
between the release of primary and secondary nitrogen at different spatial
regions of the galaxy but with coincident galactocentric radii
\citep{VilaCostas:1993p3908}. 
The distribution of the N/O ratio in the inner ($\rho < \rho_{25}$) part of
\ngc\ decreases linearly with galactocentric radius, with a derived slope of the
N/O gradient ($-0.88$ dex $\rho_{25}^{-1}$) being two-times {\em steeper} than
that of the corresponding O/H gradient ($-0.43$ dex $\rho_{25}^{-1}$), as
shown in \autoref{tab:n628_grad2} for both the \fxf\ and \hh region catalogue
samples. The steeper N/O ratio is also present when considering the O/H
gradient based on the ff--$T_e$ abundance estimator, with differences within
0.05 dex $\rho_{25}^{-1}$. These is in agreement with a previous determination
of the N/O gradient for \ngc, obtained by
\citet{vanZee:1998p81} for 25 \hh regions within $\rho_{25}$. They derived a 
gradient $-0.57$ dex $\rho_{25}^{-1}$ for the N/O ratio and $-0.99$ dex
$\rho_{25}^{-1}$) for the O/H abundance, using the $R_{23}$-based
\citetalias{McGaugh:1991p314} calibrator. 
Nevertheless, as in the case of the oxygen
abundance gradient, the N/O ratio as a function of radius shows a clear
flattening (and a lower level of dispersion) for regions $\rho >
\rho_{25}$. This feature was also noted by
\citetalias{Ferguson:1998p224}, who inferred that the outer \hh regions are
consistent with a combination of primary and secondary production of nitrogen.
Previous observations of the outermost \hh regions in spiral galaxies have
also found similar patterns and constant N/O ratios to those found in low
metallicity \hh regions in dwarf galaxies, where the production of nitrogen is
mainly due to massive stars \citep[e.g.][]{Garnett:1990p3906,Thuan:1995p3907}.

\begin{figure}
  \centering
  \includegraphics[width=\hsize]{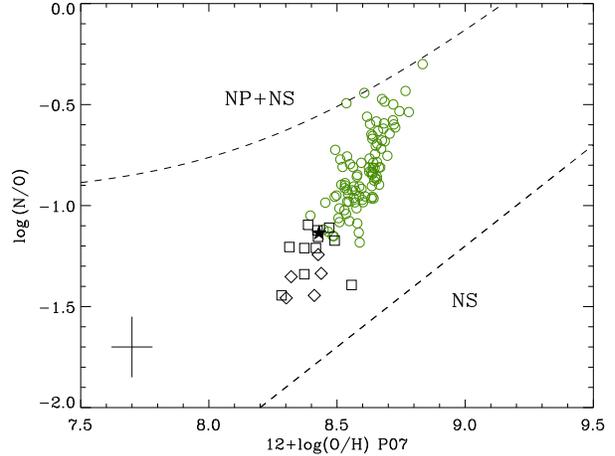}
  \caption[]
  {
    Observed N/O ratio as a function of the oxygen abundance
    12~+~log(O/H) as calculated using the P07 calibrator. Colour-symbol coding
    as in \autoref{fig:no}. The dashed lines show the relative abundance when N
    is secondary (NS), or a combination of primary and secondary (NP+NS).
    \label{fig:no2}
  }
\end{figure}

\begin{table}
\centering
\caption[N/O abundance gradient]{
  Radial gradient of the N/O ratio for \ngc, based on the results from the P07
  calibrator for both the \fxf\ and \hh region catalogue samples.
  The rows correspond to: the central value at galactocentric
  radius $\rho = 0$; the {\em characteristic} ratio at  $\rho =
  0.4\rho_{25}$; and the slope of the gradient in
  dex~$\rho^{-1}_{25}$ and dex~kpc$^{-1}$, respectively.
  The linear fit was only performed with those regions with $\rho <
  \rho_{25}$.
  \label{tab:n628_grad2}
}

{\small

\begin{tabular*}{\hsize}{@{\extracolsep{\fill}} lrr }\hline\\[-5pt]
&\multicolumn{2}{c}{P07}\\[2pt]
\cline{2-3} \\[-5pt]
& \multicolumn{1}{c}{\em Fibre-by-fibre} & \multicolumn{1}{c}{\em H~{\tiny II} regions} \\[2pt]
\hline\\[-5pt]

log(N/O)$_{\rho=0}$          &  -0.59~$\pm$~0.06 & -0.50~$\pm$~0.03  \\[5pt] 

log(N/O)$_{\rho=0.4\rho_{25}}$ &  -0.92~$\pm$~0.06 &  -0.86~$\pm$~0.03 \\[5pt]

log(N/O)$_{\rm (dex~\rho^{-1}_{25})}$ & -0.83~$\pm$~0.13 &  -0.88~$\pm$~0.06 \\[5pt]  

log(N/O)$_{\rm (dex~kpc^{-1})}$ &  -0.059~$\pm$~0.010 & -0.062~$\pm$~0.004 \\[5pt] 
\hline
\end{tabular*}

}
\end{table}

%
%
%

Further comparisons between the radial gradients of oxygen and nitrogen
can provide us with clues regarding the chemical evolution of the galaxy. 
If nitrogen is a secondary element, the N/O ratio is expected to vary linearly
with O/H with a slope $\geqslant 1$. In that case, the N/O gradient should be
somewhat steeper than that shown by O/H \citep{Henry:2000p3913}. On the other
hand, if the production of primary nitrogen is important, then galaxies tend
to have shallower O/H gradients \citep{vanZee:1998p81}. 
As shown in \autoref{fig:no2}, in the case of \ngc\ the N/O ratio increases
linearly with oxygen abundance, as expected in the high metallicity regime of
\ngc, indicating that nitrogen is predominantly of secondary nature
\citep{Molla:2006p3851}. However, the innermost regions of the galaxy (with
larger O/H) present N/O ratios which are larger than those expected by the
pure secondary nature of nitrogen, resembling the N/O behaviour of some
circumnuclear star-forming regions \citep[CNSFR][]{Diaz:2007p102}.
This might be a signature of time delay between the release
of oxygen and nitrogen, deviations from closed-box chemical evolution models, or
even the presence of dynamical processes such as gas infall or outflow. 
In fact, as previously indicated by \citet{Molla:2006p3851} using theoretical
evolutionary models, the star formation efficiency plays a relevant role in
the evolution of the relative abundance N/O, i.e. that differences in the star
formation history of a particular galaxy (and/or regions within them),
can explain the dispersion of the data of the N/O vs. O/H (and therefore
vs. $\rho$) plane. Further improvements in the methods to calculate abundances
and in the theory of galaxy chemical evolution are needed to better understand
these trends.

\subsection{Reliability of the O/H abundance gradient}
\label{sec:scatter}

\begin{figure}
  \centering
  \includegraphics[height=21cm]{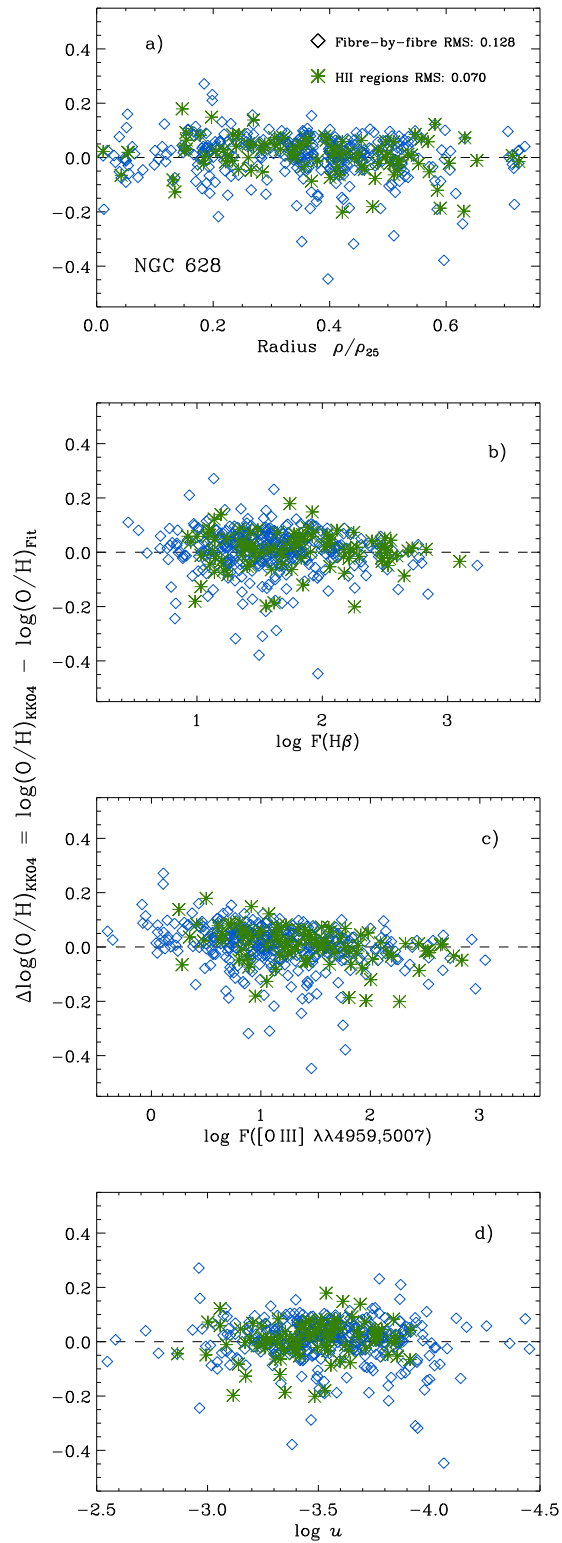}
  \caption[]
  {
    Scatter of the KK04 metallicity as measured by the
    $\Delta$log(O/H)$_{\rm KK04}$ value, calculated with respect to the
    linear fit of the \fxf\ sample (blue symbols) and the \hh region catalogue
    (green symbols). Panel a): as a function of galactocentric radius; Panel
    b): as a function of the observed flux in \hb; Panel c): as a function of
    the observed flux in \oiii\ \lam\lam4959,5007; Panel d): as a function of
    the ionization parameter $\log u$.
    \label{fig:scatter}
  }
\end{figure}

The linear fits for the inner region ($\rho < \rho_{25}$) of \ngc\ shown in
\autoref{fig:grad_HII} are based on a relatively large number of \hh regions,
considering that typical spectroscopic studies are performed with a handful of
\hh regions from which previous abundance gradients have been derive. 
As stated by many studies \citep[e.g.][]{Diaz:1989p3307}, most of the \hh
regions observed in external galaxies fall under the category of {\em Giant
  Extragalactic \hh regions} (GEHR). These correspond to very large \hh
regions with dimensions up to 1\,kpc, and with a substantial number of ionizing
stars. Being such extended regions, the nebulae would include zones of
different physical conditions, gas in different degrees of ionization and
different amounts of reddening. Therefore, the observations of these regions
would lead to systematic errors of the total derived abundance, if the latter is
inferred from calibrators based on photoionization models
which fail to provide adequate ionization correction factors, ICF (i.e. the ratios
between the total abundances of the various elements and the abundance in a
single state of ionization), and/or if they do not consider the 3D geometrical
distribution of the ionizing sources.
As suggested by EBS07, if compact clusters (small \hh regions) and loose
associations (GEHRs or \hh complexes) are randomly distributed throughout
a given galaxy, the systematics errors on the observed metallicities at a
given radius due to the effects mentioned above would only cause a larger
scatter at a given galactocentric distance, and given a sufficient number
statistics they would not affect the overall metallicity distribution.
On the other hand, if the location of compact clusters and large
associations are somewhat dependent on the galactocentric radius, then the
systematic errors may introduce a bias on the measured metallicity gradient if
the abundances are obtained from methods that do not take into account the
differences on the excitation between small and large \hh regions.

One way of assessing this issue is by comparing the results from metallicity
indicators based on photoionization models (e.g. KK04), and those based
on purely empirical methods. As discussed above, the most evident
differences between the abundance estimators consists in the absolute
metallicity scale and difference in the steepness of the gradient.
However, considering that we benefit from a nearly complete spectroscopic
coverage of small and large \hh regions over the area of the galaxy
where a gradient is present, give us the possibility to explore the systematics
between different ways of obtaining the abundance gradients. 
One possibility is to select different sets of spectra in order to explore
variations and systematics of the derived abundance gradients. This has been
partially done, by considering the different spectra samples analysed so far.
Nevertheless, the \hh region catalogue offer us the possibility to
discriminate between \hh regions of different brightness and sizes, and
therefore we can simulate for example, the abundance gradient determination
based only on a few bright \hh regions of a galaxy, like performed in most
``classical'' studies, and compare the results with the abundance gradients
obtained from the full distribution of \hh regions.

In order to perform this exercise, we chose 15 large (characteristic aperture
$>$ 8 arcsec) and bright (log(L(\ha)) $> 39$) \hh regions
from the PINGS catalogue of \ngc, distributed over a good range of
galactocentric distances, and selected from the four quadrants of the
galaxy as defined in \autoref{fig:n628_id}. 
The minimum galactocentric distance of these sample is $\rho/\rho_{25} \sim
0.18$, and the maximum is  $\rho/\rho_{25} \sim 0.9$.
As the derived gradients using these regions may be biased by their somewhat
arbitrary selection, a statistical approach was followed in order to derive
the abundance gradients from a {\em bright} \hh region subsample. 
Eight regions were extracted randomly from the sample of 15 large/bright
regions; the abundance gradients were then calculated from this \hh region
subsample. The process was repeated over 50 times for each abundance
calibrator. Then we obtained a gradient, central oxygen and characteristic
abundance from the mean values of the 50 realizations.
The derived gradients resulted in very similar trends for all the considered
calibrators, with variations in the central and {\em characteristic} oxygen
abundances of the order of $\pm$\,0.05 dex. In terms of the derived slope, we
find relatively low differences of the order of $\pm$\,0.1 dex $\rho_{25}^{-1}$,
but consistent in average with {\em steeper} slopes and {\em higher} central
oxygen abundances, compared to the abundance gradients obtained from the full
catalogue of \hh regions. The reason for this variation resides mainly in the
spatial sampling of the {\em bright} \hh regions, as in the innermost part of
the galaxy there are practically no \hh regions which fall under that category,
and therefore, the \hh regions at $\rho/\rho_{25} \sim 0.2$, which happen to
posses a higher O/H abundance than the inner \hh regions, have a higher weight
in the linear fitting of the {\em bright} subsample, causing a higher central
oxygen abundance and the steeper slope. However, in the case of \ngc\ these
differences are negligible, considering the intrinsic systematic errors of the
strong-line abundance determinations. Therefore, it seems that --to a
first-order-- the spatial distribution of compact clusters and large association
of the galaxy do not affect significantly the measured metallicity gradients.

Another feature that might influence the determination of the metallicity
gradient with a large number of spectroscopic points is the scatter in the
metallicity for a given radius. By examining \autoref{fig:grad_fxf} and
\autoref{fig:grad_HII}, we note that different abundance estimators present a
different dispersion in the O/H values for a given $\rho$, with the KK04
$R_{23}$-based calibrator presenting (statistically) the higher level of
scatter. One relevant question in the 2D IFS study of \ngc\ is whether this
dispersion in metallicity is real or is just an artifact, maybe produced by the
failure of a given metallicity calibrator in estimating the true abundance, due
either to spectra of low S/N or spectra that do not correspond to the physical
conditions for which the strong-line calibrator was modelled or calibrated.

\begin{figure*}
  \centering
  \includegraphics[width=0.49\textwidth]{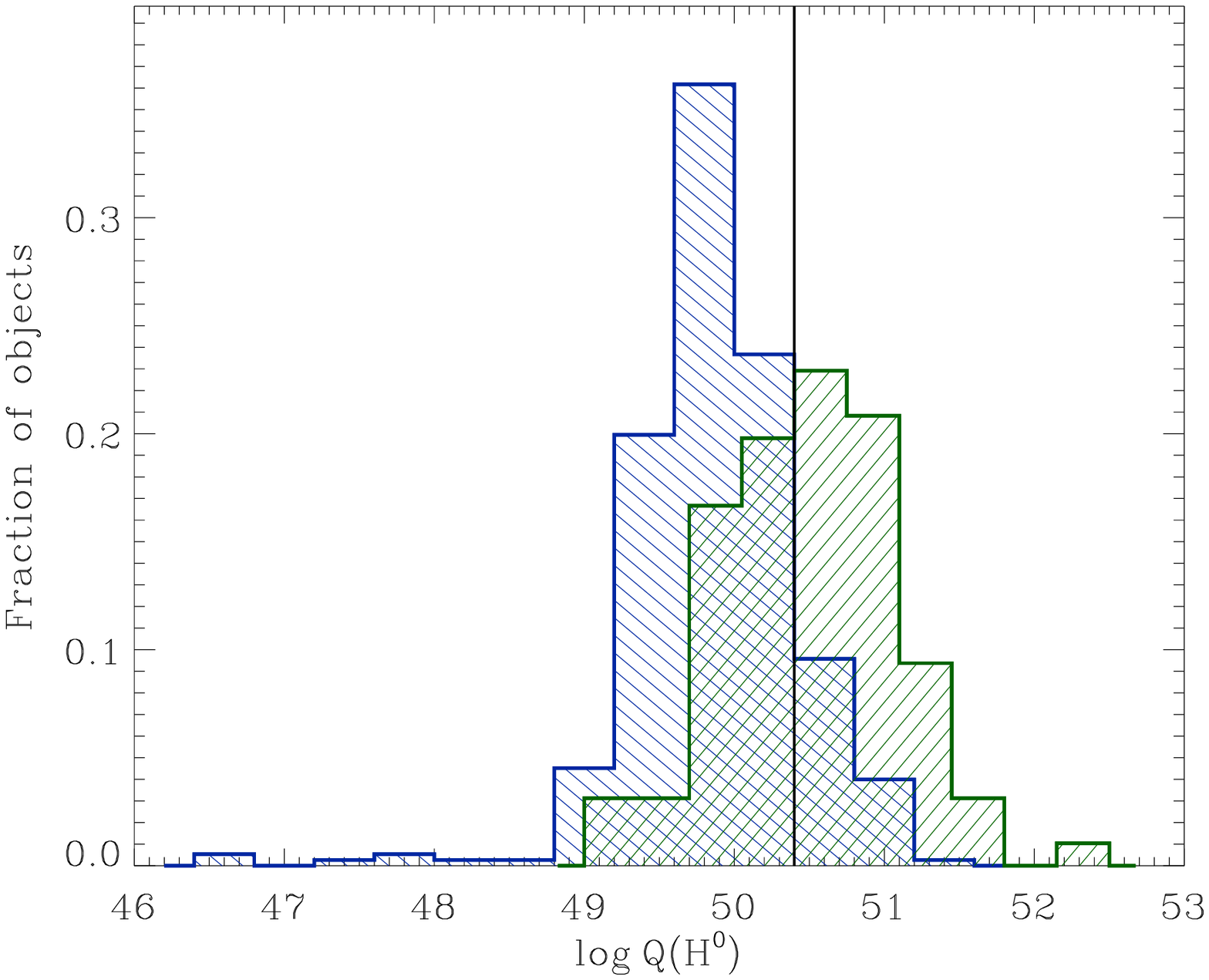}\hspace{0.2cm}
  \includegraphics[width=0.49\textwidth]{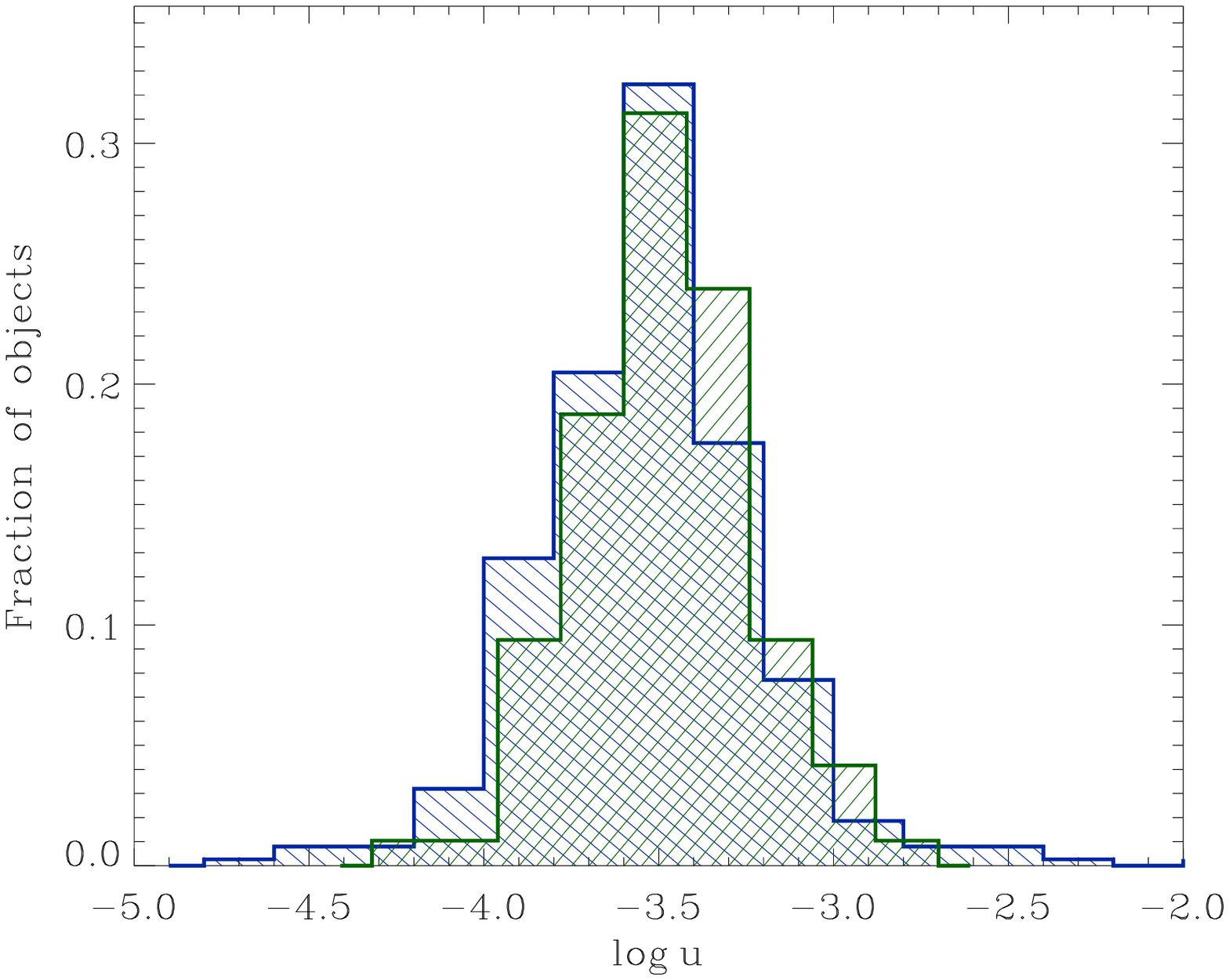}
  \caption[]
  {
    {\em Left}: Histogram of the number of ionizing Lyman continuum photons
    Q(H$^0$), as derived from the \ha\ luminosity of the individual spectra in
    the \fxf\ sample (blue) and \hh region catalogue (green). The thick
    vertical line corresponds to the log~Q(H$^0$) value for an ionizing
    cluster of 10$^4$ M$_{\odot}$ which is free of stochastic effects. 
    {\em Right}: Histogram of the {\em ionization parameter} $\log u$ for
    both spectroscopic samples.
    \label{fig:q0}
  }
\end{figure*}

A possible explanation of the high level of scatter found in the \fxf\ analysis
may be statistical variations due to spectra that, despite the quality selection
criteria, are of relatively low signal-to-noise compared with the integrated
spectra of the \hh regions catalogue. In this scenario, the line intensity
variations would be just reflecting spurious effects due to the relatively
weakness of those emitting regions.
Panel a) of \autoref{fig:scatter} shows the value 
$\Delta$log(O/H)$_{\rm KK04}$ for each spectrum of the \fxf\ sample (blue
symbols) and the \hh region catalogue sample (green symbols), calculated with
respect to the log(O/H) value derived from the corresponding linear fit (at
the same $\rho$ than the observed spectrum), i.e. the {\em scatter} of the
log(O/H)$_{\rm KK04}$ value as a function of radius. From this figure we can
note that the level of scatter is higher for the \fxf\ sample ($rms = 0.128$)
than for the \hh region catalogue ($rms = 0.070$), but that in general terms
(with the exception of a few outliers), the dispersion in metallicity does not
vary as a function of radius, i.e. there is not a preferential spatial position
at which the scatter in the oxygen abundance is larger. The reason of the higher
scatter for the \fxf\ sample is most probably due to the smaller spatial
sampling of the individual fibres compared to the integrated spectra of the \hh
regions, where the differences in the line emission has been averaged. Panels b)
and c) of \autoref{fig:scatter} show the $\Delta$log(O/H)$_{\rm KK04}$ value as a
function of the observed flux in \hb\ and \oiii\ \lam4959 + \lam5007
respectively. If the scatter in the metallicity was due to the low S/N of
the observed spectra, then we would expect a ``butterfly'' pattern (e.g. see
\autoref{fig:n628_ratios}),
consistent with higher dispersion in $\Delta$log(O/H)$_{\rm KK04}$ for lower
line strengths (where a lower S/N is expected), and vice versa. However, we do
not see that pattern in panels b) and c) of \autoref{fig:scatter}, the level of
scatter is relatively similar over more than two orders of magnitude in each
case, and where the largest $\Delta$log(O/H)$_{\rm KK04}$ values correspond to
relatively high line intensities (especially in the \oiii\ lines).

On the other hand, if the observed dispersion in metallicity was due to the
physical conditions of the emitting region (e.g. gas in different degrees of
ionization), then we could expect that regions with different levels of
excitation are more prone to deviations in the calculated oxygen abundance. 
Panel d) of \autoref{fig:scatter} shows the $\Delta$log(O/H)$_{\rm KK04}$ value
as a function of the ionization parameter $\log u$. Again, we note that the
scatter in metallicity does not seem to depend on a certain ionization
parameter, and that the level of dispersion is similar for both spectroscopic
samples over two orders of magnitude in $\log u$. All the above results suggest
that, if the scatter in metallicity does not depend on any particular region
of the galaxy, that is not due to the possible low S/N of the observed
spectra, and that shows no systematic dependence with the ionization
conditions of the gas, then the dispersion in metallicity for a given
galactocentric radius is real, and it is reflecting a true spatial physical
variation of the oxygen content.

Another question that might be raised regarding the reliability of the
metallicity determination is related to the validity of the strong-line
abundance calibrators applied to each spectrum of both spectroscopic samples. 
The empirical methods used to derive the oxygen abundance, such as those based
on the $R_{23}$ index, were developed for \hh regions which are large enough
so that they contain a substantial number of ionizing stars in order to be
free of stochastic effects. Given the relatively small spatial sampling of the
IFS mosaic of \ngc\ --particularly that of the \fxf\ sample-- we may be in a
regime in which the observed spectra is originated in regions with a Lyman
continuum produced by only a few ionizing stars, fact that could lead to
systematic errors in the abundance derived through strong-line methods.
One way of assessing this issue is to calculate the flux of ionizing photons
as measured by the observed \ha\ luminosity for a given emitting region. 
The left-panel of \autoref{fig:q0} shows the histograms of the number of
ionizing Lyman continuum photons Q(H$^0$), for both the \fxf\ sample (blue)
and \hh region catalogue (green), calculated through the relation between
Q(H$^0$) and \ha\ luminosity, Q(H$^0$) = $7.31 \times 10^{11}$ L(\ha),
following \citet{GonzalezDelgado:1995p3909}. The \fxf\ sample shows an
unimodal distribution with a peak at log~Q(H$^0$) $\sim$ 49.8 and very few
outliers, while the \hh region sample is consistent with a slightly broader
distribution with a peak at log~Q(H$^0$) $\sim$ 50.4. The shift of the \hh
region distribution to higher log~Q(H$^0$) values can be simply explained by
the fact that the spectra of the \hh regions catalogue is the integrated value
of several fibres within an given aperture, which increases the observed \ha\
luminosity and the corresponding Q(H$^0$) value.

We can compare these distributions with theoretical predictions of 
the \ha\ luminosity emitted by a \hh region ionized by ionizing
clusters. According to \citet{GarciaVargas:1995p3910}, the minimum mass of a
stellar cluster in order to be free of stochastic effects is $\sim 10^4$
M$_{\odot}$. An ionizing cluster of 10$^4$ M$_{\odot}$ with an age of 4 Myr
and solar metallicity produces (in average) a maximum of log~Q(H$^0$) $\sim$
50.4 (\citet[e.g. Starburst99][]{Leitherer:1999p3911};
\citet{GarciaVargas:1995p3910}), which corresponds to a luminosity L(\ha) =
$3.44 \times 10^{38}$ erg s$^{-1}$. The vertical thick line in the left-panel of
\autoref{fig:q0} marks this value, which coincides nearly with the peak of the
distribution of the \hh region catalogue sample, and it is slightly higher
than the emission of most of the spectra in the \fxf\ sample. This exercise
shows that --to a first order-- the determination of the metallicity abundance
through the strong-line indicators can be considered reliable for both
samples, as the observed \ha\ luminosity corresponds to that expected for
relatively large \hh regions, in which the flux of ionizing photons is
produced by stellar clusters with the necessary mass to overcome problems due
to stochastic effects. 
On the other hand, the right-panel of \autoref{fig:q0} shows the histograms of
the ionization parameter $\log u$ for both spectroscopic samples. Both
histograms show an unimodal distribution with a nearly coincident peak at 
$\log u$ $\sim -3.5$, i.e. a relatively low level of excitation. Therefore, 
as there are not strong deviations in the excitation conditions of the
emitting gas, we can apply (and compare) with confidence any given strong-line
calibrator for both samples, knowing that the derived metallicity does not
depend importantly on the ionization conditions of the emitting gas.

As discussed in \citetalias{RosalesOrtega:2010p3836} and \citetalias{Sanchez:2011p3844},
it has been argued that several factor may
prevent an accurate determination of the chemical abundance of a \hh region,
some of them related to the intrinsic geometrical structure of the emitting
region, e.g. the existence of electron temperature fluctuations, the
presence of chemical inhomogeneities, the geometrical distribution of the
ionization sources, the possible depletion of oxygen on dust grains, etc.
If true, some or all of these effects may account for the large scatter seen
in the determination of metallicities, even if the electron temperature can be
estimated directly from the spectra.

If we take as a premise that for a sufficiently large \hh
region, the emission line measurements are aperture and spatial dependent,
i.e. that the light is emitted under different physical conditions, by gas in
different degrees of ionization, and modified by different amounts of
reddening (and therefore providing different elemental ionic abundances), 
the scatter seen in the \fxf\ analysis may be due to the
intrinsic distribution of the ionizing sources, gas content, dust
extinction and ionization structure within a given region, i.e. we would be
sampling real point-to-point variations of the physical properties within a \hh
region. Two factors may give support to this hypothesis; first, the
high level of consistency between the results from the \fxf\
sample and the \hh region catalogue in the spectroscopic analysis of this (and
other) galaxies. Second, the fact that the aperture of the PPAK fibres is
relatively large compared with typical widths employed in long-slit
spectroscopy ($\sim$ 1--3 arcsec), and therefore, at the assumed distances of
\ngc, each fibre samples regions larger than 100 pc, i.e. a
physical size which in principle, would subtend a large-enough volume of
ionized gas capable of emitting a detectable, physical spectrum.
These factors, combined with the fact that the \fxf\ sample was
selected after applying a relatively strict quality selection criteria, and the
sanity checks performed in this section,  may
suggest that, {\em in the scenario of intrinsic physical and geometrical
variations in an emitting nebula}, the observed dispersion in the derived
line ratios and chemical abundances might be due to real variations of these
properties within a single \hh region.

Therefore, considering that the distribution of compact and large \hh regions
does not seem to depend on the galactocentric radius, that the observed
scatter in metallicity is due to intrinsic spatial variations within the
emitting nebulae, that the observed spectra in both samples is originated (in
average) by ionizing clusters which are free of stochastic effects (regime in
which the strong-line abundance calibrators can be considered reliable), and
that there is an unimodal distribution of the ionization conditions between
the two samples, we can conclude that given a sufficient number statistics of
\hh regions in the galaxy, the measured galactic metallicity gradient is not
biased by any systematic error introduced by the effects mentioned above, and
that the IFS abundance gradient of \ngc\ can be considered reliable.


\section{Asymmetries in the nebular properties of NGC\,628}
\label{sec:angle}

\begin{figure*}
  \centering
  \includegraphics[width=0.49\hsize]{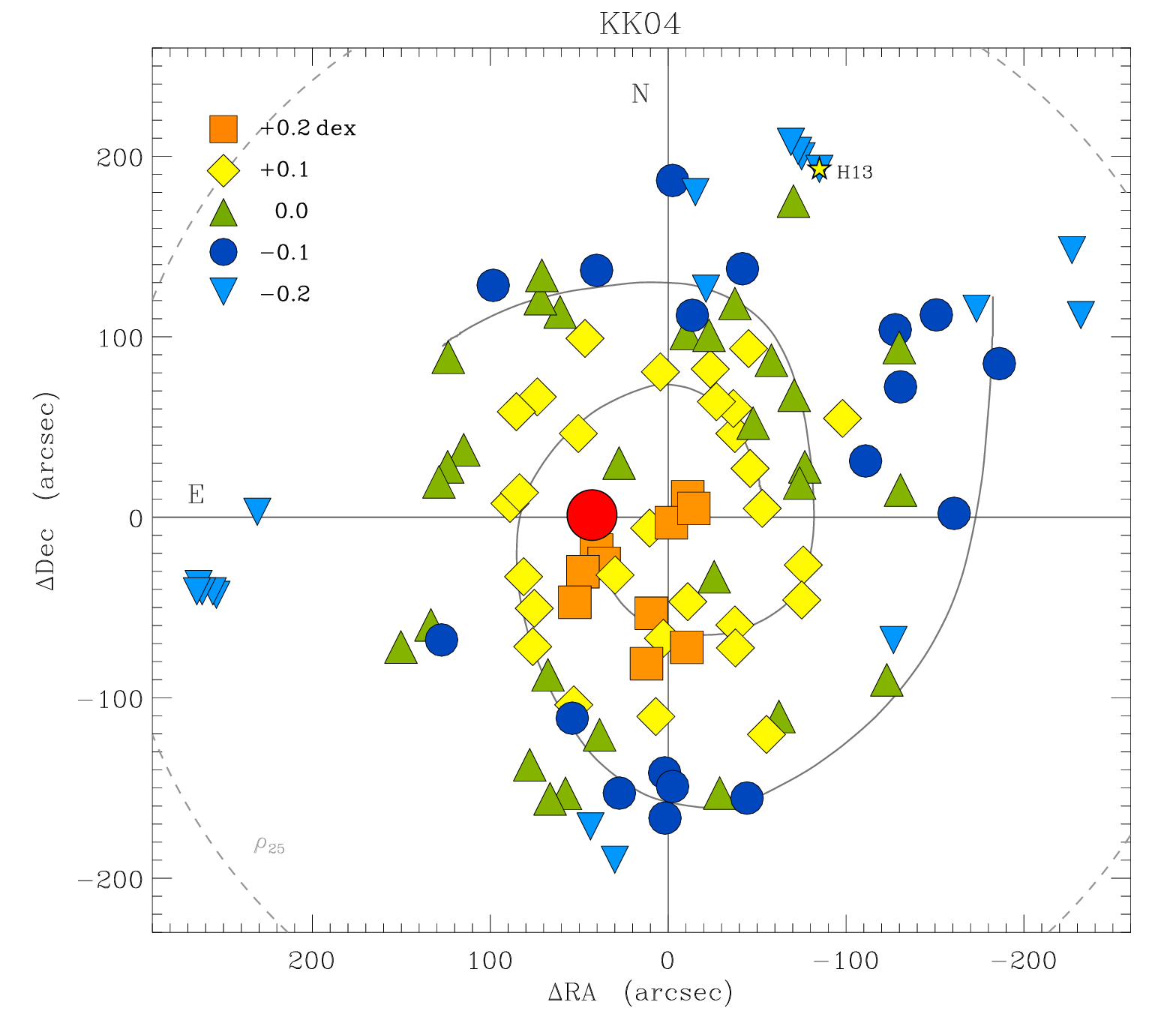}\hspace{0.1cm}
  \includegraphics[width=0.49\hsize]{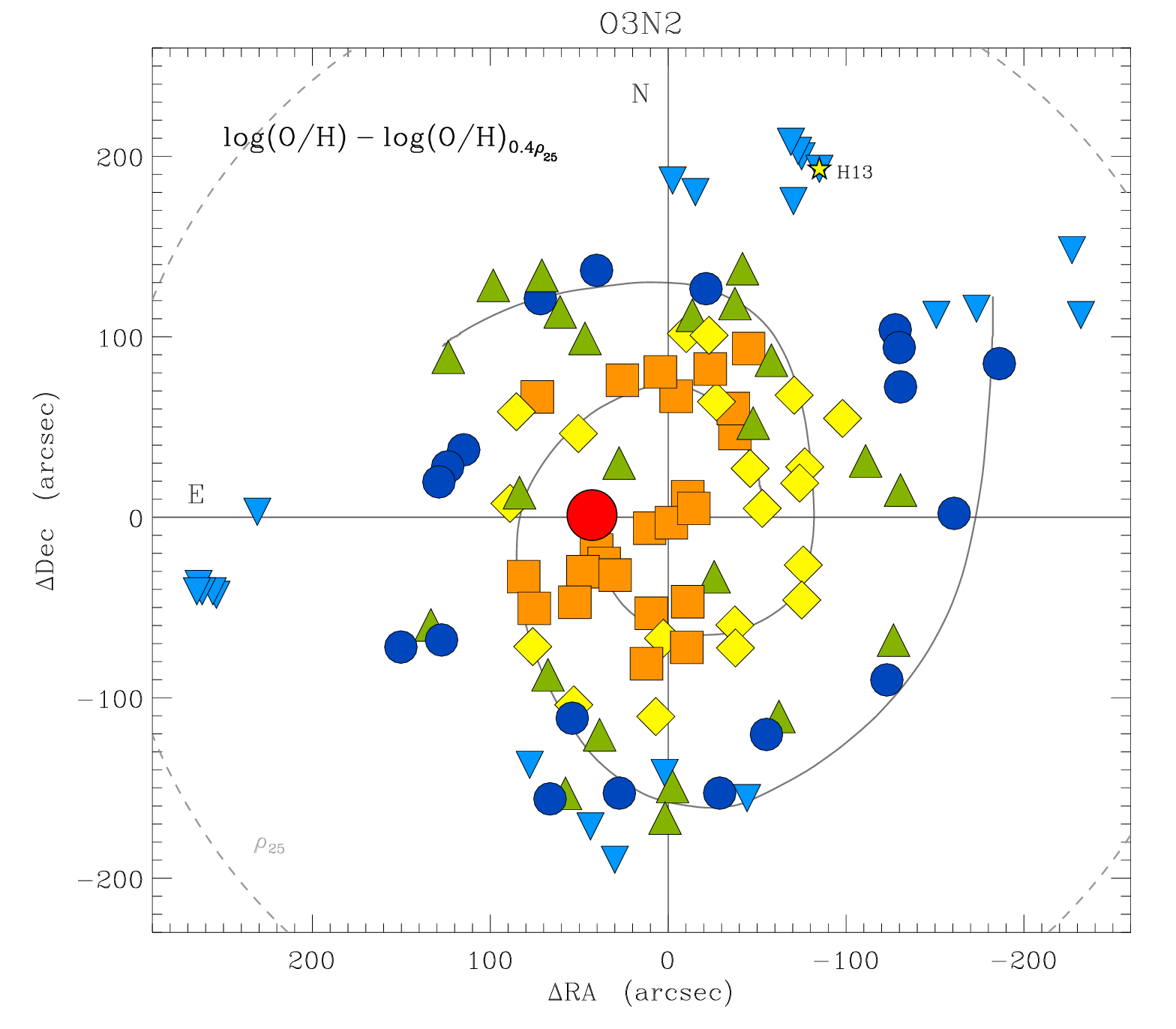}\vspace{0.1cm}
  \includegraphics[width=0.49\hsize]{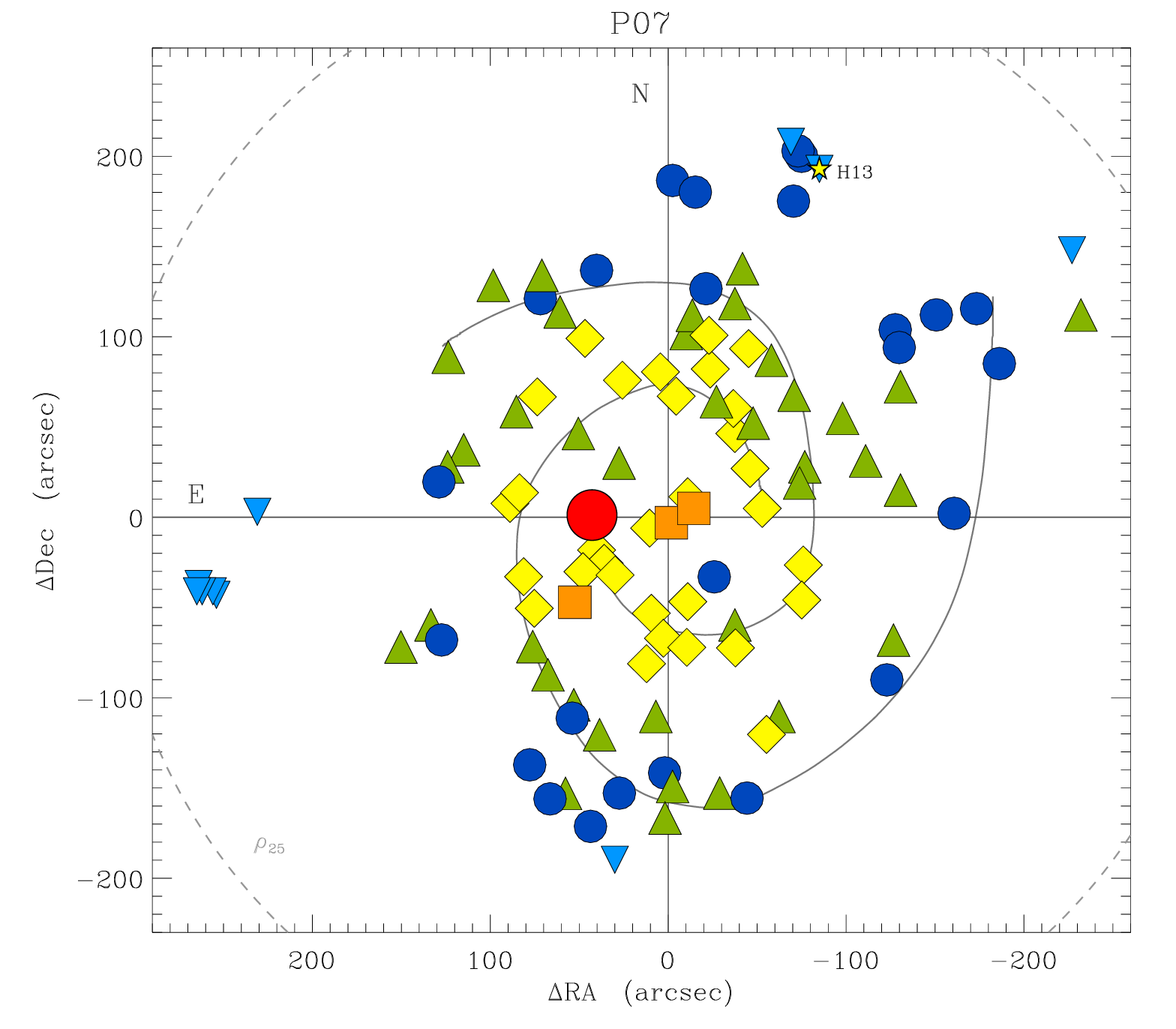}\hspace{0.1cm}
  \includegraphics[width=0.49\hsize]{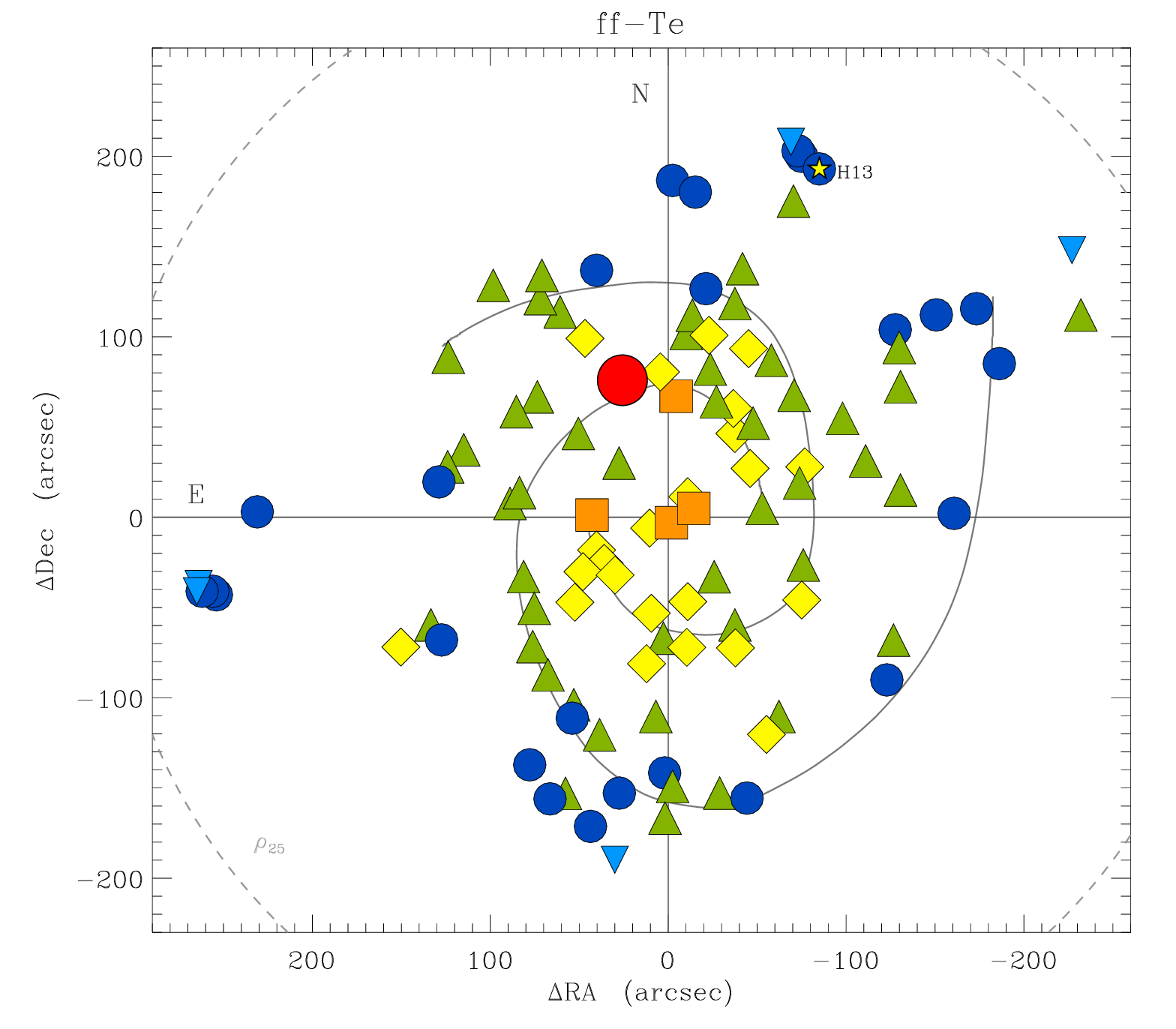}

  \caption[]
  {
    2D distribution of the oxygen abundances derived from the
    IFS \hh regions catalogue of \ngc\ (plus selected \hh regions from the
    literature), for the KK04 (top-left), $O3N2$ (top-right), P07
    (bottom-left) and ff-Te (bottom-right) metallicity calibrators.
    The shape and colours of the symbols correspond to the difference
    $\Delta$[12\,+\,log(O/H)]  $\equiv \Delta\log(\rm{O/H})$ between the abundance obtained
    on each \hh region with respect to the {\em characteristic} abundance 
    $12 + \log (\rm{O/H})_{\rho = 0.4\rho_{25}}$ of the same calibrator, grouped in bins
    of $0.0, \pm0.1, 0.2$ dex (e.g. +0.1 dex = $0.05 \leq \Delta\log(\rm{O/H}) < 0.15$).
    The large symbol in red colour stands for the location of the \hh region
    with the maximum amount of 12\,+\,log(O/H) measured for that calibrator. 
    The position of the H13 region mentioned in the text is shown as a
    yellow-star.
    The grey-thick lines define the {\em operational} spiral arms of the
    galaxy. The dotted-circle corresponds to the size of the optical radius
    $\rho_{25}$. See the text for more explanation.
    \label{fig:map1}
  }
\end{figure*}


The 2D spectroscopic view of \ngc\ offer us the opportunity to go beyond
a simple radial abundance gradient, we are now in the position to test
whether the metal abundance distributions in the disc are axisymmetric (factor
which is usually taken for granted in chemical evolution models), to locate
chemical enhanced regions, areas in which the physical conditions of the gas
differ with respect to morphologically similar or axisymmetric regions, and the
possible relations of these properties with dynamical features.
\ngc\ is a nearly face-on galaxy ($i=24$\degree) with a well-defined disc,
geometric variations of its physical and chemical properties might be sought
by partitioning the area of the galaxy in different sectors and/or by
comparing the trends and results of the different abundance indicators at
difference sectors.

For this analysis, we took as a base the spectroscopic data corresponding
to the IFS \hh regions catalogue of \ngc\ plus \hh regions from the literature
with $\rho \gtrsim \rho_{25}$. The positions of the \hh regions were
deprojected in order to account for any geometric bias imposed by the small
inclination of the galaxy.
\autoref{fig:map1} shows a 2D distribution of the oxygen abundances 
for the four strong-line abundance calibrators considered in
the previous section. The shape-colour coding corresponds to the difference
$\Delta$[12\,+\,log(O/H)]  $\equiv \Delta\log(\rm{O/H})$ between the abundance
measured on each \hh region with respect to the {\em characteristic} abundance
$12 + \log (\rm{O/H})_{\rho = 0.4\rho_{25}}$ of the same calibrator as listed in
\autoref{tab:n628_grad1} (which nearly coincides with the metallicity derived
from the integrated spectrum of the galaxy), grouped in bins of $0.0 \pm0.1, 0.2$ dex
(e.g. +0.1 dex = $0.05 \leq \Delta\log(\rm{O/H}) < 0.15$).
The large circle in red colour stands for the location of the \hh region
in which the maximum metallicity was measured for that calibrator.
The figure shows the {\em operational} spiral arms as defined in
\autoref{fig:n628_id} (grey-thick lines), the optical size of the galaxy
$\rho_{25}$ (dotted-circle) and as a yellow-star the position of the H13
region described above.
This representation of the abundance distribution might be particularly useful in
order to examine the 2D metallicity structure as derived by a particular
abundance calibrator, to detect chemical enhanced regions, or to compare the
metallicity distribution with the morphology of the galaxy and with other
metallicity indicators, bearing in mind that this can be performed only in
statistical terms, given the well-known uncertainties of the strong-line
abundance indicators. Nevertheless, this rendering may still be useful when
describing the overall structure of the 2D metallicity distribution for a
given calibrator.

Direct comparison between these abundance renderings show that the 2D
metallicity structure varies between the different metallicity calibrators. 
Whereas we find many regions with $\Delta\log$(O/H) = +0.2 dex with the KK04 and
$O3N2$ indicators, the P07 and ff--$T_e$ calibrations show very few (which
explains the flatter slopes of the two latter distributions on
\autoref{fig:grad_HII}).
An important point to note is that in all cases the maximum log(O/H) value is
not found at the centre of the galaxy, in the case of KK04, $O3N2$ and
P07, the location of the maximum abundance it is displaced $\sim45$ arcsec
($\sim2$ kpc) to the east with respect to the morphological centre of the
galaxy, while for the f-$T_e$ calibrator the point of maximum abundance is
different, being found some $\sim60$ arcsec ($\sim3$ kpc) north of the centre.
The distribution of regions with similar $\Delta\log$(O/H)
are also different, in the case of the KK04 calibrator the regions of
enhanced abundance are located at the central south-east region of the galaxy,
while in the case of the $O3N2$ calibrator, the enhanced regions are
morphologically linked to the central sections of both spiral arms. 
In the case of the P07 and ff--$T_e$ calibrators, the 2D abundance structures
are somewhat similar (with the exception of the maximum log(O/H) \hh region).
Note that the 2D rendering of the abundance distribution of \ngc\ has
shown information that is usually lost in a simple radial abundance gradient,
and that might be relevant when constructing a chemical evolution model based
on a particular abundance determination.

As a second exercise using the 2D information of \ngc, we divided the galaxy
in four quadrants with respect to an arbitrary axis, considering the abundance
analysis of \autoref{sec:hii} for each sector separately. Then, we looked
for variations or dissimilar trends between the different sectors in diagnostic
diagrams, line-ratios as a function of deprojected galactocentric radii, and metal
abundance gradients for different calibrators. We performed this exercise for
different values of the position angle (PA$_0$) of the arbitrary axis from which the
galaxy quadrants were defined, ranging from PA$_0$=0 to 90\degree, in 5\degree
steps.

\begin{figure*}
  \centering
    \includegraphics[width=\textwidth]{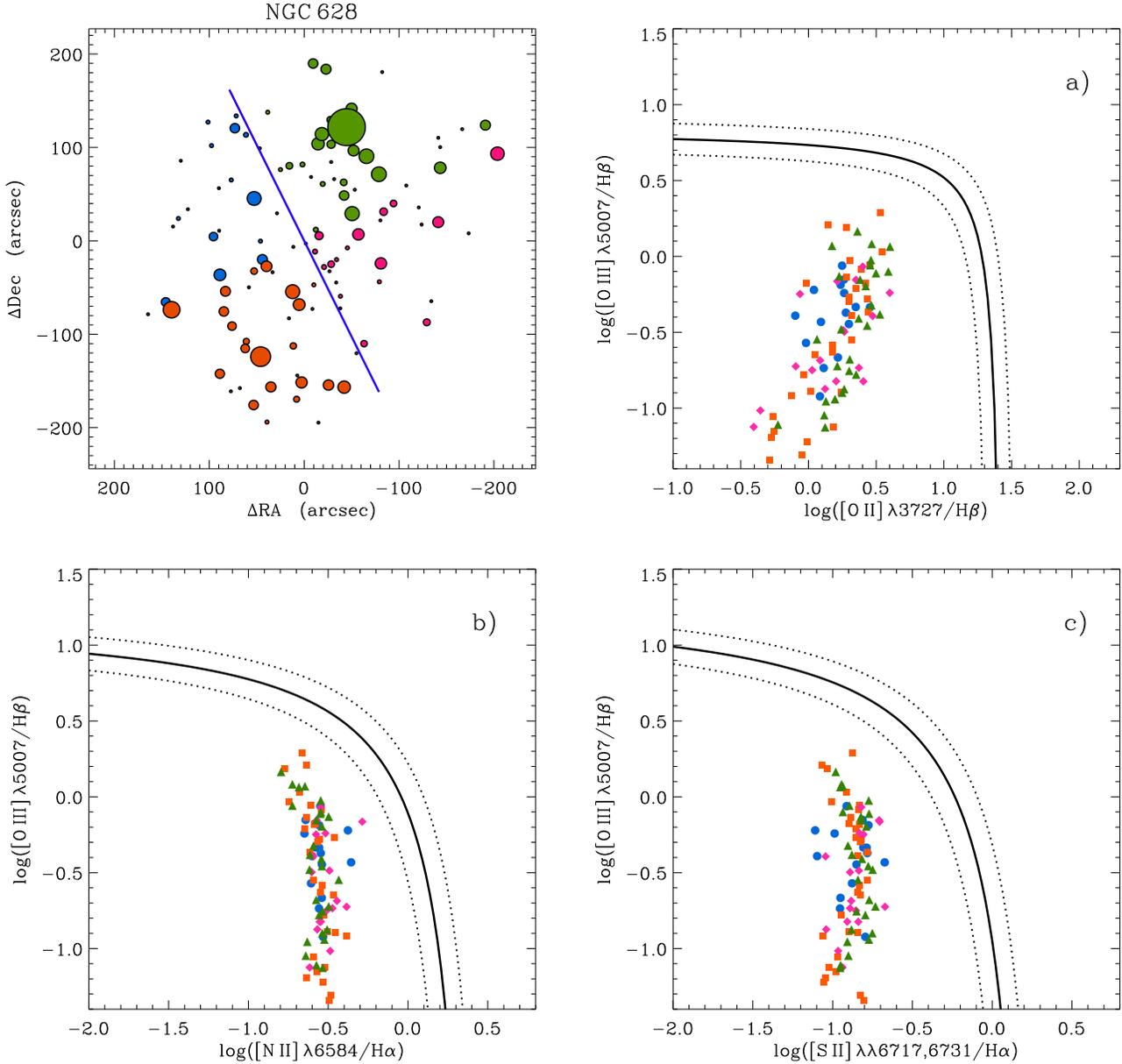}
    \caption[]{
     Top-left panel: deprojected spatial distribution of the selected \hh
     regions of \ngc\ grouped according to their geometric position
     (quadrant) with respect to the blue axis drawn across the galaxy
     surface (PA$_0$ = 25\degree). \hh regions belonging to the 1st quadrant (i.e. 0\degree $\le$
     PA $<$ 90\degree with respect to the main axis) are coded as blue, regions
     in the 2nd quadrant (90\degree $\le$ PA $<$ 180\degree) are drawn in
     orange, regions of the 3rd quadrant (180\degree $\le$ PA $<$ 270\degree)
     in magenta, and those of the 4th quadrant (270\degree $\le$ PA $<$
     360\degree) in green. The size of the circles defining an individual \hh
     region corresponds to the ``aperture'' dimension as defined in
     \autoref{sec:grad}.
     Panels a), b) and c) show diagnostic diagrams for these \hh regions
     coded according to the quadrants as follows: 1st: blue-circles, 2nd:
     orange-squares, 3rd: magenta-diamonds, 4th: green-triangles.
     \label{fig:angle1}
    }
\end{figure*}

The top-left panel of \autoref{fig:angle1} shows the deprojected spatial
distribution of the selected \hh regions of \ngc. The size of the circles
defining an individual \hh region corresponds to the ``aperture'' dimension as
defined in \autoref{sec:grad}. The \hh regions have been grouped in different
colours corresponding to the geometric position (quadrant), with
respect to the axis at which the more noticeable variations between the physical
properties of the different sectors were found (see below). The PA$_0$ of this
particular axis is 25\degree (drawn across the galaxy surface). \hh regions
belonging to the first quadrant (i.e. 0\degree $\le$ PA $<$ 90\degree with
respect to the main axis PA$_0$) are coded as blue, regions in the second quadrant
(90\degree $\le$ PA $<$ 180\degree) are drawn in orange, regions of the third
quadrant (180\degree $\le$ PA $<$ 270\degree) in magenta, and those of the
fourth quadrant (270\degree $\le$ PA $<$ 360\degree) in green.

Panels a), b) and c) of \autoref{fig:angle1} show similar diagnostic diagrams as
those introduced in \autoref{sec:fxf}. The colour of the regions correspond to the
different quadrants as explained above, different symbols have been used in
order to easily discriminate between the different sets of data (1st quadrant:
circles, 2nd quadrant: squares, 3rd quadrant: diamonds, 4th quadrant:
triangles). In the case of these diagnostic diagrams, the locus of different
sectors do not show a clear trend or do not populate a clearly visible region on
any diagram, compared to the rest of the quadrants. Points from all the regions
are equally distributed within the cloud of points on each diagnostic diagram.
However, some minor features are noticeable in some cases, for example in
Panel a) the regions belonging to the 1st quadrant (blue-circles) do not seem to
show gradient in the value of \oii\ \lam3727/\hb\ as the rest of the sectors,
i.e. the log(\oii/\hb) ratio for the 1st quadrant shows a
restricted range of values ($\sim0.0$ to 0.5 dex), while for the rest of
the sectors, the same ratio ranges for more than an order of magnitude
(log(\oii/\hb) $\sim -0.5$ to 0.7 dex). The range of \oiii
\lam5007/\hb\ values for the 1st quadrant is also shorter, compared to the rest
of the sectors, comprising only values between log(\oiii/\hb)
$\sim -0.9$ to 0.0 dex, while for example, the 2nd and 4th quadrants range from
$\sim -1.5$ to 0.3 dex. One possible reason of the restricted range in the
\oii\ and \oiii/\hb\ ratios for the 1st quadrant might be a simple statistical
bias, considering the relative low number of \hh regions sampled within that
sector (16), compared to the regions in the 2nd and 4th quadrants (33 and 30
respectively), and that the range of galactocentric distances of the \hh regions
within this sector is restricted, excluding regions close to the nucleus of the
galaxy (with typically low \oii\ and \oiii/\hb\ ratios) and far for it (with
higher ratios of the same lines).

\begin{figure*}
  \centering
    \includegraphics[width=\hsize]{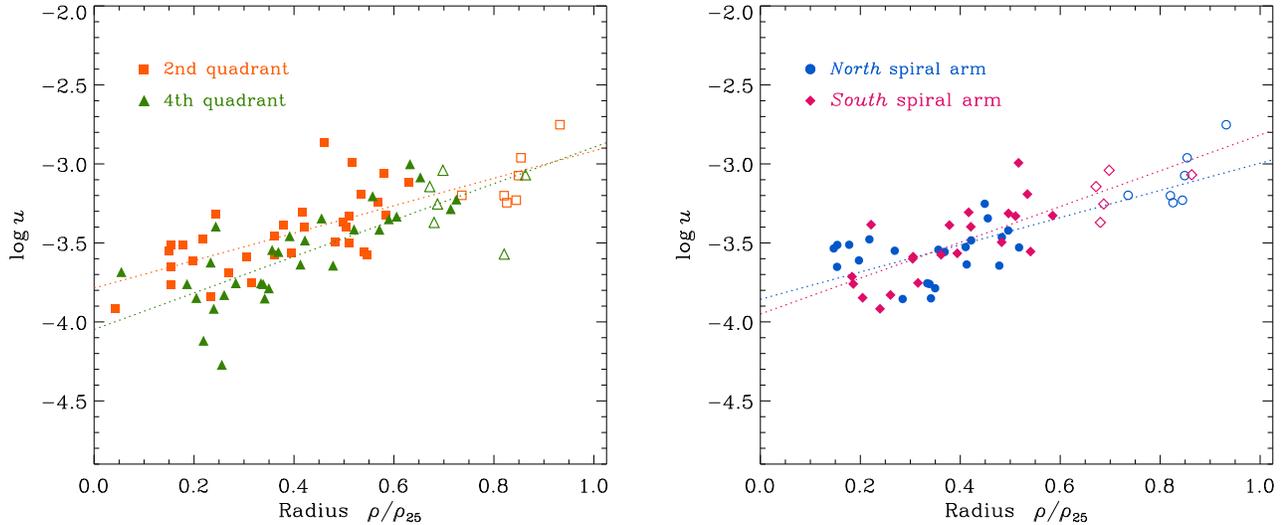}
    \caption[]{
      Radial gradients of the ionization parameter $\log u$
      for geometrical and morphologically selected \hh regions of
      \ngc. The left panel shows the $\log u$ vs. $\rho$ relation for the regions
      of the 2nd and 4th quadrants of the galaxy as defined in
      \autoref{fig:angle1}. The right-panel shows the same relation for the regions
      belonging to the {\em North} and {\em South} spiral arms as defined in the
      text.
      \label{fig:arms1}
    }
\end{figure*}

As mentioned above, the \nii\ \lam6584/\ha\ and \sii\ \lam\lam6717,31/\ha\ ratios
shown in Panels b) and c) do not show any clear trend among the different
sectors, apart from the difference in the oxygen ranges discussed before. Other
typical diagnostic diagrams were explored with similar finding, i.e. the locus
of the different sectors are spread among the overall scatter showing no clear
trends. Following the course of action of the previous sections, we explored
possible variations among the different sectors in the emission line intensities
and derived properties as function of galactocentric radius, similarly to
\autoref{fig:radial_fxf} and \autoref{fig:radial_HII}. 
The most discernible variation was found in terms of the ionization parameter
$\log u$ (which is equivalent to the {\em excitation diagnostic} \oiii/\oii).
The left-panel of \autoref{fig:arms1} shows the values of the ionization
parameter as function of galactocentric radius for those \hh regions within
the 2nd and the 4th quadrant only. The other sectors were not included for the
sake of clarity and in order to stress the differences between these two samples.
Furthermore, the 2nd and 4th quadrants include most of the largest and brightest
\hh regions in the galaxy, they are the most populated sectors in terms of the
total number of individual \hh regions, they contain the whole range of
galactocentric distances of both the IFS observations (filled symbols) and the
\hh regions from the literature which were identified to belong
to a certain section (open symbols), sampling geometrically symmetric regions
of the galaxy [see Panel a) of \autoref{fig:angle1}].

Visually, the values of the ionization parameter found in the 2nd quadrant seem
to be higher than those found in the 4th sector, and with a different level of
variation as function or galactocentric radius, i.e. the gradient of the
ionization parameter on the 4th quadrant is steeper than one found at the
opposite side of the galaxy. The dotted lines in
\autoref{fig:arms1} represent a linear fit to each of the samples, in the
case of the 2nd sector, the value of $\log u$ at $\rho_0$ is $-3.8 \pm 0.06$,
with a slope of $0.87 \pm 0.12$ dex~$\rho_{25}^{-1}$, while for the 4th
quadrant, $\log u$ at $\rho_0$ is equal to $-4.1 \pm 0.08$, with a slope of
$1.15 \pm 0.16$ dex~$\rho_{25}^{-1}$. Although the differences in the central
$\log u$ and slope of the gradients for the two samples are not extreme, they
do indicate a different behavior of the ionization parameter in these two
opposite regions of the galaxy. In fact, a two-sided Kolmogorov-Smirnov test
applied to the distributions of the two samples yields a probability of 11\%
that two arrays of data values are drawn from the same distribution. 
The two innermost points of the figure at $\rho < 0.1\rho_{25}$ do not seem to
visually follow this behavior. However, given their proximity between them and
to the centre of the galaxy, the segregation of these regions into different 
``geometrical quadrants'' might not be relevant. Nevertheless, the inclusion
of these points does not affect the general results of the linear fitting and
the trends discussed above.

\begin{figure*}
  \centering
    \includegraphics[width=0.9\hsize]{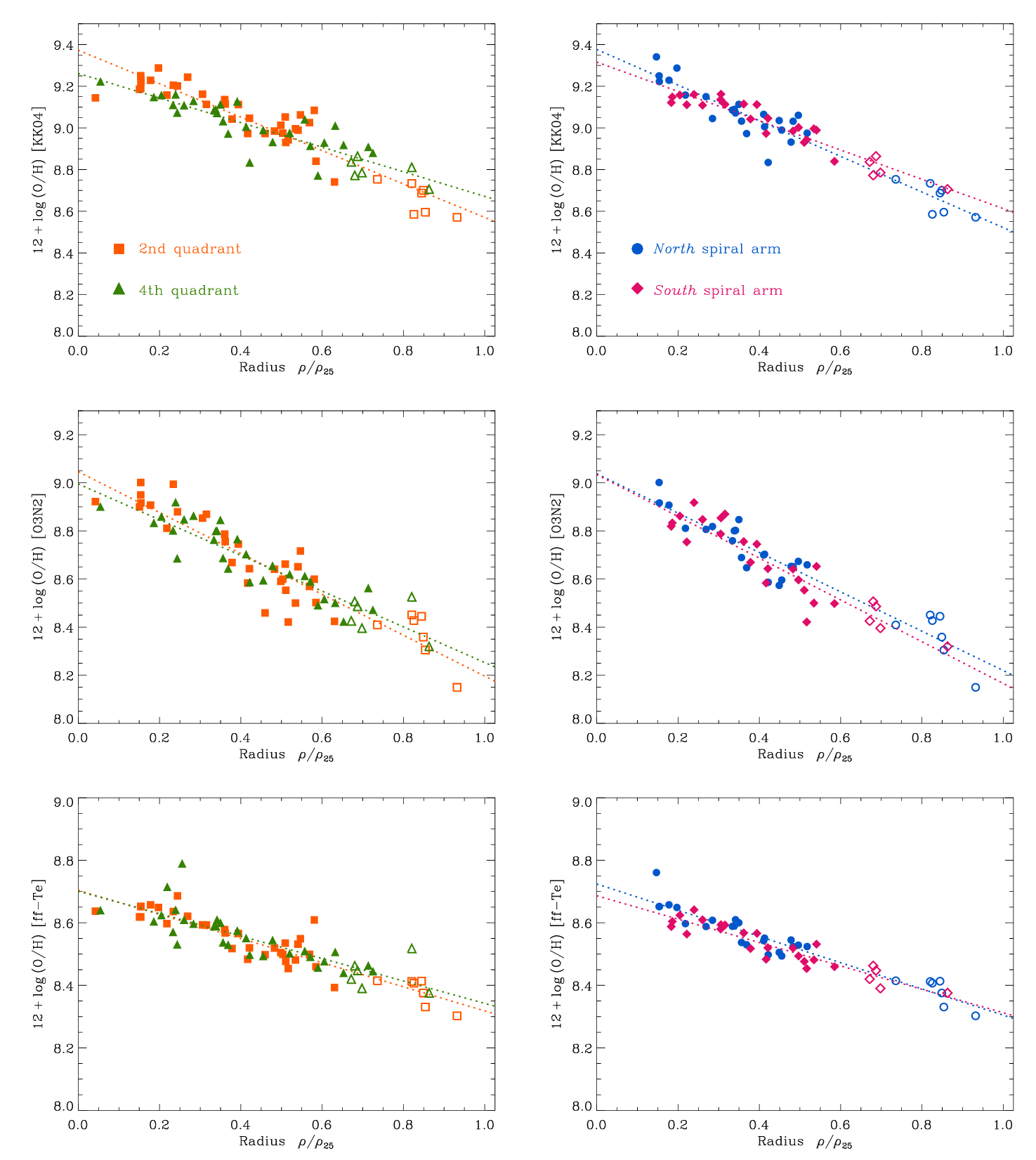}
    \caption[]{
      Radial abundance gradients of \ngc\ based on geometrical and
      morphologically selected \hh regions. Top-panels: Oxygen gradient
      derived after the $R_{23}$-based KK04
      calibrator. Middle-panels: Oxygen gradients
      based on the \citetalias{Pettini:2004p315} calibrator. Bottom panels:
      Abundance gradients based on the ff--$T_e$ method.
      In all panels, the dotted-lines correspond to linear fits to the data. 
      \label{fig:arms2}
    }
\end{figure*}

The \hh regions included in the 2nd and 4th quadrants trace a good portion of
the spiral arms of \ngc, but each of these sectors includes regions from
both of the most prominent arms of the galaxy. One logical step in the 2D
spectroscopic analysis of this galaxy is to test for variations in the abundance
and physical properties between the spiral arms of \ngc. For this purpose,
we selected the largest and brightest \hh regions along the two most prominent
spiral arms sampled by the FOV of the IFS mosaic of the galaxy, as defined in
\autoref{fig:n628_id}.
For the {\em North} arm, we chose the \hh regions N628~6-9, 26-31, 35-43, 45, 56, 68 and
69, while for the {\em South} arm, we selected regions
N628~12-15, 20, 22, 60, 65, 66, 74, 80-85, 87-88, 90-91 and 94. 
The regions in each spiral arm were selected to sample the closest coincident range in
deprojected galactocentric distances, we also excluded regions near the centre
of the galaxy in order to avoid confusions in the association to any specific arm. 
We also considered \hh regions from the literature found to belong to a
certain spiral arm, although we only included those beyond
the FOV of the IFS data (open symbols).
The results of the abundance analysis for each spiral arm were compared in
a similar way as the analysis for the four quadrants defined above. Diagnostic
diagrams, emission line ratios and derived properties were explored, seeking for
any variation between the two new samples. Again, the most notorious difference
was found in the values of the ionization parameter as a function of radius, as
shown in the right-panel of \autoref{fig:arms1}. The regions belonging to the 
{\em North} spiral arm are coded as blue-circles, while regions from the 
{\em South} arm are drawn as reddish-diamonds.

The difference between the two spiral arms clearly resides in the slope of the
gradient of $\log u$, for the {\em North} arm, the values of $\log u$ increase
moderately with galactocentric distance, while for the {\em South} arm, the
ionization parameter increases with a steeper slope. The dotted-lines
represent linear fits to the data, for the {\em North}
arm the $\log u$ at $\rho = 0$ is $-3.86 \pm 0.06$ , with a slope of $0.86 \pm 0.12$
dex~$\rho_{25}^{-1}$. On the other hand, the $\log u$ at $\rho_0$ for the {\em South}
arm is $-3.95 \pm 0.08$, with a slope of $1.13 \pm 0.17$
dex~$\rho_{25}^{-1}$. Coincidently, the linear fits of the two spiral arms
intersect at a radius $\rho \sim 0.4\rho_{25}^{-1}$, with $\log u \approx
-3.5$, i.e. the $\log u$ value of the integrated spectrum of \ngc. A two-sided
Kolmogorov-Smirnov test reports a probability of 36\% for the two samples to
be originated from the same distribution.


The positive gradient of the ionization parameter (or the \oiii/\oii\
emission-line ratio) shown in \autoref{fig:arms1} reflects an increase of the
nebular excitation with galactocentric distance. These emission-line ratios
radial trends are known since the pioneering works by \citet{Aller:1942p3955}
and \citet{Searle:1971p1962} and are subject of different interpretations. In
the scenario that the stellar excitation per nebula is more or less uniform
across the galaxy (which is unlikely given the large range of physical
parameters within extragalactic \hh regions, e.g. density, temperature, size
and the spectrum of ionizing radiation, etc.), the increase in the excitation
with radius has been interpreted as an effect of the shift of the principal
cooling mechanism, from infrared fine-structure transitions to optical \oiii\
emission in regions of lower metallicities found at larger radii, i.e. as a
consequence of the metallicity gradient \citep[see][and references
therein]{Shields:1990p3958}.
An alternative explanation of the observed gradients in excitation (and in the
EW(\hb) emission) is a gradient of the mean stellar temperature of the
ionizing source $T_*$, with the ionizing radiation becoming harder with
increasing radius
\citep{Shields:1974p1960,Shields:1976p2006,Shields:1978p3959}. 
Indeed, the effect of increase $T_*$ with decreasing abundance has been
confirmed by different authors
\citep[e.g.][]{Stasinska:1980p3960,Campbell:1986p2185,Vilchez:1988p2011}. However,
some other authors argue that this observed trend is consistent with a single
ionizing temperature and a varying nebular geometry \citep{Evans:1985p3962},
or that there is metallicity-dependent IMF which translates into a
differential star-formation at varying radii \citep{Shields:1976p2006}, etc.
More recent synthesis models include evolving clusters which
predict a softening of the stellar ionizing spectra with increasing abundance
\citep[e.g.][]{Maeder:1990p3963}, or that the variation on the strength of the
\oiii\ emission-line and the ionization parameter (which is not a
free-parameter) is due to the changes in the cluster mass and in the
cluster age, which determines the number of ionizing photons, the overall
ionization spectrum hardness and the ionized region size
\citep{MartinManjon:2010p3846}. Although is beyond the scope of this paper,
comparison of these models with the 2D IFS data of \hh regions is
necessary in order to discriminate the different possible scenarios.


The top-panels of \autoref{fig:arms2} show the abundance
gradient of \ngc\ for the 2nd and 4th sectors as defined in
\autoref{fig:angle1}, and the two spiral arms \hh regions samples introduced
previously for the $R_{23}$-based KK04 calibrator, following the same style
prescriptions as before. Linear fits
to each data sample are represented by dotted-lines. The effect of the different
gradient of the ionization parameter is evident in these two panels. In the
first case, the metallicity gradient derived for the 2nd quadrant of \ngc\
shows slightly higher oxygen values than those calculated for the 4th
quadrant. The central oxygen value for 2nd sector is 12~+~log(O/H) = 9.37 
$\pm 0.03$, while for the 4th sector is 9.26 $\pm 0.02$. The slope for the 2nd
quadrant is much steeper ($-0.82 \pm 0.05$ dex~$\rho_{25}^{-1}$) than for the
4th sector ($-0.58 \pm 0.06$ dex~$\rho_{25}^{-1}$). In the case of the spiral
arms, the independent metallicity gradients for each sample are also clearly
distinct. The {\em North} spiral arm shows a steeper gradient ($-0.85 \pm
0.05$ dex~$\rho_{25}^{-1}$), consistent with a higher central oxygen abundance
(12~+~log(O/H)$_0$ = 9.38 $\pm 0.02$), while the abundance gradient for the
{\em South} arm shows a flatter slope ($-0.70 \pm 0.04$ dex~$\rho_{25}^{-1}$),
a lower central abundance (12~+~log(O/H)$_0$ = 9.31 $\pm 0.03$), but most
importantly, a practically flat oxygen distribution for regions at $\rho <
0.3\rho_{25}$.

The middle and bottom-panels of \autoref{fig:arms2} show the oxygen
metallicity gradients of the samples discussed above for the $O3N2$
calibration and the ff--$T_e$ method, respectively. 
In the case of the $O3N2$ calibration, the left, middle-panel shows 
different slopes for the 2nd and 4th quadrants, with the same behavior than the
KK04 calibrator, i.e. a steeper gradient for the 2nd sector; while in the right,
middle-panel the \hh regions corresponding to the {\em North} spiral arm are
consistent with a well-defined gradient with a steep slope at all
galactocentric distances, with a coincident central abundance a nearly equal
slopes. In the case of the ff--$T_e$ method, the distribution of data points in both
cases are equally scattered at all galactocentric distances in each panel
the distributions of the \hh regions show a tight similarity, evident also
through the linear fits applied to the individual samples.
However, the {\em South} arm sample show the same behavior as in the case of
the KK04 index, i.e. a decreasing gradient for the outer regions, and a flat
distribution for the inner regions ($\rho \leq 0.35\rho_{25}$), suggesting a
flat oxygen abundance distribution of 12~+~log(O/H) $\sim$ 8.6.

In summary, although the variations between the central oxygen abundances and
slopes for both, the geometrical (quadrants) and morphological (arms) regions of the
galaxy fall within the expected errors of strong-lines empirical calibrators,
the radial trends in the ionization parameter and metallicity abundances are
somewhat distinct, indicating that to a certain extent, the physical conditions
and the star formation history of different-symmetric regions of the galaxy have
evolved in a slightly different manner.

\section{Summary and conclusions}
\label{sec:fin}

The IFS analysis of \ngc\ was taken as a case study in order to explore 
different spectra extraction and analysis methodologies, taking into account the
signal-to-noise of the data, the 2D spatial coverage, the physical meaning of the
derived results, and the final number of analysed spectra.
The proposed methods differ mainly in the way to select a subsample of
spectra from the IFS mosaics, from which a spectroscopic analysis can be later
performed.
The selection criteria of the first method, the \fxf\ sample, proved to select
regions with good quality spectra from which different physical properties of
the gas were derived, although with a high level of scatter. This method
was also able to identify regions of interstellar {\em diffuse} emission.
The second extraction method consisted in creating a catalogue of
``classical'' \hh regions from a purely geometrical principle, i.e. by
co-adding (by visual selection) fibres considered to belong to the same
morphological region.
Quality and sanity checks were applied in both methods, seeking for only
those regions with meaningful spectral features. 
A combination of these methods was used in order to perform the 2D
spectroscopic analysis of \ngc. The locus of the selected spectra was
explored by employing typical diagnostic diagrams of reddening corrected
emission line ratios.


General results inferred from the different spectroscopic selection methods are
the following:

\begin{enumerate}
\item 
Despite the large number of spectra contained in the original observed mosaic of
\ngc, the final number of fibres containing analysable spectra of enough
signal-to-noise for a spectroscopic study of the ionized gas
represents only a reduced percentage of the total number of fibres contained
in the full IFS mosaic. For this particular case, less than 10\% of the total
area sampled by the IFU observations is considered of sufficient quality.

\item
The ``non-physical'' \oiii\ \lam5007/\lam4959 line ratios found in the final
samples of the \fxf\ and \hh region catalogues are due to
errors introduced by the subtraction of the stellar population contribution to the data,
the effect is especially important in regions of weak emission in the \oiii\
lines (e.g. the centres of the galaxies). Therefore, although the \oiii\ ratio
may not be close to the theoretical value, the combination of their derived
line ratios are still representative of the physical conditions and
metallicity abundance.

\item
The comparison between the line-ratio trends derived from the \fxf\
sample and those of the \hh region catalogue shows that, the selection criteria
applied to the former case resulted in regions that follow, in statistical
basis, exactly the same patterns and trends provided by the more
``refined'' \hh region sample, although with a larger level of scatter.

\item
In both cases, the selected spectra were consistent with emission
produced by ionization due to a thermal continuum, i.e. hot OB stars. The radial
variation of the $R_{23}$ metallicity indicator, the ionization parameter and
the \nii\ \lam6584/\ha\ ratio were also analysed, showing clear trends.

\end{enumerate}

It is important to mention that, during the analysis process described in this
paper, special attention was paid in order to avoid an over-automatization that
could lead to errors in the derived physical parameters from the resultant
spectra. Given the large amount of data, some problems might be
expected for individual fibres, pointings or regions, or in the derived
quantities from the measured line intensities. However, all the quality checks
performed to the data (and the results derived from them) suggest that the
methodology implemented in this work is very advantageous for a statistical and
comparative study, and when dealing with a large number of spectra.

Once the quality and physical meaning of the spectra samples were verified, we
proceeded to perform a chemical abundance analysis of the nebular emission of
the galaxy. A suite of strong-line metallicity indicators were employed in order to
derive the oxygen abundance radial gradients of \ngc\ for the different sets of
spectra. The results were compared among them in a statistical basis. \hh
regions from the literature were incorporated to the PINGS \hh region
catalogue in order to increase the baseline for the derivation of the abundance gradients.
Likewise, the emission line ratios and metallicity abundance of 9 \hh regions
from the literature within the FOV of the IFS mosaic of \ngc\ were compared to
the observations of this work, with an overall agreement.
Furthermore, a ``classical'' abundance gradient determination was also performed, by
selecting large and bright \hh regions, the results are consistent (within the
errors) with the values derived from the analysis of the full \hh region
distribution.

The most important results derived from the 2D IFS emission-line analysis of
\ngc\ are the following:

\begin{enumerate}

\item
By comparing the results of the different abundance calibrators, we found that
the metallicity distribution of \ngc\ is consistent with a nearly
flat-distribution in the innermost regions of the galaxy ($\rho/\rho_{25} <
0.2$), a steep negative gradient for $0.2 \lesssim \rho/\rho_{25} < 1$, and a
shallow or nearly-constant distribution beyond the optical edge of the galaxy,
i.e. implying a  {\em multi-modality} of the abundance gradient of \ngc. The
same feature is observed for the N/O vs. $\rho$ distribution.
The existence of this feature may be related to the differences in the 2D gas
surface density and star formation rate between the inner and outer disc which
inhibits the formation of massive stars in the outer regions, causing a lack of
chemical evolution in the outer disc compared with the inner regions.

\item
The observed dispersion in the metallicity at a given radius for the \fxf\
sample is neither a function of spatial position, nor due to low S/N of the
spectra, and shows no systematic dependence with the ionization
conditions of the gas, implying that the dispersion is real, and is reflecting
a true spatial physical variation of the oxygen content.

\item
Considering that the observed \ha\ luminosity corresponds to that expected for
relatively large \hh regions (in which the flux of ionizing photons is
produced by stellar clusters with the necessary mass to overcome problems due
to stochastic effects), the determination of the metallicity abundance
through the strong-line indicators can be considered reliable for both the
\fxf\ and \hh region samples, as the derived metallicity does not
depend importantly on the ionization conditions of the emitting gas.

\item
The values of the oxygen abundance derived from the integrated
spectrum for each calibrator equal the abundance 
derived from the radial gradient at a radius 
$\rho \sim 0.4\rho_{25}$, confirming for this galaxy
previous results obtained for other objects, i.e. that the integrated
abundance of a normal disk galaxy correlates with the {\em characteristic}
gas-phase abundance measured at $\rho \sim 0.4\rho_{25}$.

\item
By constructing 2D maps of the oxygen abundance distributions, we
found that the 2D metallicity structure of the galaxy varies depending on the
metallicity calibrator employed in order to derive the oxygen
abundance. Different calibrators find regions of enhanced log(O/H) at spatial
positions which are not coincident among them.
This implies that the use of different empirical calibrations do not only
reflect in a linear scale offset, but may introduce spurious inhomogeneities.
This information is usually lost in a simple radial abundance gradient,
and that might be relevant when constructing a chemical evolution model based
on a particular abundance determination.

\item
While trying to find axisymmetric variations of the metallicity content in the
galaxy, we found slight variations between the central oxygen abundances and
slopes for both, the geometrical (quadrants) and morphological (arms) regions of the
galaxy. Although these small variations fall within the expected errors
involved in strong-line empirical calibrations, the radial trends in the ionization
parameter and metallicity abundances are somewhat distinct, indicating that to
a certain extent, the physical conditions and the star formation history of
different-symmetric regions of the galaxy have evolved in a different manner.

\item
The metallicity of the gas seems to follow the trend marked by previous IFS
studies of the stellar populations at the centre of the galaxy (see \citetalias{Sanchez:2011p3844}).
This result, together with the presence of circumnuclear star-forming regions
might indicate a scenario in which the gas is being radially transferred,
inhibiting the enhancement of the gas-phase metallicity at the innermost regions
of the galaxy.

\end{enumerate}

Hitherto, most spectroscopic works have obtained abundance gradients by
observing only a few, large and bright \hh regions over the surface of the
galaxies. However, the body of results presented in this work were obtain (to a
good extent) thanks to the fact that we benefit from a nearly complete coverage
of small and large \hh regions observed over the surface of the galaxy, which
gives us the possibility to explore the systematics between different ways of
obtaining the abundance gradients.
In particular, the results from this paper give support to the existence of flat
metallicity distribution in the inner and outer parts of the discs of
nearby spiral galaxies, although the definite existence (and interpretation) of
these attributes will require further observations and a larger statistical
sample. The capability to detect these features represent one of the powers of
IFS observations. Likewise, the results also contribute to reaffirm (in a
statistical way) the robustness of the strong-line methods applied to
individual regions sampled by different fibre apertures, and to the integrated
spectra of the whole galaxy.

The study of the nebular properties of \ngc\ presented in this paper has been
restricted to include a 2D abundance analysis, focused on the spectra selection
method, on the abundance gradients and variations of nebular derived properties
across the surface of the galaxy. More detailed studies regarding the 2D
distribution of the physical and chemical properties of this galaxy will be
addressed in future papers of this series.

\section*{Acknowledgments}

FFRO would like to acknowledge the Mexican National Council for Science and
Technology (CONACyT), the Direcci\'on General de Relaciones Internacionales
(SEP), Trinity College, the Cambridge Philosophical Society, and the Royal
Astronomical Society for the financial support to carry out this research.
AD and SFS thank the Spanish Plan Nacional de Astronom{\'i}a programme
AYA\,2007-67965-C03-03 and AYA\,2010-22111-C03-03 respectively.
SFS acknowledges the ICTS-2009-10 MICINN program.
We would like to thank the referee Christophe Morisset for the very valuable
comments and suggestions which improved the final content of this paper.

\bibliographystyle{mnras}
\bibliography{mnras}

\appendix

\section{On the spectra extraction methodology}
\label{app:methods}


In this appendix, we give further information regarding the \fxf\ spectra
extraction technique that was implemented during the IFS analysis of \ngc.
This method considers that the fibre aperture of the PPAK instrument samples a
region which is large-enough in physical scale to be the source of an 
{\em analysable} spectrum.
As explained in \autoref{sec:fxf}, the spectra extraction for the \fxf\ method
was performed on {\em residual} mosaic obtained after discarding all those
fibres with low S/N and foreground objects. The sample selection was split in
three steps, each with different quality criteria conditions based on
different assumptions, explained as follows:

\begin{enumerate}

\item An analysable spectrum would need to include several detected lines in
  order to perform a basic analysis. One obvious line is \hb, as the
  typical line ratios used in any spectroscopic analysis are normalised to the
  flux intensity of this line, and as it is required to derive a first-order
  correction for interstellar extinction based on the H$\alpha$/\hb\ ratio.
  Furthermore, a well-defined region from which we could derived physical
  meaningful properties would have to include both the \oiii\ \lam4959
  and \lam5007 lines. The \lam4959 line is weaker than the \lam5007 by a
  theoretical factor of 2.98 \citep{Storey:2000p3365}, therefore the detection of
  the \oii\ \lam4959 line would also assure the detection of the \lam5007
  line. It is important to note that the detection of the \lam4959 and
  \lam5007 lines does not imply necessary their correct measurement, as
  discussed in \autoref{sec:fxf}. Given the theoretical constrain on the observed
  ratio of these two lines, the requirement of the detection of \oiii\ \lam4959
  (and consequently of \lam5007) will help to characterise the quality of the
  subsample spectra and their physical meaning.
  Therefore, the first criterion applied to the {\em residual} mosaic was to select
  those fibres where both, the \hb\ and \oiii\ \lam4959 line intensities were
  greater than zero, i.e. meaning that the lines are detected in the automated
  line intensity calculation.
  In the case of \ngc, the total number of fibres for which this criterion
  was fulfilled is 2659, i.e. 38\% of the 6949 fibres contained in the {\em
  clean} mosaic and 20\% of the original number of fibres in the observed,
  unprocessed mosaic.

\item During the data processing of the line intensities and their subsequent
  manipulation into reddening corrected line flux ratios, there were found
  problems of non-floating numbers among the thousands of derived figures.
  A careful inspection showed that the reason of these values was due to an
  incorrect determination of the logarithmic extinction coefficient
  c($H\beta$) (calculated from the H$\alpha$/\hb\ ratio, accordingly to the
  prescriptions describe in \citetalias{Sanchez:2011p3844}), 
  which produced Not-a-Number (NaN) or infinite values in those cases in which
  the intensity of the \hb\ line was very close to zero.
  Therefore, the second selection criterion considered only those fibres for
  which the calculated c($H\beta$) value was a finite-floating number,
  regardless of its value (including negative, non-physical ones).
  The number of fibres with non-finite c($H\beta$) values in \ngc\ accounts
  for 97 spectra, reducing the number of selected fibres after this step to 2562.


\item After applying the previous selection criteria, the data
  subsample consisted in a set of fibres with line intensities greater than zero
  for \hb\ and \oiii\ \lam4959 and finite values of the derived c(H$\beta$).
  Based on the experience learned from the emission line maps analysis, an
  additional signal-to-noise cut had to be applied in order to obtain a
  subsample of spectra from which meaningful information could be obtained. As
  in the previous case, the flux threshold was based on the line
  intensity of the \hb\ line. The \hb\ flux cut was chosen instead of
  the probably more common \ha\ because, as experience with the data
  manipulation proved, in some cases a certain line intensity threshold on
  \ha\ did not mean the correct detection and measurement of \hb,
  and a high cut in \ha\ would eliminate many regions of low intensity
  but with physical meaning.
  Furthermore, as the main focus is to characterise the chemical abundance of
  the galaxy, we required the presence on the spectrum of typical strong lines
  from which we could obtain information on the abundance for a given
  region. Thus, a further requirement applied at this point to the spectra was
  the detection of the \oii\ \lam3727 line, given that many of the most important
  abundance calibrators are based on the line strength of this line
  (e.g. $R_{23}$, see \autoref{sec:grad}).
  Therefore, the last selection criteria applied to the spectra sample obtained
  in the previous steps was, to select those fibres where the line intensity of
  the \hb\ line was greater than or equal to a given flux limit threshold,
  {\em and} that the line intensity of \oii\ \lam3727 was greater than zero,
  i.e. the emission line was detected. 
  The flux limit applied in \hb\ was equal
  to 8 $\times$ 10$^{-16}$ \flux. The final number of spectra after applying
  the last selection criteria was 376 fibres, which constitutes only $\sim$ 3\%
  of the original spectra in the IFS mosaic of \ngc.

\end{enumerate}

Even though all the selection criteria discussed above could had been merged
into a single pipeline, we decided to separate the different steps for the
following reasons: 1) In order to check the number of fibres kept and removed at
each step; 2) For an easier manipulation of the data in computational terms; 
3) The different data sets could be analysed independently in order to check for
systematic errors and trends due to the different quality selection criteria;
4) The subsample obtained after the second selection criteria was very likely
to include regions of line emission of different physical properties of
classical \hh regions, e.g. regions of DIG.

\section{Online material}
\label{app:online}

In this appendix, we list the data products that are made publicly available
in online version as part of the IFS analysis of \ngc\ presented in this
paper. The data can be accessed through the PINGS project webpage: 
\url{http://www.ast.cam.ac.uk/ioa/research/pings}, under the {\sc Public datasets}
section. The data release includes: 

\begin{enumerate}

\item The \hh region catalogue of \ngc, including the number identification
  (ID) as shown in \autoref{fig:n628_id}; the Right Ascension and
Declination of the central reference point of each \hh region (in sexagesimal
and degree units), for the 2000 equinox; the offsets of each region, in arcsec,
with respect to the reference central point of the IFS mosaic, as reported in
Table 1 of \citetalias{RosalesOrtega:2010p3836}, in the standard configuration
with north-east positive; a ``Method'' column corresponding to the extraction
method employed during the selection of the \hh region (see
Appendix~\ref{app:methods}): {\bf A} stands for an {\em aperture} extraction
(i.e. a circular aperture of given size in arcsec), and {\bf M} stands for
{\em manual} extraction (i.e. fibres selected by hand); an ``Aperture'' value
standing for the real (aperture) or equivalent (manual) extraction aperture
diameter in arcsec; a ``Size'' column corresponding to the physical size of
the \hh region in parsecs, at the assumed distance to the galaxy, and the
total number of fibres from which the spectra of the \hh regions were
extracted.

\item Reddening-corrected emission line intensities for the \hh region
  catalogue, normalised to \hb\ (with 1$\sigma$ errors),
  including: \oii\ \lam3727, \hg\ \lam4340, \hb, \oiii\ \lam4959, \oiii\ \lam5007,
  \hei\ \lam5876, \nii\ \lam6548, \ha, \nii\ \lam6584, \sii\ \lam6717.

\item Derived properties of the \hh region catalogue, including: A$_V$, the 
  logarithmic extinction coefficient c(\hb) derived from the Balmer decrement,
  \hb\ flux in units 10$^{-16}$ erg s$^{-1}$ cm$^{-2}$, the \ha\ luminosity,
  the ionization parameter $\log u$, the $R_{23}$ value, and the 12~+~log(O/H)
  values for the four strong-line metallicity calibrators (KK04, $O3N2$,
  ff--$T_e$, P07).

\end{enumerate}

%





\label{lastpage}

\clearpage

\end{document}